%% file: ms.tex
\documentclass[12pt,preprint]{aastex}
\begin{document}
\shorttitle{$E_{\rm p}$ Evolution Patterns of GRBs} \shortauthors{Lu et al. }
\title{A Comprehensive Analysis of Fermi Gamma-ray Burst Data: II. $E_{\rm p}$-Evolution Patterns and Implications for the Observed Spectrum-Luminosity Relations}
\author{Rui-Jing Lu\altaffilmark{1}, Jun-Jie Wei\altaffilmark{1}, En-Wei Liang\altaffilmark{1,2,3}, Bin-Bin Zhang\altaffilmark{4}, Hou-Jun L\"{u}\altaffilmark{3}, Lian-Zhong,  L\"{u}\altaffilmark{1}, Wei-Hua Lei\altaffilmark{5,3}, and Bing Zhang\altaffilmark{3}}
\altaffiltext{1}{Department of Physics and GXU-NAOC Center for
Astrophysics and Space Sciences, Guangxi University, Nanning 530004,
China; lew@gxu.edu.cn} \altaffiltext{2}{The National Astronomical
Observatories, Chinese Academy of Sciences, Beijing 100012, China}
\altaffiltext{3}{Department of Physics and Astronomy, University of
Nevada, Las Vegas, NV 89154; zhang@physics.unlv.edu}
\altaffiltext{4}{Department of Astronomy and Astrophysics,
Pennsylvania State University, University Park, PA 16802}
\altaffiltext{5}{School of Physics, Huazhong University of Science and Technology,
Wuhan, 430074, China}
\begin{abstract}
We present a time-resolved spectral analysis of 51 long and 11 short
bright GRBs observed with the {\em Fermi}/GBM, paying special
attention to $E_{\rm p}$ evolution within a same burst. Among 8
single-pulse long GRBs, 5 show hard-to-soft evolution, while 3 show
intensity-tracking. The multi-pulse long GRBs have more complicated
patterns. Statistically, the hard-to-soft evolution pulses tend to
be more asymmetric than the intensity-tracking ones, with a steeper
rising wing than the falling wing. Short GRBs have $E_{\rm p}$
tracking intensity exclusively with the 16ms time resolution
analysis. We performed a simulation analysis, and suggest that at
least for some bursts, the late intensity-tracking pulses could be a
consequence of overlapping hard-to-soft pulses. However, the fact
that the intensity-tracking pattern exists in the first pulse of
multi-pulse long GRBs and some single-pulse GRBs suggest that
intensity tracking is an independent component, which may operate in
some late pulses as well. For the GRBs with measured redshifts, we
present a time-resolved $E_{\rm p}-L_{\gamma, \rm iso}$ correlation
analysis and show that the scatter of the correlation is comparable
to that of the global Amati/Yonetoku relation. We discuss the
predictions of various radiation models regarding $E_{\rm p}$
evolution, as well as the possibility of a precessing jet in GRBs.
It seems that the data pose great challenge to all these models, and
hold the key to unveil the physics of GRB prompt emission.

\end{abstract}

\section{Introduction}
The origin of prompt gamma-ray emission of cosmic gamma-ray bursts
(GRBs), the most luminous events in the universe, is still a great
puzzle since the discovery of this phenomenon. The GRB spectrum is
usually well fit with a smoothly-joint broken power-law with the
low- and high-energy photon spectral indices $\alpha$ and $\beta$,
breaking at $E_{\rm p}$ in the $\nu f_\nu$ spectrum, the so-called
Band function (Band et al. 1993). Broadband observations with the
Large Area Telescope (LAT) and Gamma-Ray Burst Monitor (GBM) onboard
the {\em Fermi} mission, which covers an energy band from 8 keV to
hundreds of GeV (Atwood et al. 2009), reveal that the spectra of
most GRBs are still well fit with the Band function up to the GeV
energy band, and only a small fraction of GRBs are detected with LAT
(L\"{u} et al. 2010; Goldstein et al. 2012). In the first paper of
this series (Zhang et al. 2011, Paper I), we have presented a
comprehensive analysis of 17 LAT GRBs. By performing a time-resolved
spectral analysis, we identified three elemental spectral components
(Band component, thermal component, and power law component) that
constitute GRB spectra. We found that except for some cases (e.g.
GRB 090902B, GRB 090510, Abdo et al. 2009; Ryde et al. 2010;
Ackermann et al. 2010), the Band function spectral component indeed
dominates the GRB spectra in most LAT GRBs. In Paper I, we have also
studied the evolution of spectral parameters in each burst.

The peak energy of the $\nu f_\nu$ spectrum ($E_{\rm p}$) is one of
the most interesting parameters of GRBs. It dramatically evolves
with time. The physical radiation mechanism that defines $E_p$ and
its evolution is unclear (e.g., Zhang \& M\'{e}sz\'{a}ros 2002,
2004). In general, the spectral properties carry the key to
understand the physics of GRBs, such as energy dissipation
mechanism, radiation mechanism, jet structure, as well as the
properties of the central engine. Some observed
energy/luminosity-spectrum relations involving $E_{\rm p}$ have been
widely discussed in the literature. Amati et al. (2002) discovered a
relation of the isotropic gamma-ray energy ($E_{\gamma, \rm iso}$)
to $E_{\rm p}$ in the burst frame. Ghirlanda et al. (2004) replaced
$E_{\gamma, \rm iso}$ with collimation corrected gamma-ray jet
energy ($E_{\gamma, j}$), and claimed a tighter correlation between
$E_{\rm p}$ and $E_{\gamma,j}$. Liang \& Zhang (2005) introduced the
optical temporal break time $t_{\rm b,opt}$, and discovered a
$E_{\rm p} - E_{\gamma, \rm iso} - t_{\rm b,opt}$ ``fundamental
plane'' correlation. These spectrum-energy correlations have been
proposed to be plausible probes of cosmological parameters (e.g.
Bloom et al. 2003; Schaefer 2003; Dai et al. 2004; Ghirlanda et al.
2004; Liang \& Zhang 2005, 2006). Similarly, the isotropic peak
luminosity ($L_{\rm p, iso}$) is also found to be correlated with
$E_{\rm p}$ in the burst frame among bursts (Wei \& Gao 2003;
Yonetoku et al. 2004), and within a same GRB (Liang et al. 2004).
This $E_{\rm p}-L_{\gamma, \rm iso}$ relation is even tighter within
a GRB pulse, especially during the decay phase (Lu \& Liang 2010; Lu
et al. 2010). The tight $E_{\rm p}-L_{\gamma, \rm iso}$ relation
within a GRB and/or a pulse of GRB may be the origin of the global
Amati/Yonetuku relation (Lu \& Liang 2010; Lu et al. 2010; Ghirlanda
et al. 2010; Firmani et al. 2009; Ohno et al. 2009), which suggests
that the Amati/Yonetuku relation may not be caused by an
observational selection effect (c.f., Nakar \& Piran 2005; Band \&
Preece 2005; Shahmoradi \& Nemiroff 2009). The evolution of $E_{\rm
p}$ within a burst would then be a key to reveal the origin of these
relations.

Two evolution patterns of $E_{\rm p }$ have been seen in GRBs,
i.e., hard-to-soft evolution and intensity-tracking (Liang \& Kargatis
1996; Ford et al. 1995; Kaneko et al. 2006; Lu et al. 2010; Peng et al.
2010). Two-thirds of the smooth GRB pulses in the BATSE sample have
$E_{\rm p}$ showing a hard-to-soft evolution, while in the others
a strong intensity-tracking was observed (Lu et al.
2010). In this paper, we focus on the spectral evolution patterns of
multiple-pulse GRBs in a selected bright GBM GRB sample, and investigate
the possible origins of $E_{\rm p}$-evolution and the
Amati/Yonetoku-relation. Our sample selection and data reduction
are described in Section 2. The time-resolved spectral analysis results
are shown in Section 3. In section 4, we present the detailed temporal
evolutions in 51 long and 11 short GRBs. A simulation about
overlapping hard-to-soft evolution pulses making an intensity-tracking
pattern is also presented. In Section 5, we discuss
a correlation between flux and
$E_{\rm p}$ and its implications for the Amati/Yonetoku relation.
We summarize our findings and discuss the
physical implications of our finding in Section 6.

\section{Sample Selection and Spectral Fits}

\begin{figure*}
\resizebox{16cm}{!}{\includegraphics{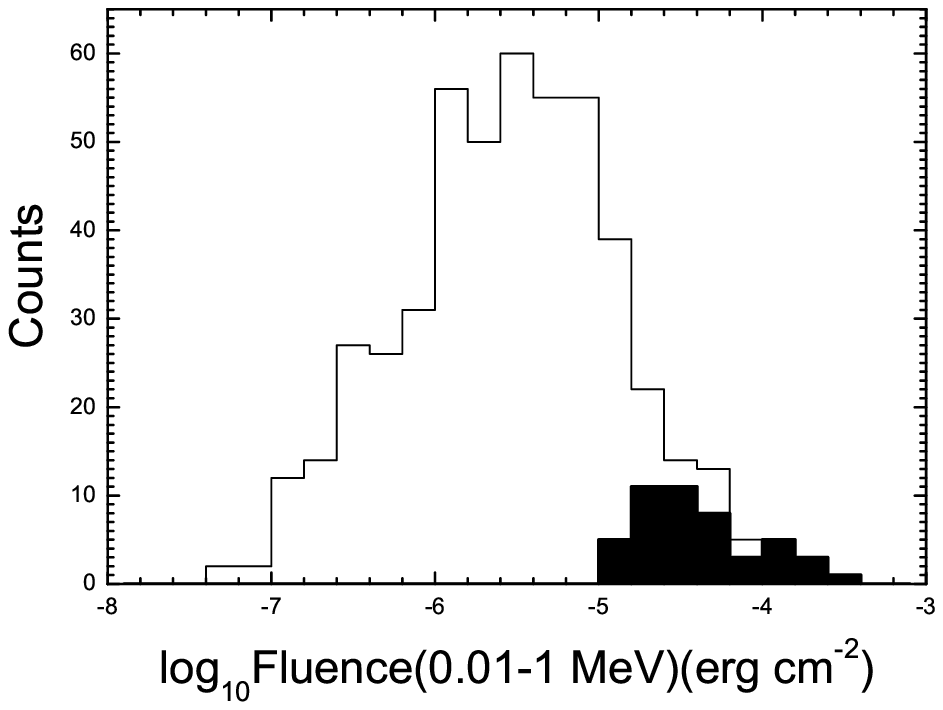}}\caption{Gamma-ray
fluence distribution of the GRBs in our sample (solid histogram) in
comparison with that of the first 2-year GBM Catalog (open
histogram, William et al. 2012).}\label{fluence}
\end{figure*}

GBM has 12 sodium iodide (NaI) detectors covering an energy range
from 8 keV to 1 MeV, and two bismuth germanate (BGO) scintillation
detectors sensitive to higher energies between 200 keV and 40 MeV
(Meegan et al. 2009). The signals from all the 14 GBM detectors are
collected by a central Data Processing Unit, which packages the
resulting data into three different types: CTIME, CSPEC, and TTE.
The TTE event data files contain individual photons with time and
energy tags. We download data from the NASA {\em Fermi} web
site\footnote{ftp://legacy.gsfc.nasa.gov/fermi/data/}, and use the TTE
data to make spectral fits with the software package RMFIT (version
3.3pr7). User-defined intervals before and after the prompt emission
phase are selected to obtain the background spectrum. The Band
function is adopted for the spectral fits. We make an extensive
time-resolved spectral analysis for the GBM GRBs as of 31 August
2011. We make joint fits to the spectra collected by the NaI and BGO
detectors. The time slices are normally selected to assure a signal-to-noise
ratio of 35. The reduced $\chi^{2}$ of our fits are normally $\sim
0.9-1.1$.  In order to present robust analyses for the spectral
evolution, our sample includes only those GRBs that at least five
time-resolved spectra are obtained from the data. These GRBs are
bright, with a fluence in the GBM energy-band larger than
$10^{-5}$ erg cm$^{-2}$ for long bursts and $8\times10^{-7}$ erg
cm$^{-2}$ for short bursts. We finally get a sample of 51 long
GRBs and 11 short GRBs. We show a comparison of the fluence
distribution of the GRBs in our sample with that of the first
GBM catalog in Figure \ref{fluence}. It is clearly seen that
our sample belongs to the brightest sub-sample of GBM GRBs.

\section{Time-Resolved Spectral Results}

Temporal evolution of $E_{\rm p}$ along with the lightcurve of each
GRB\footnote{The
time-resolved spectral data are available in the electronic version
only.} are shown in Figs. \ref{long} and \ref{short}.
We derive the bolometric energy flux ($F$) in the $1-10^4$ KeV
band and the spectral parmaters for each time slice. The $E_{\rm
p} - F$ plots for all the GRBs are also shown in Figs.
\ref{long} and \ref{short}. We show the $E_{\rm p}$, $\alpha$, and
$\beta$ distributions of oue sample in comparison with those of the
bright {\em CGRO}/BATSE GRBs (8459 time-resolved burst spectra of
350 GRBs;  Kaneko et al. 2006) in Figure
\ref{Distribution}. It is found that the $\alpha$ distributions are
well consistent with each other. The $E_{\rm p}$ of the Fermi/GBM
GRBs tends to be lower than that of BATSE GRB sample. This may be
due to an instrumental selection effect. At the low-energy end, the
GRM energy band extends to 8 keV, being lower than the low end of
the BATSE band. The trigger efficiency of X-ray flashes would be higher
than BATSE, similar to HETE-2 (e.g., Liang \& Dai 2004). The
photon indices in the high energy band tends to be shallower than
that observed with BATSE. Note that the BGO detector extends the
efficient energy band up to 40 MeV. Therefore, the photon indices of
the high-energy end may be better constrained.

\section{Evolution of $E_{\rm p}$ in Long and Short GRBs}

As shown in Figs. \ref{long} and \ref{short}, $E_{\rm p}$
dramatically evolves with time. Two types of $E_{\rm p}$ evolution,
hard-to-soft evolution and intensity-tracking, are observed, as
reported previously by Liang \& Kargatis (1996), Ford et al. (1995),
Kaneko et al. (2006), Preece et al. (2000), Peng et al. (2010), and
Lu et al. (2010) with the CGRO/BATSE data.

\subsection{Long GRBs}

Our analysis of 51 long GRBs are presented in Fig.\ref{long}.

There are 8 single-pulse long GRBs. They clearly fall into two categories.
The hard-to-soft evolution pattern appears in 5 GRBs: 081224, 090809B,
100612A, 100707A, and 110817A. The intensity tracking pattern appears in
3 GRBs: 081207, 090922A, and 100528A. Inspecting the lightcurves, a
general trend is that the hard-to-soft pulses tend to be asymmetric,
with the rising wing much steeper than the falling ring (except
GRB 100612A), while the intensity-tracking pulses tend to be more
symmetric (but see GRB 090922A).

The rest 43 long GRBs have multiple pulses. The $E_{\rm p}$
evolution patterns become more complicated. In a good fraction of
GRBs, the first pulse shows a clear hard-to-soft evolution, while
the rest pulses show the tracking behavior. On the other hand, a
good fraction of GRBs have all pulses (including the first one)
showing the intensity tracking behavior. In one case, i.e. GRB
090131 that shows at least 3 high-spike pulses, it is interesting to
see that the {\em second} pulse shows a clear hard-to-soft
evolution, even though the first pulse shows a nice tracking
behavior. The general message from such a rough inspection is that
mixed $E_{\rm p}$ evolution patterns can co-exist in a same burst,
with a variety of combined patterns. We investigate Fig.\ref{long}
in detail, and identified following groups:

\begin{itemize}

\item Intensity-tracking in all pulses (17/43 GRBs): 080825C,
080916C, 081009, 081222, 090323, 090424, 090804, 090820A,
090828, 090829, 090902B, 090926A, 091020, 091127, 100724B,
110123A, 110301A;

\item Hard-to-soft evolution in the first pulse followed by
intensity-tracking (11/43 GRBs):  080916A, 081215A, 081221,
090618, 090626, 090718B, 100728A, 100814A, 100906A, 101023A,
110721A;

\item The evolution pattern of the first pulse is unclear, while
late pulses show intensity-tracking (7/43 GRBs): 090328, 090524,
091003, 091120, 100116A, 100122A, 101014A;

\item Clear hard-to-soft evolution throughout the burst (3/43 GRBs):
081125, 090719, 100701B;

\item First pulse tracking, second pulse hard-to-soft evolution,
third-pulse unclear (1/43 GRBs): 090131;

\item First pulse hard-to-soft, second pulse also show hard-to-soft
trend (1/43 GRBs): 090530B;

\item Data quality not good enough to give clear conclusion
(3/43 GRBs): 090516A, 101123A, 110731A.
\end{itemize}

To investigate whether all hard-to-soft evolution pulses tend to be
more asymmetric than the intensity-tracking pulses, we selected 30
pulses (15 hard-to-soft pulses and 15 tracking pulses excluding the
pulses from short GRBs) that have clearly identified either pattern,
fit each pulse with the function (Kocevski et al. 2003)
\begin{equation} \label{KRL}
F(t)={F_m}\left(\frac{t+t_0}{t_m+t_0}\right)^r\left[\frac{d}{d+r}+\frac{r}{d+r}\left(\frac{t+t_0}{t_m+t_0}\right)^{(r+1)}\right]^{-\frac{r+d}{r+1}},
\end{equation}
and investigate the distribution of the parameter $t_r/t_d$, the
ratio between the rising time scale and the falling time scale (Fig.
\ref{trtd}). Indeed such a trend is revealed in the histogram,
although some opposite examples also exist.

We have also searched for other possible differences between the GRB
samples that have dominant hard-to-soft and intensity-tracking
patterns. No statistically significant correlations are found. This
may be due to the small sample effect, but it is possible that the
two patterns are related to fundamental emission mechanisms that do
not depend on the global properties of GRBs. Indeed the fact that
both patterns co-exist in a same burst also suggests this.

\subsection{Short GRBs}

The $E_{\rm p}$-evolution of 11 short GRBs in our sample are shown
in Fig.\ref{short}. Interestingly, clear intensity-tracking patterns
are observed in all of them whose data quality is good enough.
Notice that $E_{\rm p}$ does not always exactly track the intensity.
The maximum $E_{\rm}$ may lag behind or come before the peak of the
corresponding pulse (e.g. GRBs 090227B, 090510, and 090228A). We'd
like to caution that the time resolution of the light curves is 16
ms in this analysis. The light curves of short GRBs are usually
highly variable in shorter time scales. The time bins of our
spectral analysis are usually larger than variability time scales of
short GRBs. On the other hand, Guiriec et al. (2010) presented a
time-resolved spectral analysis for 3 bright short GRBs with a time
scale as short as 2 ms. They showed that the $E_{\rm p}$ still
tracks intensity, with significant fluctuation (see also Ghirlanda
et al. 2011 for a sample of 13 short GRBs).

\subsection{Superposition of adjacent hard-to-soft pulses as the origin
of the intensity-tracking?}

Since most late time pulses have intensity-tracking, a natural
question is whether they can be due to the superposition of
hard-to-soft evolution pulses. Hakkila \& Preece (2011) argued that
all correlated pulse characteristics can be explained by the
hard-to-soft $E_{\rm p}$ evolution, and the intensity-tracking is
merely the result of superposition of two or more hard-to-soft
pulses. In order to test such an effect, we perform a simulation
analysis with the RMFIT package. We take GRBs 081224 and 100707A,
both having a single pulse with hard-to-soft evolution, as a
template to perform the simulations. The simulation procedure is as
follows.

\begin{itemize}
\item Extract the TTE data of the brightest NaI and BGO detectors of
the GRB (GRB 081224 or GRB 100707A),
and shift the arrival time of each photon with a delay
timescale of 10 s, 5 s, and 3 s (case A, B, and C, respectively).

\item Co-add the original TTE data with the TTE data of Case A, B, and
C, respectively.

\item Make time-resolved spectral analysis for the co-added TTE data,
and report $E_p$ evolution of the mock light curves.
\end{itemize}

Our results are shown in Figure \ref{mockabc}. We take the
simulation using GRB 081224 as an example. In case A, since the
separation between the two pulses is wide, the $E_{\rm p}$ evolution
of each pulse is not significantly contaminated in the mock GRB,
although in the bridge region, the $E_{\rm p}$ evolution becomes
less significant. In case B, the superposition effect becomes more
significant around $4.5\sim 6$ second. In case C, the $E_p$
evolution behavior of the second pulse now turns into intensity
tracking, due to the close superposition of the two pulses. The
simulation using GRB 100707A reached the similar conclusion. From
these simulation, one can tentative draw the conclusion that
superposing two hard-to-soft evolution patterns could indeed
generate an intensity-tracking pattern under certain conditions.
Whether or not this is possible depends on the competition between
the flux contrast near the transition regions in the overlapping
pulses. If the transition has a sharp dip (corresponding to the
wide-separation case), then the hard-to-soft evolution pattern is
hardly altered. However, if the transition is smooth with a shallow
dip, the tracking behavior is more obvious. Some GRBs with
complicated lightcurves show fluctuating $E_p$ features. This may be
also a consequence of superposition. This effect also potentially
explains the irregular spectral variation in some GRBs with highly
variable light curves, such as in GRBs 101123A and 110731A.

To examine the superposition effect on the spectral shape, we
illustrate a mock spectrum of Case C of GRB 081224
in the time interval [2.473,
3.911] seconds in Figure \ref{mock_c}, which corresponds to the
onset time of the second pulse of the mock GRB. The spectrum is
roughly the overlapped spectra of GRB 081224 in the time intervals
[-0.51, 1.002] seconds and [2.473, 3.911] seconds. It is found that
the superposition significantly modifies the spectra in the two time
intervals. However, it is still well fit with the Band function. In
the low energy end, the spectrum is dominated by the low-$E_{\rm p}$
component, while it is dominated in the high-$E_{\rm p}$ component
in the high energy end. The $E_{\rm p}$ of the superimposed spectrum
is similar to that of the high-$E_{\rm p}$ component\footnote{Note
that the composed $\nu f_\nu$ spectrum may show a plateau or a
two-hump feature if $E_{\rm p}$ of the two spectral components are
well separated and have comparable energy fluxes. The spectrum may
not strictly be a single Band function. This may cause confusion in
identifying different emission components in an observed spectrum.}.

We do not claim that all the intensity-tracking pulses after the
first pulse are due to superposition of hard-to-soft pulses. This is
because some single pulse bursts, and the first pulse of about half
of the GRBs in our sample (for which superposition effect does not
exist) indeed show the intensity-tracking
behavior. As shown in Figure \ref{mockabc}, the $E_{\rm p}$
evolution pattern can be changed to tracking from hard-to-soft
evolution only when the two pulses are highly overlapped. As a
result, if one sees intensity tracking from a pulse that is well
separated from the proceeding one, it is very likely intrinsic and
is not due to the superposition effect.

\section{Time-resolved $E_{\rm p}$-$F$ Correlation and Implications
for the Amati/Yonetoku Relation}

With the time-resolved spectral analysis results, we also show the
$E_{\rm p}-F$ relation for the bursts in our sample in Figures
\ref{long} and \ref{short} for the long and short GRBs. For each
GRB, we fit the $E_{\rm p}-F$ correlation with a simple power-law
function, $E_{\rm p}\propto F^{\kappa}$,  for all the time bins, and
record the correlation index $\kappa$. For comparison, we take the
time bins during the decaying wing of each pulse, and perform the
same fits, and derive the corresponding index $\kappa_d$. In Figure
\ref{kd}, we show the distributions of $\kappa$ and $\kappa_d$ for
both long and short GRBs, and found that they are consistent with
each other, with a mode at $0.55\pm 0.22$ for both long and short
GRBs. We measure the scatter of the $E_{\rm p}-F$ relation with the
distance of the data points from the best fit line as done by
Ghirlanda et al. (2005). As shown in Fig. \ref{kd}, the scatter of
the $E_{\rm p}-F$ relation in the decay phase is much tighter than
that for the entire GRBs, with a dex of $0.070\pm0.049$ comparing to
$0.17\pm 0.08$ for GRB. The large dispersion of the $E_{\rm p}-F$
relation in the entire GRB would be due to the variation of
$\kappa_{\rm d}$ in different pulses, and the rising wing of the
pulses during which hard-to-soft spectral evolution may happen. As
shown in Lu et al. (2010), the data points of the rising wing of a
hard-to-soft pulse usually deviate from the $E_{\rm p}-F$ relation
observed in the decaying wing. Although the $E_{\rm p}-F$ relation
each pulse is tight, the slope varies among pulses. The mix of
different pulses would then enlarge the dispersion of the $E_{\rm
p}-F$ correlation in a burst.

Fifteen GRBs in our sample have redshift measurements. Among them 14
GRBs are long and 1 is short (GRB 090510). We calculate
time-resolved isotropic luminosity of these GRBs and correct $E_{\rm
p}$ to the burst rest frame. We show the $E^{\rm rest}_{\rm
p}-L_{\gamma, \rm iso}$ correlation of these 15 GRBs in Fig.
\ref{LEp}. Also plotted are the GRBs reported by Yonetoku et al.
(2010), who reported a correlation between time integrated $E^{\rm
rest}_{\rm p}$ and the peak isotropic luminosity of individual
bursts. It is found that the $E_{\rm p}^{\rm rest}-L_{\gamma, \rm
iso}$ relation for the time-resolved spectra within a GRB (our
sample) is consistent with that for the time-integrated spectra
among the GRBs (Yonetoku sample).  Our best linear fit to the
time-resolved $E_{\rm p}^{\rm rest}-L_{\gamma, \rm iso}$ relation is
$\log E_{\rm p}=-(29.854\pm0.178)+(0.621\pm0.003)\log L_{\gamma, \rm
iso}$ with a linear coefficient of $r$=0.88 (N=251) and chance
probability of $p<10^{-4}$. We measure the scatter of the data
points around the best fit and obtain dex=0.256, which is roughly
consistent with the intrinsic scatter ($\sigma_{int}$=0.195) of the
time integrated $E_{\rm p}-L_{\gamma, \rm iso}$ relation among
different GRBs (see also Ghirlanda et al. 2005). Note that the
time-resolved $E_{\rm p}-L_{\gamma, \rm iso}$ relation of the short
GRB 090510 is also consistent with that of the long GRBs, although
its $E_{\rm p}$ is significantly larger than most long GRBs in our
sample (see also Ghirlanda et al. 2011). Zhang et al. (2009a) showed
that short GRBs do not follow the $E_{\gamma, \rm iso}-E_{\rm p}$
relation (the Amati relation) of long GRBs, mostly due to their
smaller $E_{\gamma, \rm iso}$ (by 2 to 3 orders of magnitude). They
showed that in terms of $E_{\rm p}-L_{\gamma, \rm iso}$ relation,
long and short GRBs are similar (see also Ghirlanda et al. 2010 and
Zhang et al. 2012). Therefore, despite of different energy
reservoirs in long and short GRBs, their radiation physics of both
long and short GRBs may be the same (L\"{u} et al. 2010; Ghirlanda
et al. 2011).

\section{Conclusions and Discussion\label{sec:Conclusions}}

We have carried out a detailed time-resolved spectral analysis for
a bright sample of Fermi GBM bursts. By studing the $E_{\rm p}$
evolution within individual GRBs, we confirm the existence of
two evolution patterns within certain pulses: a hard-to-soft evolution
pattern and an intensity-tracking pattern. Among the 8 single-pulse
long GRBs, 5 show hard-to-soft evolution, while the other 3 show
intensity-tracking. For multi-pulse long GRBs, the patterns are more
complex. For the first pulse, the split between the two patterns
is roughly half-half. However, for later pulses the
intensity-tracking pattern becomes predominant. Through simulations,
we show that some of the late intensity-tracking pulses could be
due to the close superposition of pulses with hard-to-soft evolution.
However, this cannot account for all the late pulses, especially
those without a preceding overlapping pulse but also show the
tracking behavior. Conversely, in two bursts, the hard-to-soft
evolution pattern is observed in the second pulse of the burst,
when the pulse is well separated from the first pulse. {\em So overall,
it is clear that both patterns are intrinsic, and they can coexist
in a same GRB in different pulses.}

The situation of short GRBs is simpler. They are overwhelmingly
dominated by the intensity-tracking pattern. One caveat is that the
time resolution (16 ms) may not be fine enough to catch the possible
hard-to-soft evolution pattern in short GRBs. In any case, an
independent study with 2 ms time resolution still did not show
evidence of hard-to-soft evolution (Guiriec et al. 2010). So the
tracking behavior may be an intrinsic property of short GRBs.

We have presented the correlation between $E_{\rm p}$
and $F$ within single-pulse and multi-pulse GRBs. This
correlation is tighter in the decay phase of the GRBs, suggesting
that the decaying phase correlation may be the main source of the
global internal $F-E_{\rm p}$ correlation reported by Liang et al.
(2004) (see also Firmani et al. 2009). Fifteen GRBs
(14 long GRBs and 1 short GRBs) in our sample have redshift
measurements. We shown that the both the slope and the dispersion of
the $E_{\rm p}-L_{\gamma, \rm iso}$ relation for the time-resolved
spectra of these GRBs are well consistent with the global
Yonetoku relation derived from the pre-Fermi GRBs.

Our results suggest that the $E_{\rm p}$ evolution may hold the key
to understand the GRB radiation physics, and the origin of various
observed spectrum-energy relations (e.g. Amati et al. 2002; Yonetoku
et al. 2004; Liang \& Dai 2004; Ghirlanda et al. 2004; Liang \&
Zhang 2005). Any successful physical model of GRB prompt emission
has to able to produce two different $E_{\rm p}$-evolution patterns.
These two patterns not only operate in different bursts, but
could also operate within the same burst as well. This is
challenging. In the following, we discuss radiation models and
geometric models in turn.

\subsection{Radiation physics as source of $E_{\rm p}$  evolution}\label{sec:radiation-physics}

The prompt emission of GRBs is still a mystery (e.g. Zhang 2011 for a recent
review). The main uncertainty is the composition of the outflow (fireball
vs. Poynting-flux dominated flow), which determine the energy dissipation
mechanism (internal shocks vs. magnetic reconnection), particle acceleration
mechanism (1st- or 2nd-order Fermi acceleration), and radiation mechanism
(synchrotron vs. inverse Compton scattering). Three emission models are
widely discussed: (1) the internal shock synchrotron model; (2) the
dissipative photosphere model; and (3) the abrupt magnetic dissipation
model. These different models have different predictions regarding
$E_{\rm p}$ evolution within a burst. The data can be then used to
constrain these models.

In all the models, $E_{\rm p}$ is a function of outflow luminosity
(which is usually represented by the gamma-ray luminosity $L_{\rm
\gamma,iso}$) and the Lorentz factor $\Gamma$. Different models have
different dependences on these models (e.g. Zhang \& M\'esz\'aros
2002; Pe'er et al. 2006). One common feature for all the models is
that when emission stops abruptly, the observed emission is the
high-latitude emission from the jet due to the curvature of the
conical jet (e.g. Fenimore et al. 1996; Kumar \& Panaitescu 2000;
Dermer 2004; Liang \& Zhang 2006; Qin et al. 2008; Zhang et al.
2009b). During this curvature-effect dominated phase, an intensity
tracking behavior is expected. Observationally, both hard-to-soft
evolution and intensity-tracking show decreasng $E_{\rm p}$ with
decaying flux. This is generally consistent with this ``curvature
effect'' explanation\footnote{If the instantaneous spectrum at the
end of pulse emission is a single power law, then no spectral
evolution is expected. On the other hand, if the instantaneous
spectrum is curved, the curvature effect naturally gives rise to a
spectral evolution. Strong spectral evolution during the steep decay
phase of early X-ray afterglow of Swift GRBs has been commonly
observed (Zhang et al. 2007). This can be still interpreted within
the curvature effect model by assuming a curved instantaneous
spectrum at the end of prompt emission (Zhang et al. 2009b).}. More
specifically, during the decay phase, one has the typical frequency
$E_{\rm p}=DE_{\rm p}^{'}/(1+z)$, the specific luminosity $L_{\nu}
=D^{2} L'_{\nu'}$, and the bolometric luminosity $L_{\rm
iso}=D^{\varepsilon}L^{'}_{\rm iso}$, where $D$ is the Doppler
factor, the prime values are measured in the co-moving frame, and
the value of $\varepsilon$ takes 3 for a continuous jet and 4 for an
impulsive blob (Ghisellini et al. 1993). We have $E_{\rm p}\propto
L^{1/\varepsilon}_{\rm iso} (1+z)f(E_{\rm p}^{'}, L^{'}_{\rm iso})$,
where $f(E_{\rm p}^{'}, L^{'})=E_{\rm p}^{'}/{L^{'}_{\rm
iso}}^{1/\varepsilon}$. One then expects $E_{\rm p}\propto
L_\nu^{1/2}$ and $E_{\rm p}\propto L^{1/\varepsilon}_{\rm iso}$
regardless of the intrinsic relation between the $E_{\rm p}^{'}$ and
$L_{\gamma, \rm iso}^{'}$. Since what one measures is neither
exactly $L_\nu$ (since there is a wide band) nor $L_{\rm \gamma,
iso}$ (since the band width is limited), the expected $E_{\rm p}-F$
relation slope would be roughly between 1/2 and 1/3 (1/4). This is
roughly consistent with the data, although 1/4 would be too shallow.

The $E_{\rm p}$ evolution during the rising phase of a pulse carries
the key information to diagnose different prompt emission models.
For the standard synchrotron model (valid for internal shocks and
internal magnetic dissipation models), one can write down $E_{\rm p}
\propto \gamma_e^2 L^{1/2} R^{-1} (1+z)^{-1}$, where $L$ is the
``wind'' luminosity of the ejecta, $\gamma_e$ is the typical
electron Lorentz factor in the emission region, and $R$ is the
emission radius (Zhang \& M\'esz\'aros 2002). Naively, this would
give a tracking behavior, since $E_{\rm p} \propto L^{1/2}$.
However, considering other factors, the dependence is non-trivial.
For internal shocks (e.g. M\'esz\'aros et al. 1994; Daigne \&
Mochkovitch 1998; Daigne et al. 2011), the rising phase is related
to crossing of a shock across the colliding shells. The observed
light curve can rise even if the wind luminosity is constant. The
flux is related to the evolution of the strength of the shock during
shock crossing (and hence, $\gamma_e$), and the number of electrons
that are shocked. In general, since shock strength increases as
shock propagates, and the number of electrons tend to increase, a
rough tracking behavior is expected for the internal shock model,
even though detailed modeling is needed to give more precise
predictions. On the other hand, the internal shock model has several
issues to interpret the available data (see a full discussion in
Zhang \& Yan 2011), including its inability to account for the
Amati/Yonetoku relation in view of the recent finding of
$\Gamma-E_{\gamma,iso}$ (Liang et al. 2010) and
$\Gamma-L_{\gamma,iso}$ (L\"u et al. 2012) correlations.

Zhang \& Yan (2011) proposed a GRB prompt emission model invoking
a sudden discharge of magnetic energy through turbulent magnetic
reconnection triggered by multiple internal collisions among magnetically
dominated shells. This ICMART model also attributes GRB prompt emission
to synchrotron emission of electrons. However, an extra dependence
of $\gamma_e$ on the magnetization factor $\sigma$ is invoked.
Since during an ICMART event $\sigma$ is expected to drop with time,
the dissipated magnetic energy is expected to be shared by more and
more electrons, so that $\gamma_e$ drops with time as electron number
increases with time. As a result, a hard-to-soft evolution during
the pulse rising phase is expected, although detailed numerical
calculations are needed to validate this prediction.

Finally, the dissipative photosphere model (Rees \& M\'esz\'aros 2005;
Pe'er et al. 2006; Giannios 2008; Beloborodov 2009; Lazzati \& Begelman 2009
Ioka 2010; Toma et al. 2011; Ryde et al. 2011) attributes $E_{\rm p}$ to
the temperature of the photosphere. Naively, a quasi-thermal nature
of emission generally calls for an intensity-tracking behavior,
since a hot temperature tends to be brighter. On the other hand,
the temporal evolution of the Lorentz factor, optical depth, and
the radius of photosphere may complicate the picture, and detailed
modeling is called for (e.g. W. Deng, \& B. Zhang, 2012, in preparation).

In general, radiation models can account for both hard-to-soft
evolution (ICMART model) and intensity-tracking (internal shocks and
probably photosphere), although detailed theoretical modeling in all
these cases are desirable. The difficulty for all these models is
that both evolutionary trends coexist in different pulses of a same
burst. One therefore has to invoke multiple models to interpret
different pulses in a same burst. This may happen if the composition
of a jet varies with time in a same burst, i.e. the magnetization
parameter $\sigma$ can switch from $>1$ to $<1$ within a same burst.
This is not impossible (Zhang 2011), since given the same magnetic
field strength at the central engine, a variation in baryon loading
can cause a large fluctuation of the $\sigma$ value.

\subsection{Geometric effect as source of $E_{\rm p}$ evolution}

It was proposed that the broad pulses with a dramatical $E_{\rm p}$
evolution in GRBs may be due to the precession of GRB jets
(Portegies Zwart et al. 1999; Reynoso et al. 2006; Lei et al. 2007;
Liu et al. 2010). Fixing the observer's viewing angle, the
precession would result in the jet sweeping in and out the observer's
line of sight. This would result in a rapid evolution of the Doppler
factor. Assuming that the rest-frame emission properties remain
the same during precession, the observer would record a rapid
flux variability and spectral evolution. Since both $E_{\rm p}$
and luminosity are positively related to the Doppler factor,
this model could explain the intensity-tracking behavior of
the pulses. Since there is no preference of the precession direction,
the pulse tend to be symmetric (Liu et al. 2010).

How can a GRB jet precess? This is a great issue. The progenitors of
GRBs are massive stars or compact star binaries. In both scenarios,
the new-born central engine object (a black hole or a rapidly
rotating, highly magnetized pulsar) is expected to spin rapidly (van
Putten 2004). The anisotropic mass fall-back in a collapsar, or the
mis-alignment of angular momenta of two merged compact objects may
lead to a tilted disc. The fragmentation of the star (King et al.
2005) or disk (Perna et al. 2006) may also lead to a misaligned
disc. The Lense-Thirring precession appears for such a Kerr BH with
a tilted disc (Lense \& Thirring 1918). Current favored jet
launching models for GRBs include neutrino-annihilation mechanism
(Popham et al. 1999) and Blandford-Znajek process (Blandford \&
Znajek 1977). For both models, the jet is expected to be
perpendicular to the midplane of the disc, so that the precession of
the disc would result in precession of the jet (Lei et al. 2007; Liu
et al. 2010). It is found that the overall shape and temporal
evolution of the GRB light curves can be fit with jet procession if
the jet is narrow enough (Lei et al. 2007; Liu et al. 2010). Clear
periodic signals are not expected due to the stochastic nature of
the process (Liu et al. 2010). One issue is that for long GRBs, the
existence of a massive stellar envelope tends to stall and quench a
precessing jet, especially when the precession angle is larger than
the jet opening angle (Zhang et al. 2004). That simulation did not
include magnetic fields, which would enhance collimation of the
precessing jet. Further MHD simulations are needed to test whether
large-angle precession is allowed for a magnetically dominated jet
propagating in a stellar envelope.

The main difficulty of the precession scenario is again the coexistence of
both hard-to-soft and intensity-tracking behaviors in a same burst.
It is hard to interpret an asymmetric pulse with clear hard-to-soft
evolution during the rising wing within the jet precession model.
Other issues include the detailed energy dissipation, particle
acceleration, and emission processes of a precession jet. When these
details are considered, the simple assumption of a constant
co-moving emissivity has to be replaced by the more detailed
calculation of time-dependent co-moving spectra. The radiation
physics as discussed above (\S\ref{sec:radiation-physics}) has to be
considered, which would introduce extra complications. How the
precession hypothesis fares with the successful afterglow theory and
confronts with the afterglow data is also subject to further intense
modeling.

\acknowledgments This work is supported by NSFC (Grants No. 11025313,
10873002, 11063001, 11163001, and 10847003), the ``973"
Program of China (2009CB824800), Special Foundation for Distinguished
Expert Program of Guangxi, the Guangxi SHI-BAI-QIAN project (Grant 2007201),
the Guangxi Natural Science Foundation (2010GXNSFA013112 and 2010GXNSFC013011,
Contract No. 2011-135), and the 3th Innovation Projet of Guangxi University.
BZ acknowledges support from NASA (NNX10AD48G) and NSF (AST-0908362).

\clearpage

\newpage
\begin{figure*}
\resizebox{4cm}{!}{\includegraphics{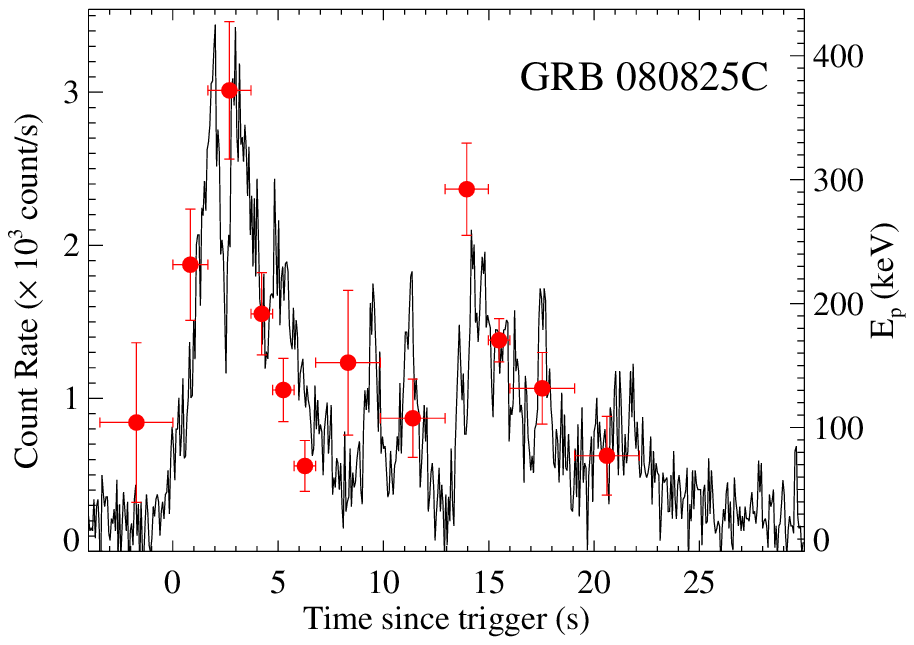}}
\resizebox{4cm}{!}{\includegraphics{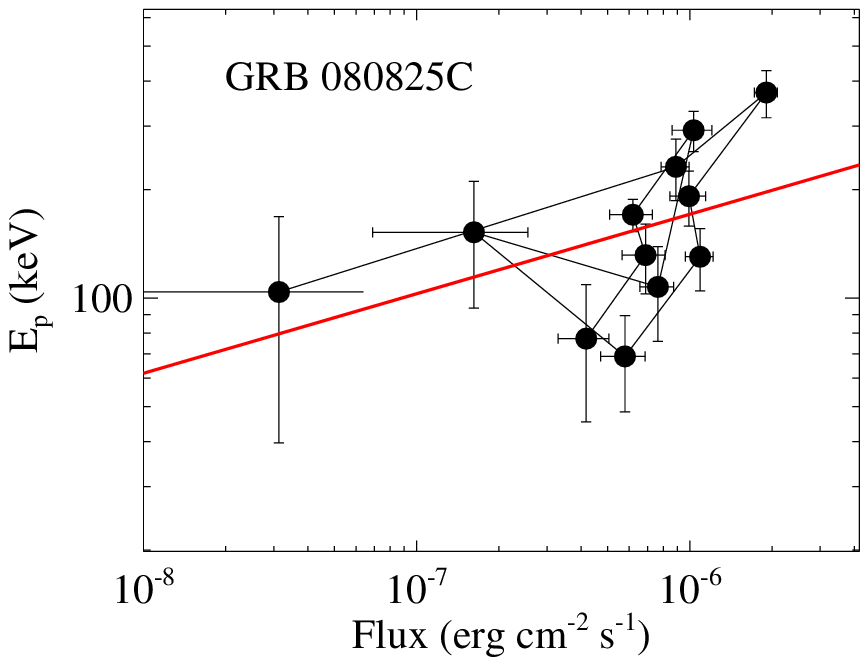}}
\resizebox{4cm}{!}{\includegraphics{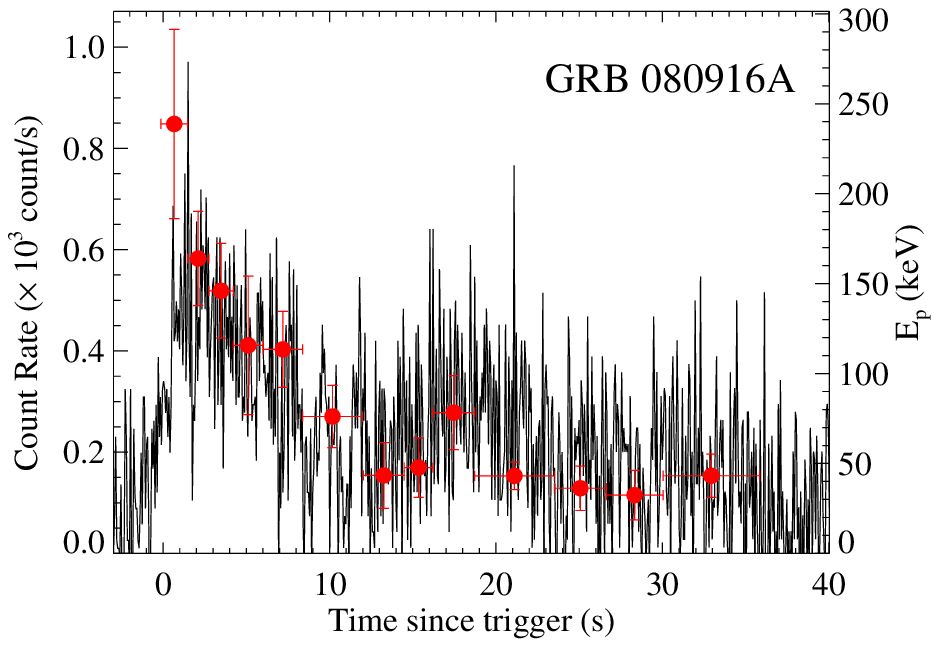}}
\resizebox{4cm}{!}{\includegraphics{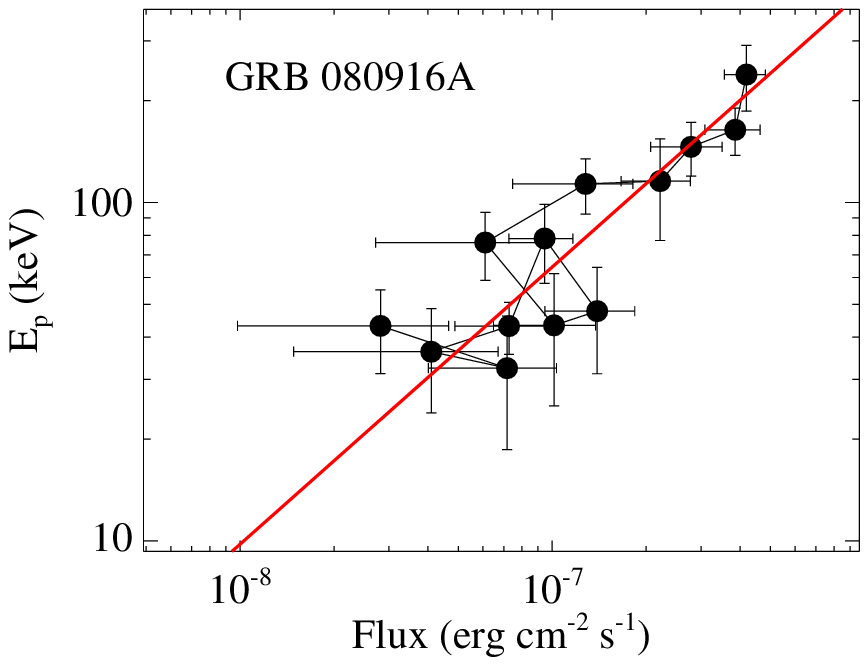}}

\resizebox{4cm}{!}{\includegraphics{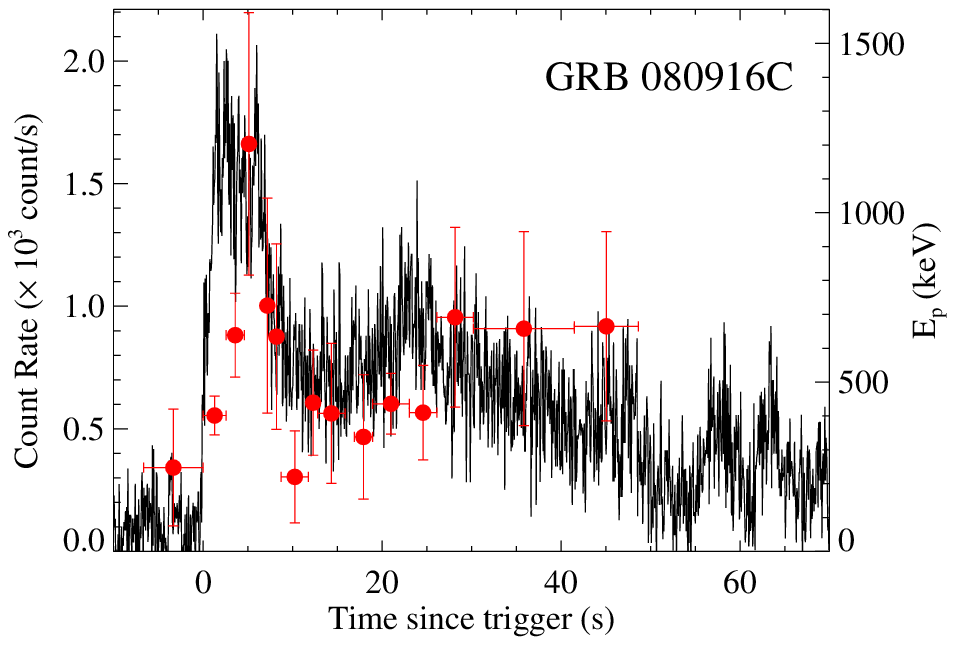}}
\resizebox{4cm}{!}{\includegraphics{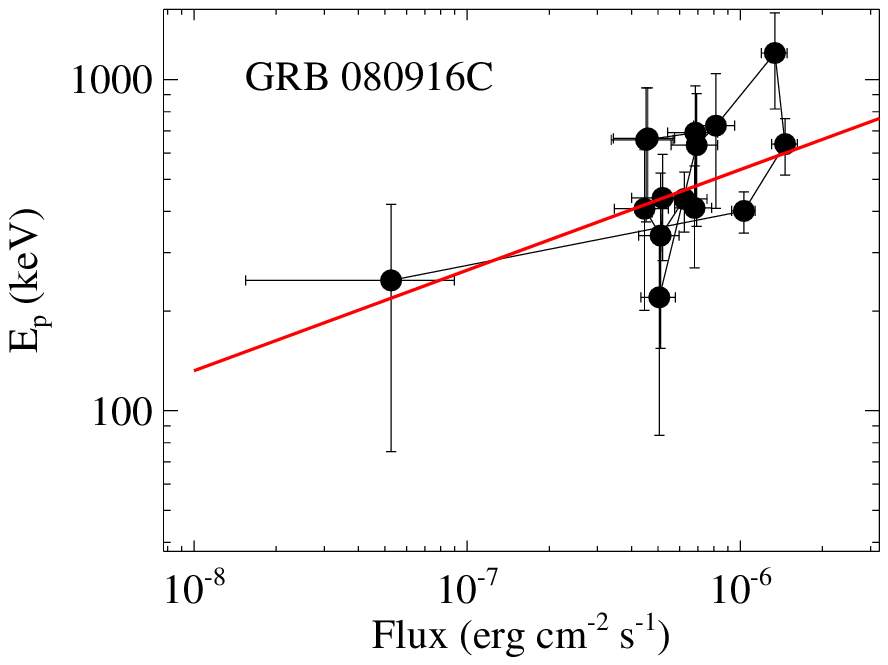}}
 \resizebox{4cm}{!}{\includegraphics{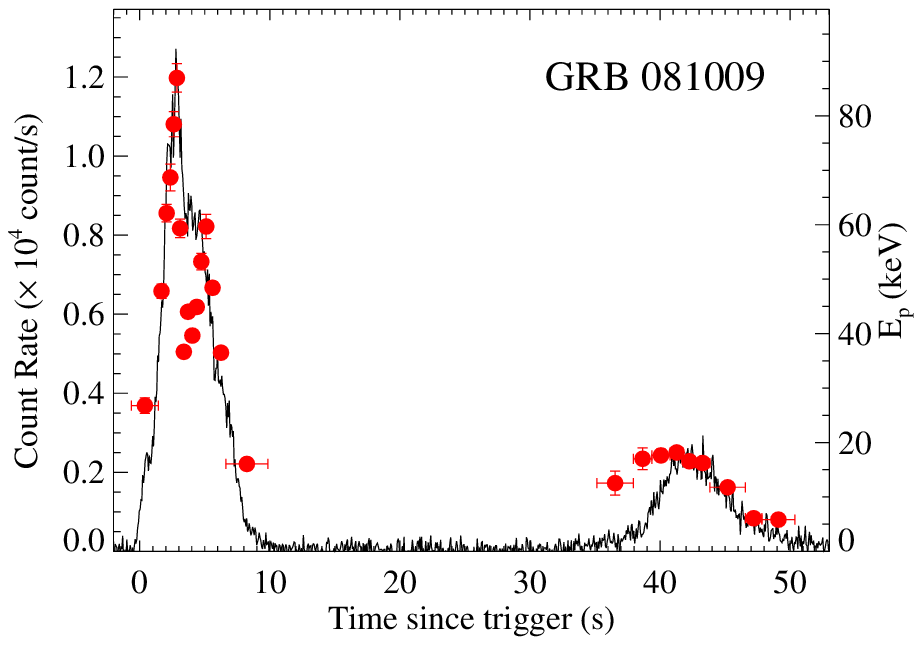}}
 \resizebox{4cm}{!}{\includegraphics{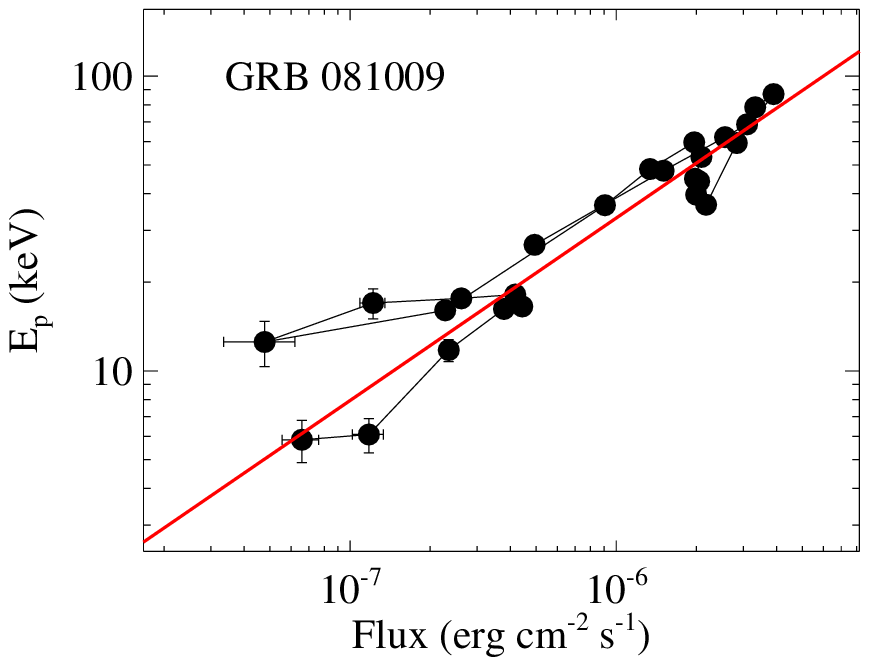}}

  \resizebox{4cm}{!}{\includegraphics{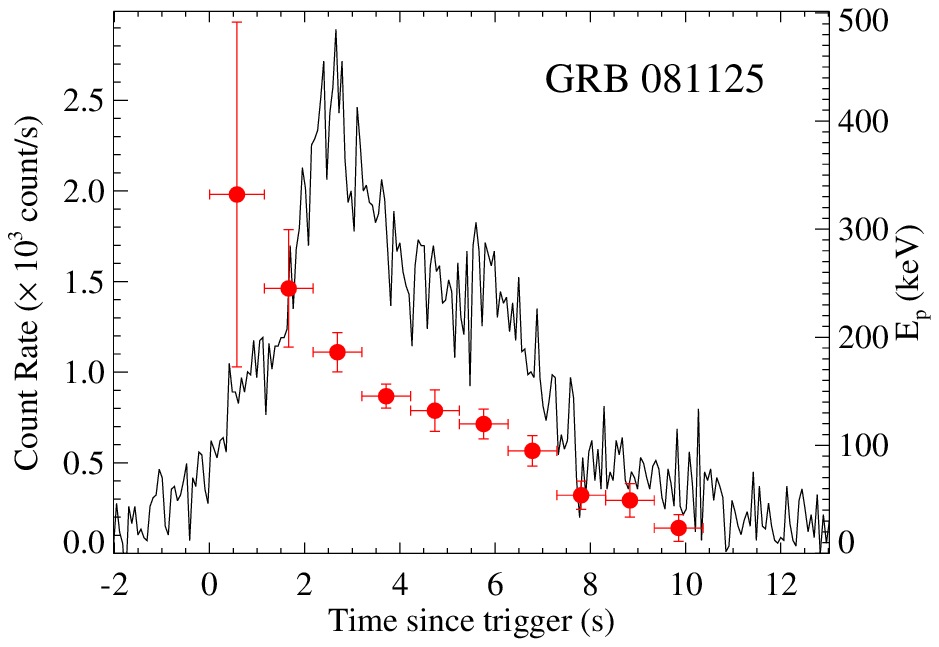}}
 \resizebox{4cm}{!}{\includegraphics{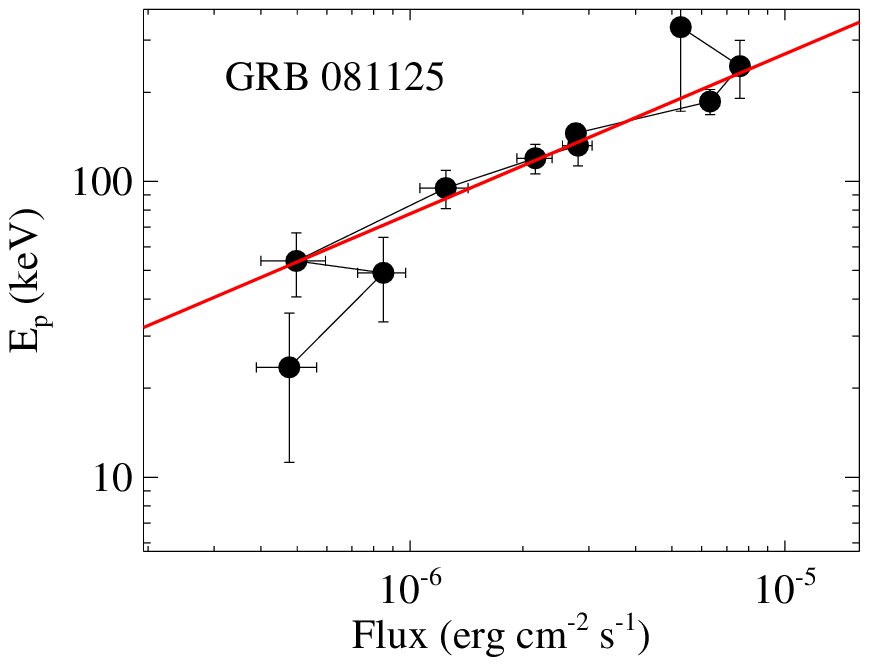}}
  \resizebox{4cm}{!}{\includegraphics{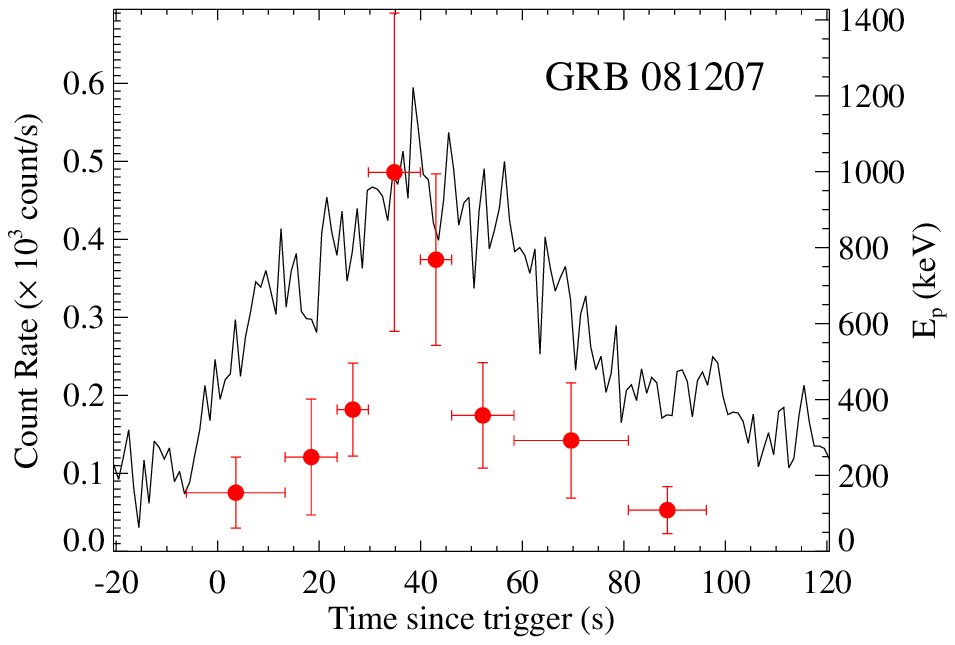}}
 \resizebox{4cm}{!}{\includegraphics{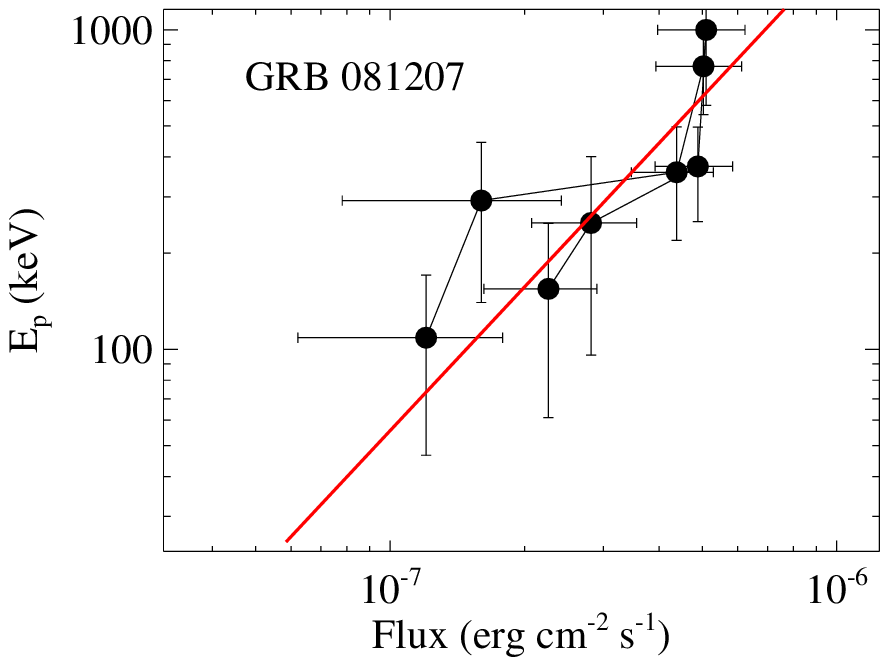}}

 \resizebox{4cm}{!}{\includegraphics{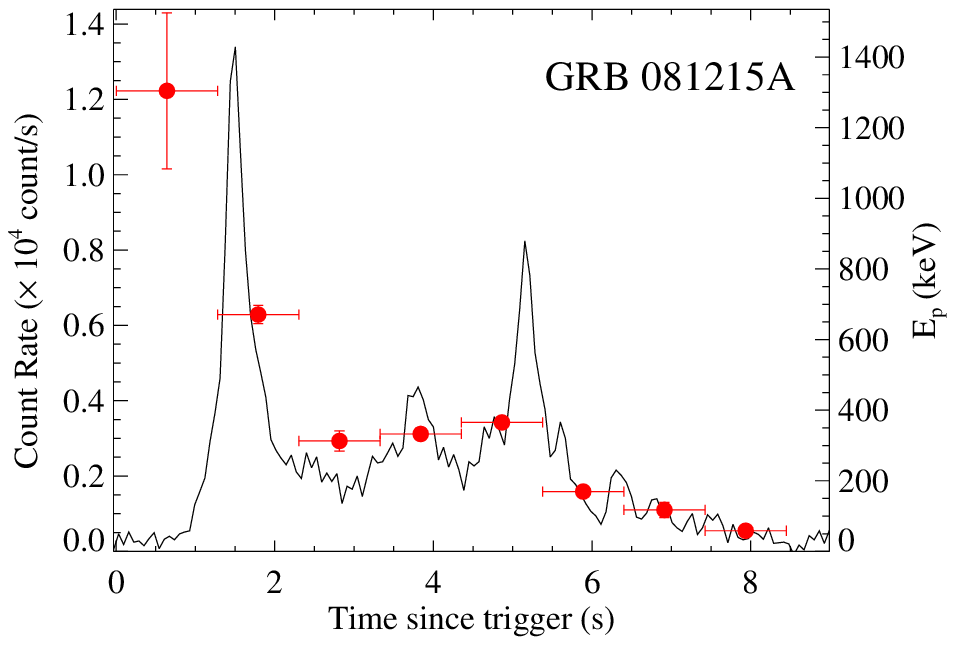}}
 \resizebox{4cm}{!}{\includegraphics{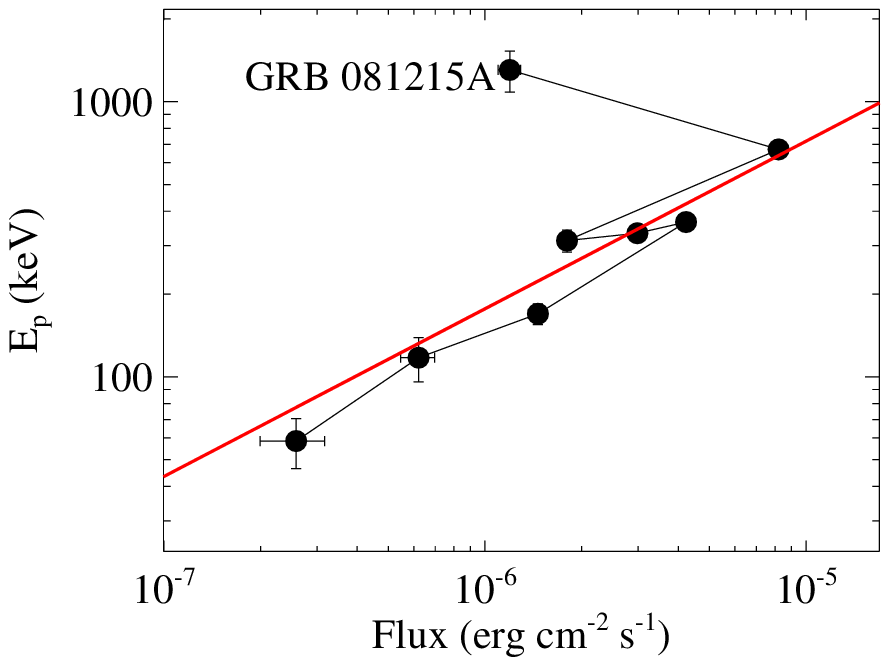}}
 \resizebox{4cm}{!}{\includegraphics{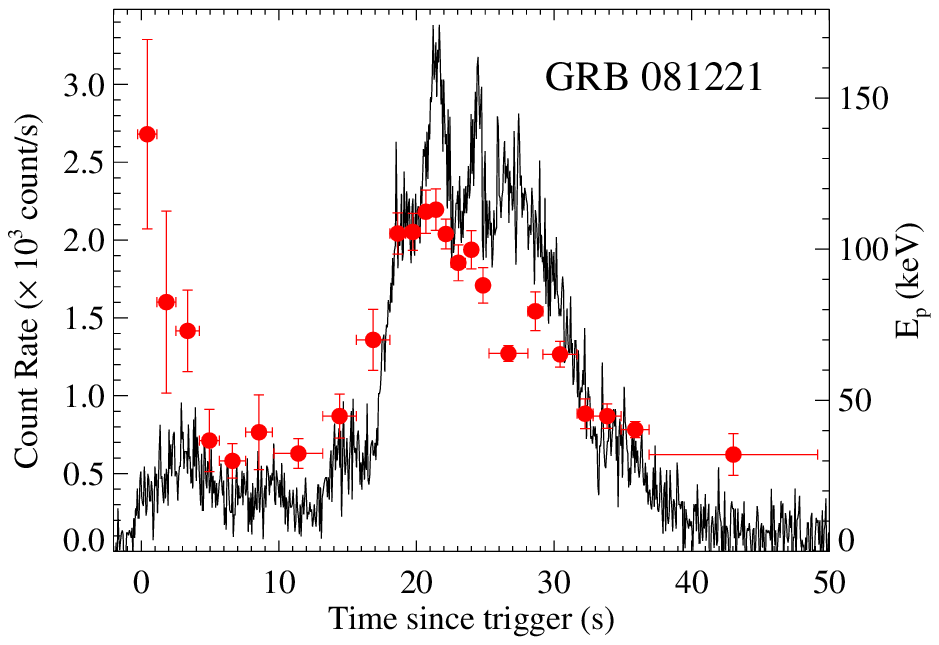}}
 \resizebox{4cm}{!}{\includegraphics{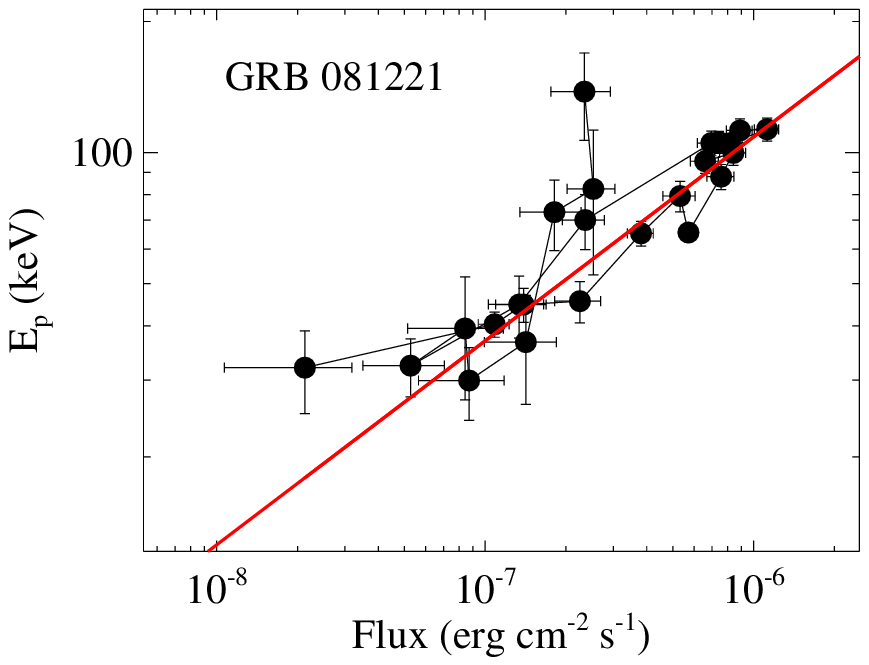}}

  \resizebox{4cm}{!}{\includegraphics{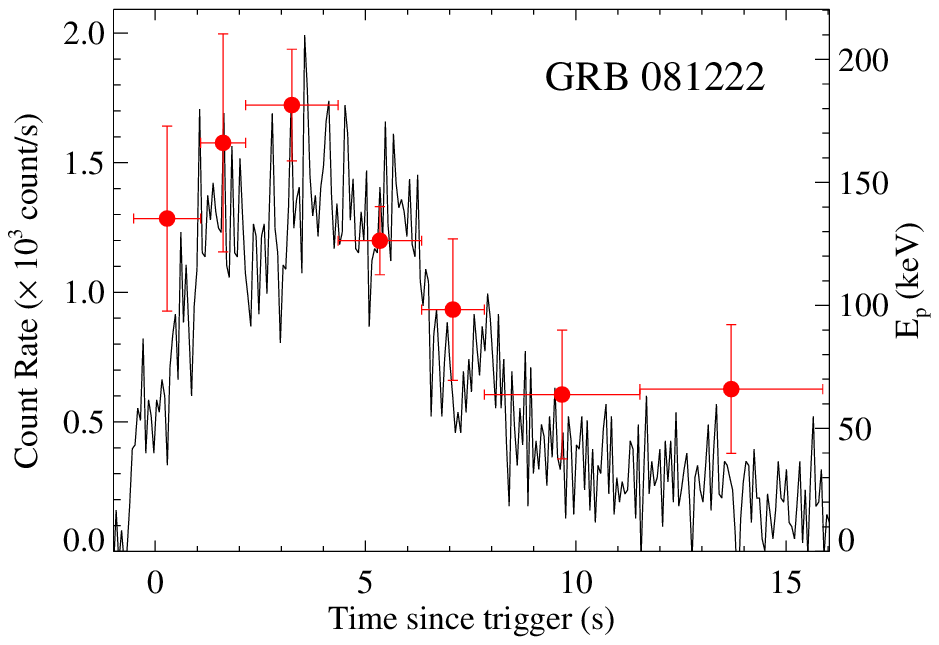}}
 \resizebox{4cm}{!}{\includegraphics{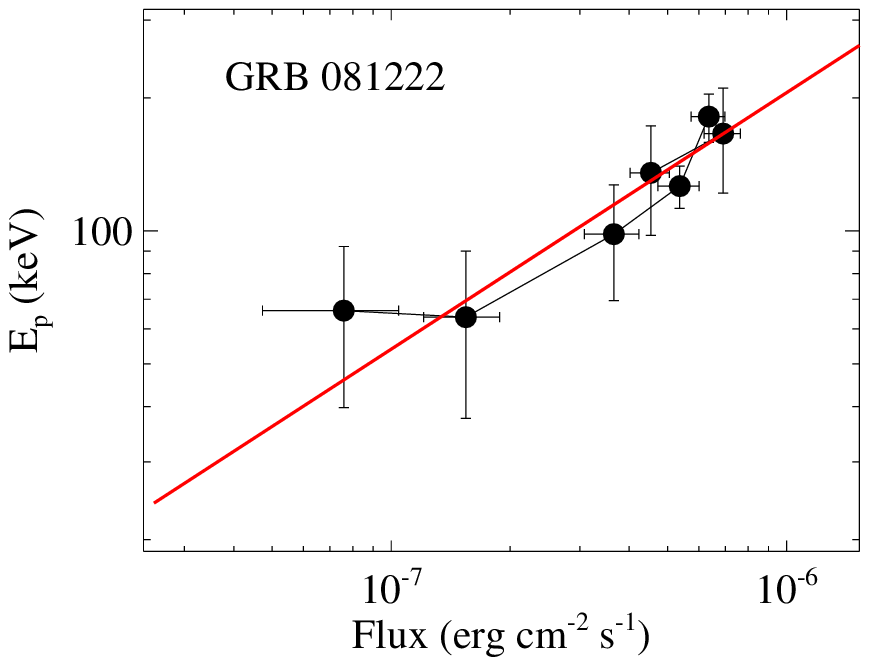}}
  \resizebox{4cm}{!}{\includegraphics{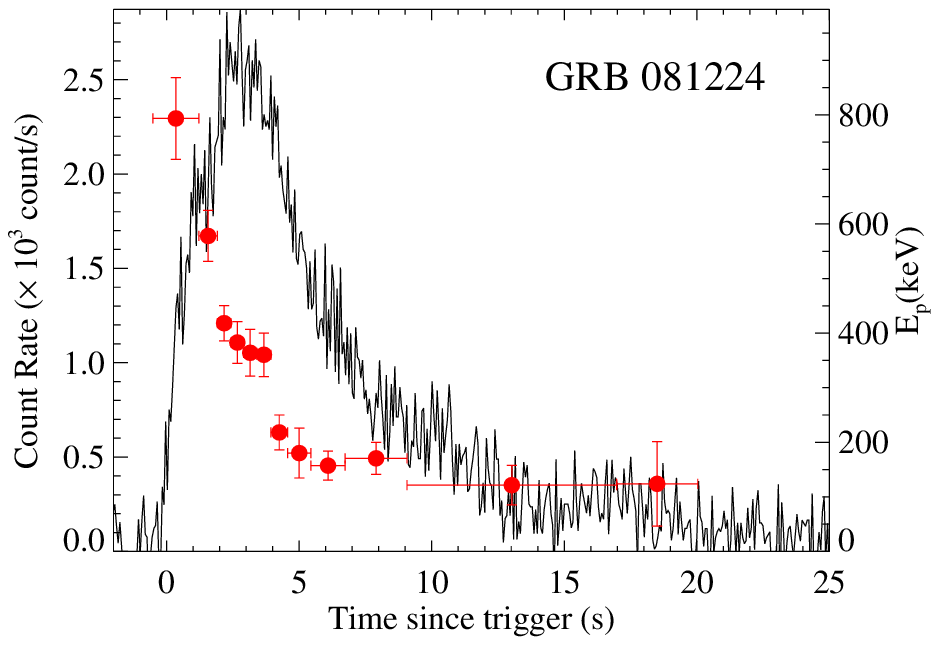}}
 \resizebox{4cm}{!}{\includegraphics{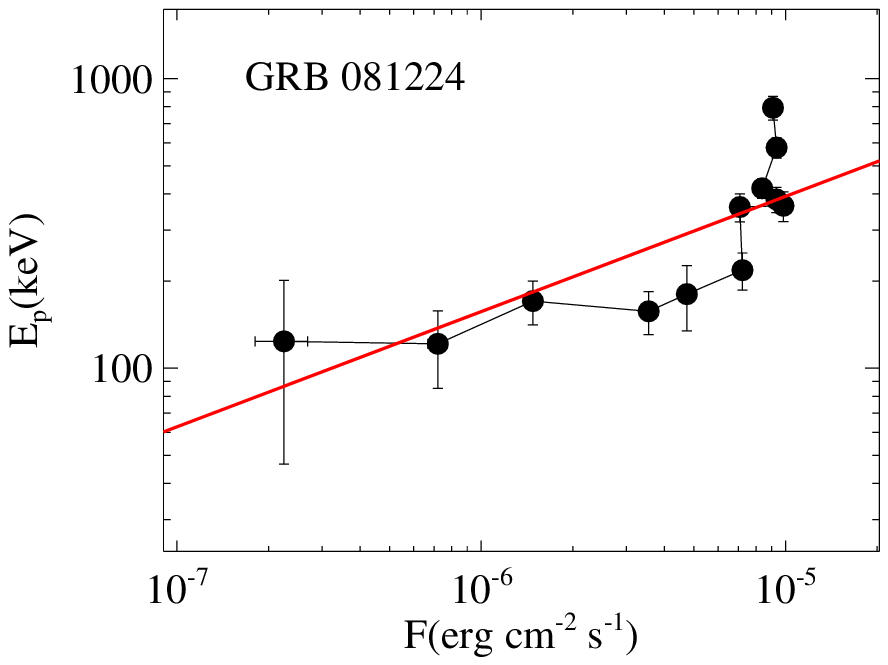}}

 \resizebox{4cm}{!}{\includegraphics{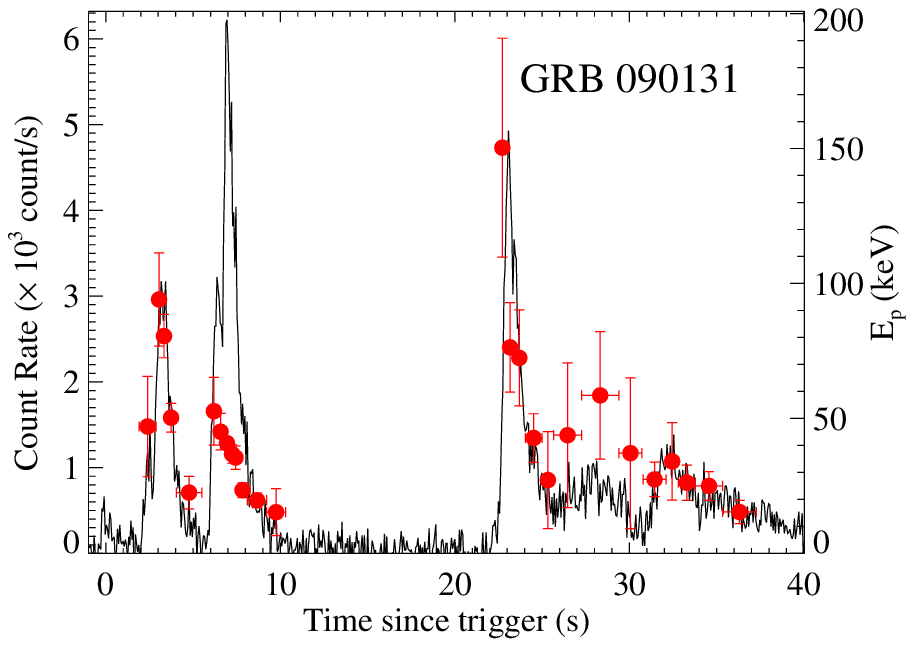}}
 \resizebox{4cm}{!}{\includegraphics{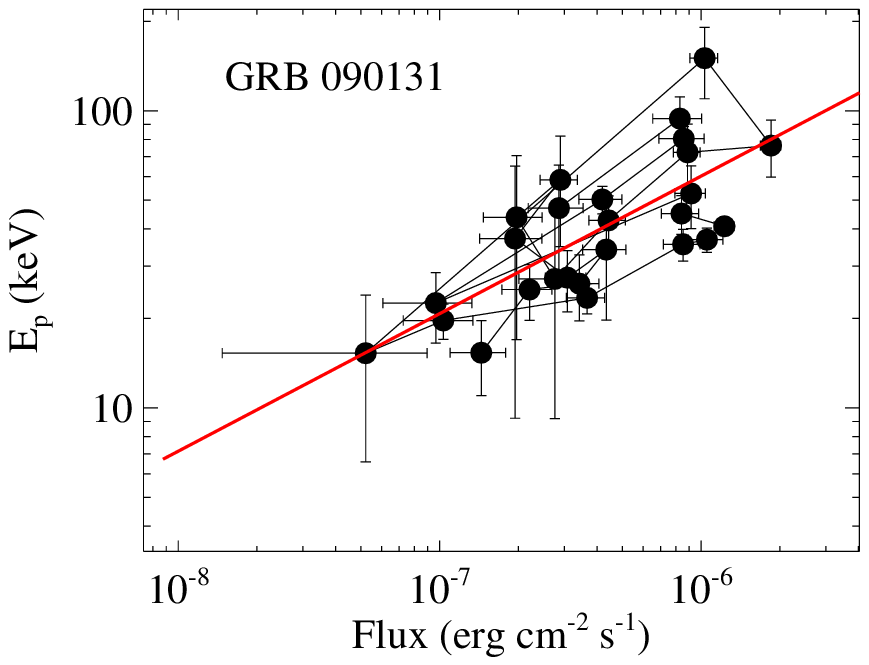}}
 \resizebox{4cm}{!}{\includegraphics{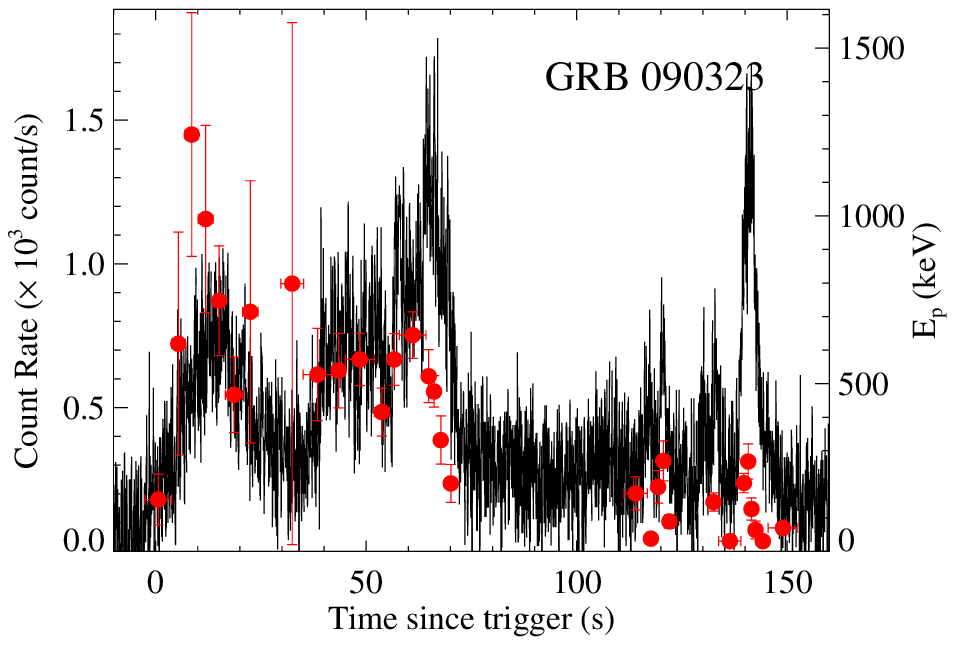}}
\resizebox{4cm}{!}{\includegraphics{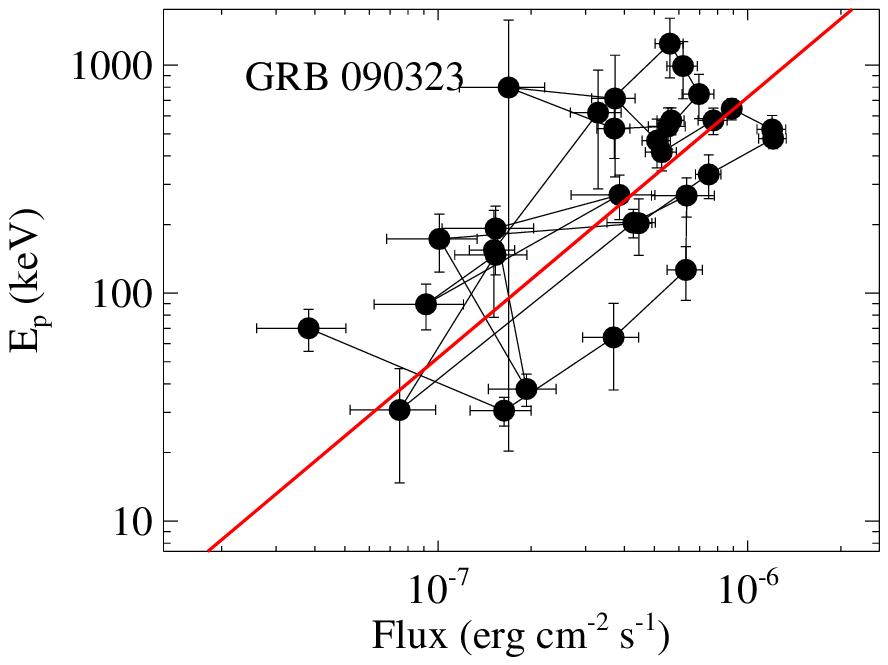}}

\resizebox{4cm}{!}{\includegraphics{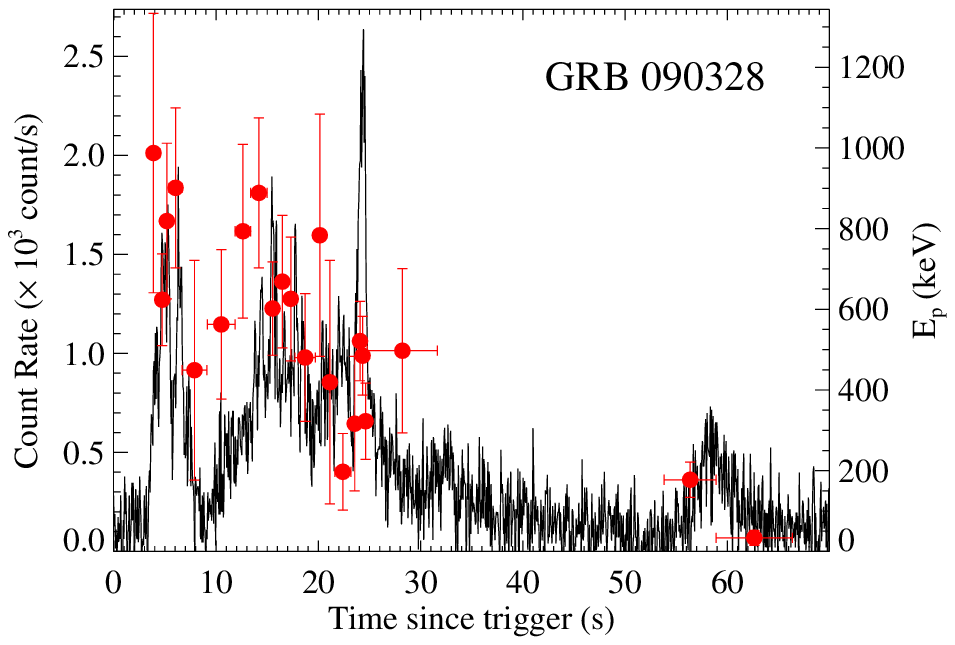}}
\resizebox{4cm}{!}{\includegraphics{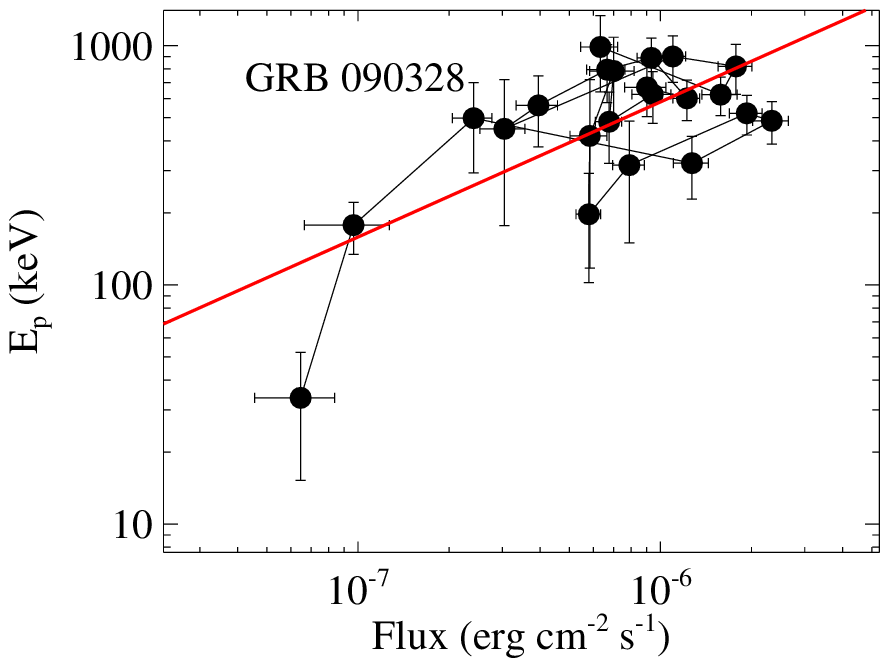}}
\resizebox{4cm}{!}{\includegraphics{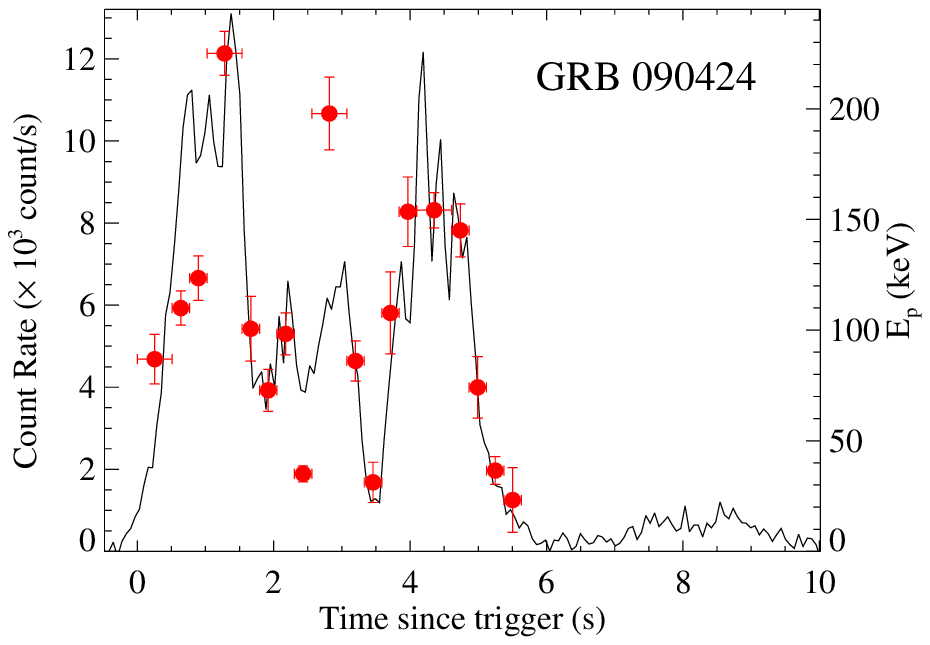}}
\resizebox{4cm}{!}{\includegraphics{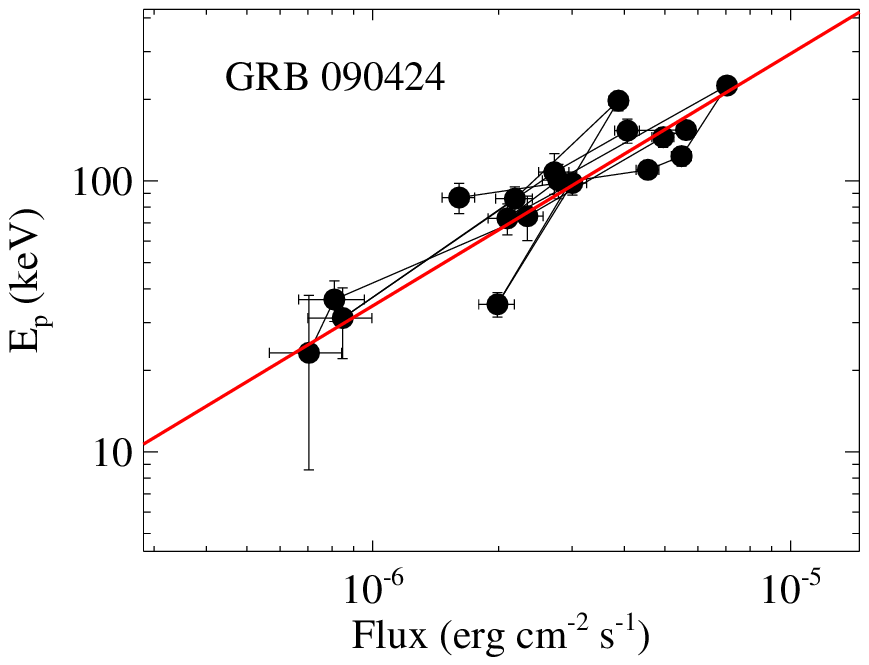}} \caption{{\em
Left:} Light curves (connected lines) along with $E_{\rm p}$
evolution (circles with error bars) of long GRBs in our sample; {\em
Right:} Time-resolved $E_{\rm p}$ as a function of flux, along with
the best fit line for the $E_{\rm p}-F$ correlation for the long
GRBs in our sample.} \label{long}
\end{figure*}

\addtocounter{figure}{-1}
\begin{figure*}
\caption{{\it-continued.} }
 \resizebox{4cm}{!}{\includegraphics{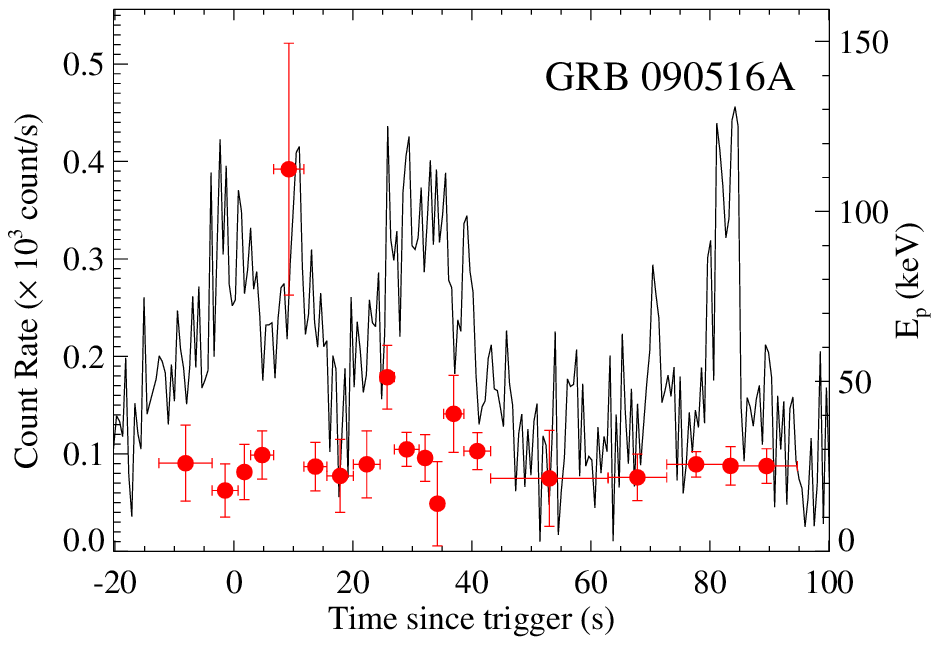}}
 \resizebox{4cm}{!}{\includegraphics{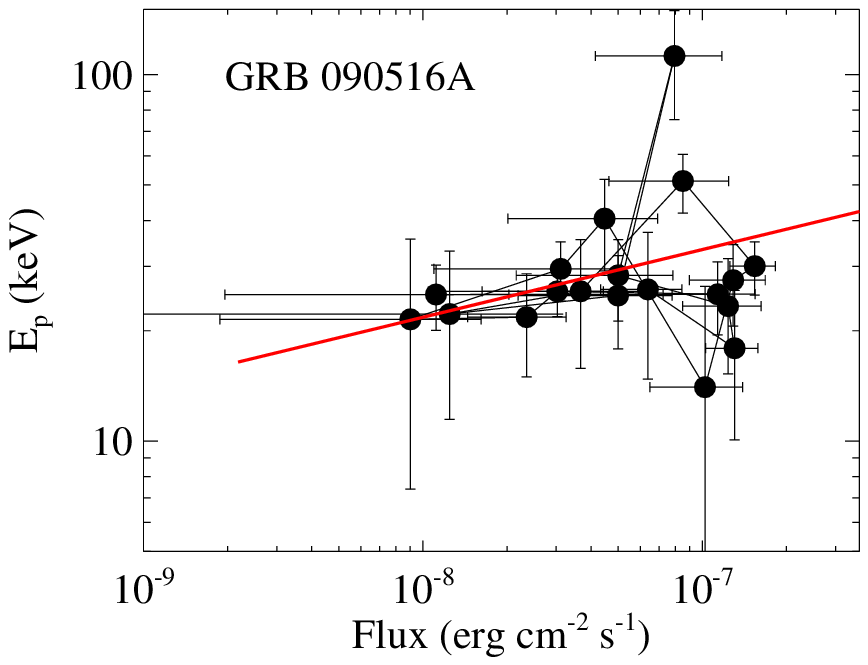}}
 \resizebox{4cm}{!}{\includegraphics{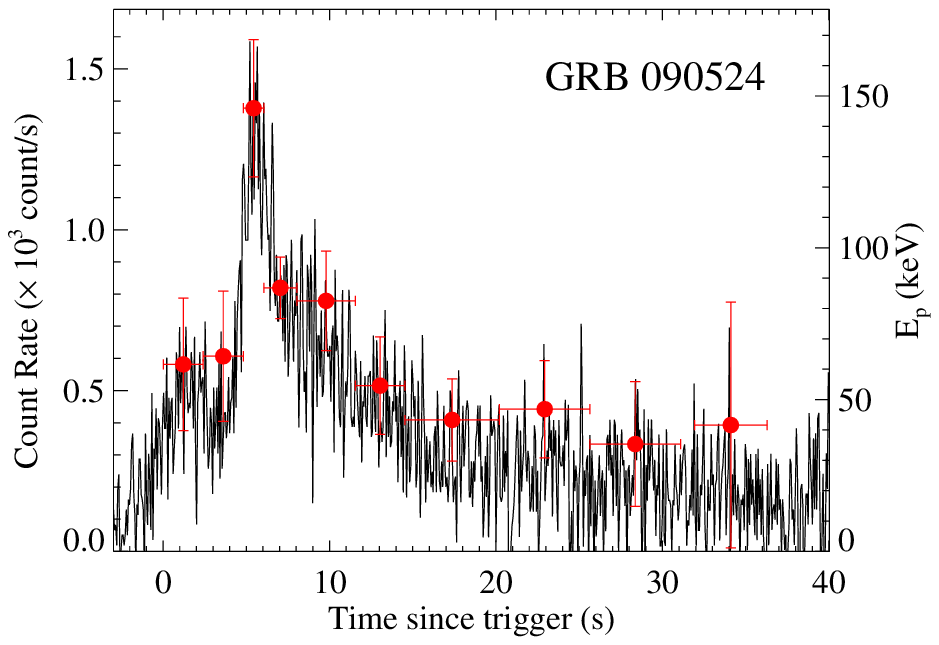}}
\resizebox{4cm}{!}{\includegraphics{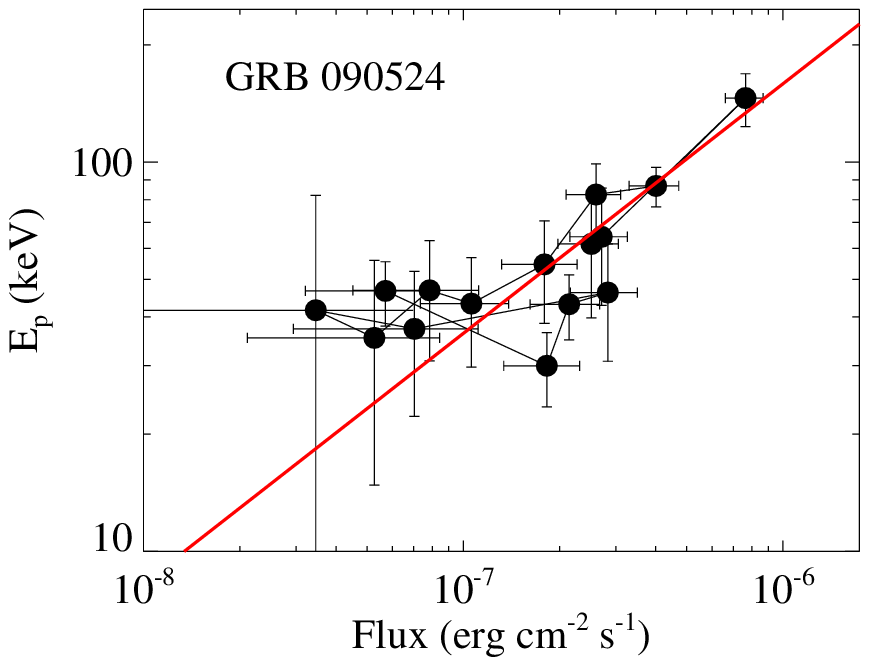}}

 \resizebox{4cm}{!}{\includegraphics{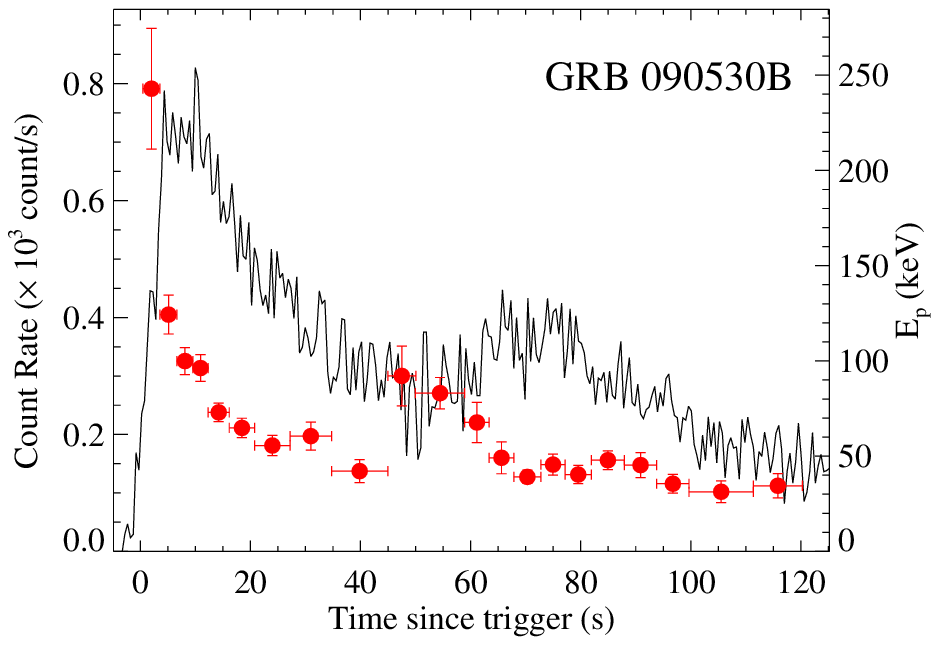}}
\resizebox{4cm}{!}{\includegraphics{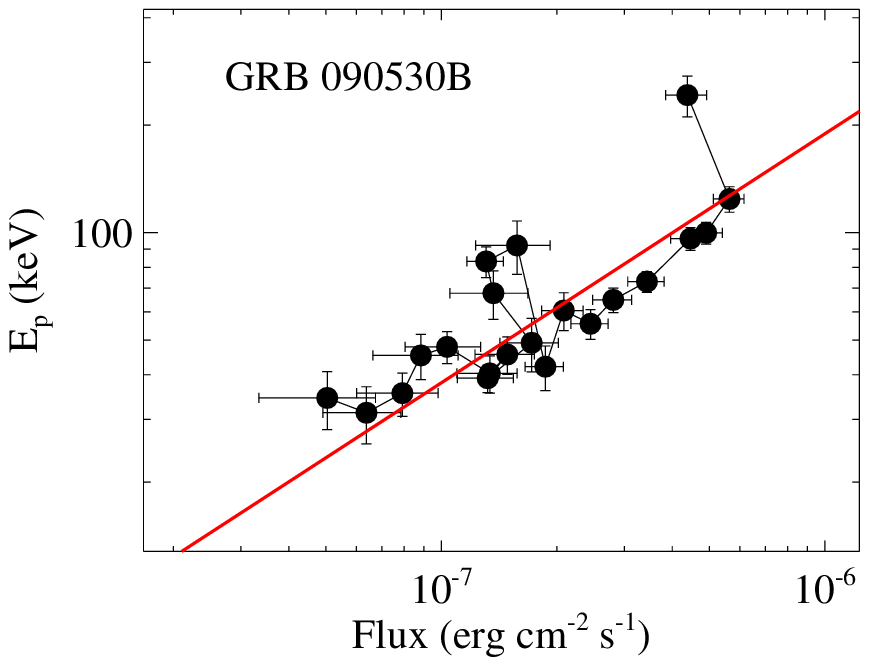}}
\resizebox{4cm}{!}{\includegraphics{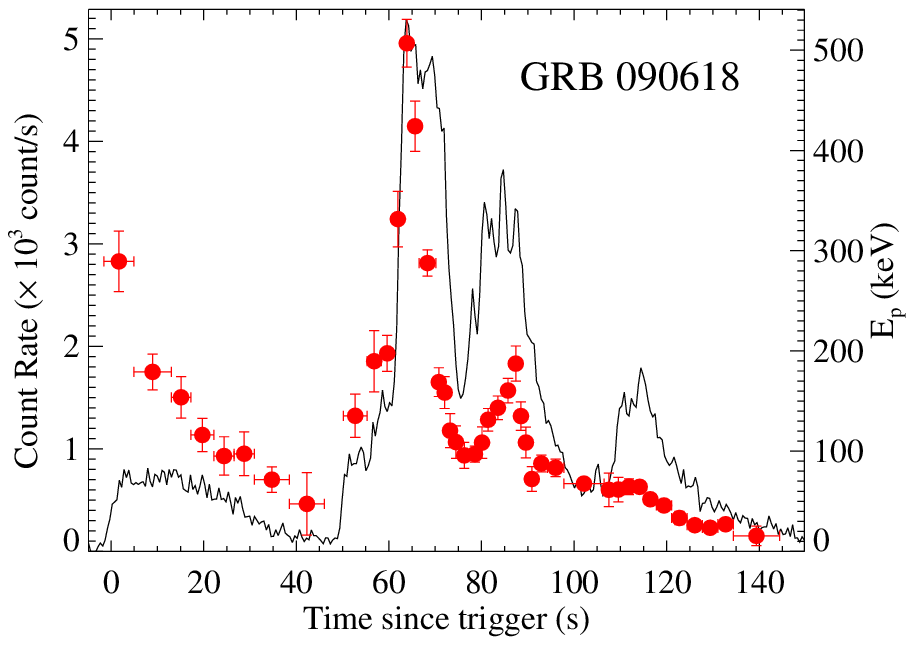}}
\resizebox{4cm}{!}{\includegraphics{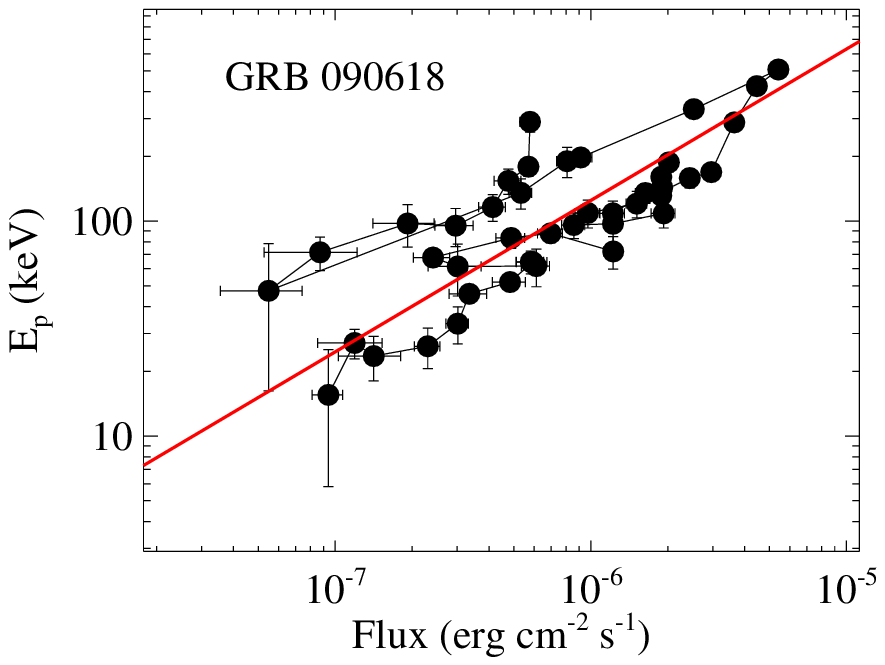}}

\resizebox{4cm}{!}{\includegraphics{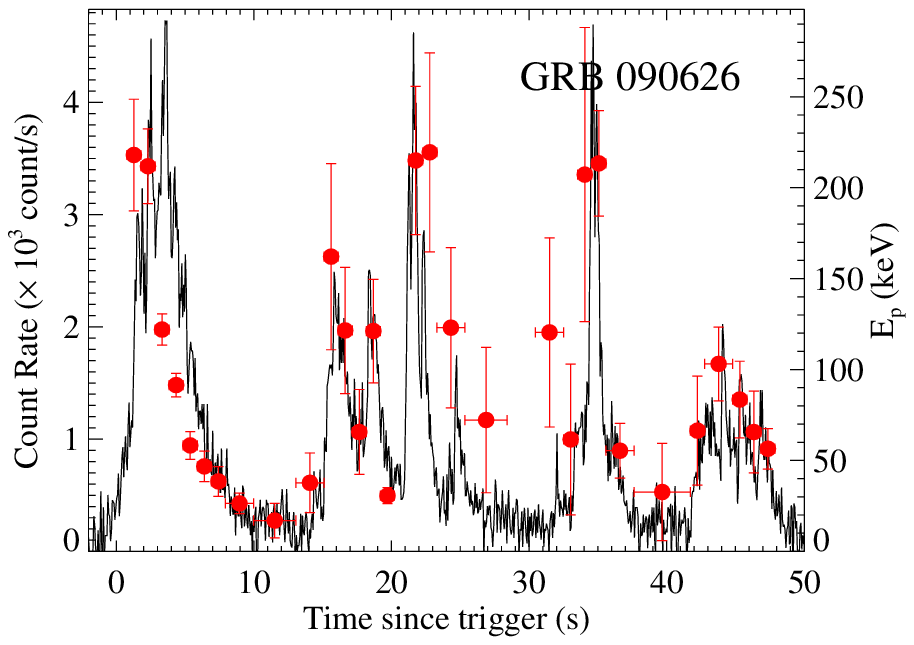}}
\resizebox{4cm}{!}{\includegraphics{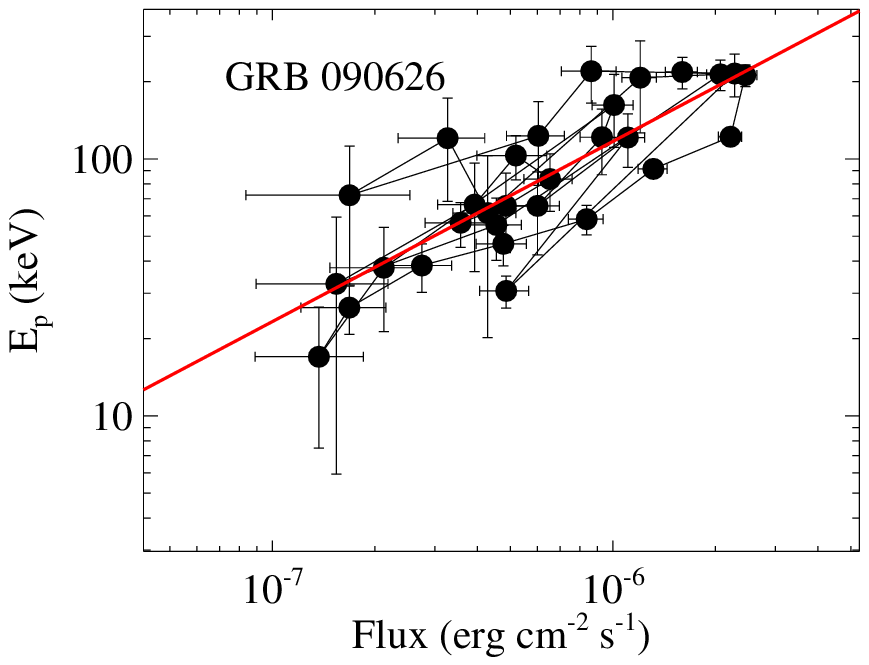}}
 \resizebox{4cm}{!}{\includegraphics{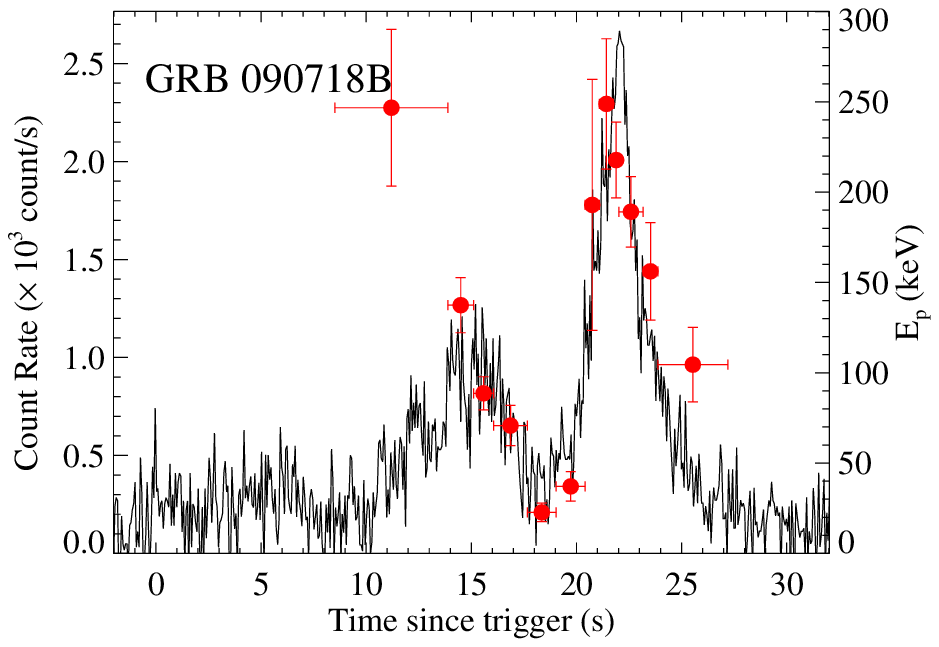}}
\resizebox{4cm}{!}{\includegraphics{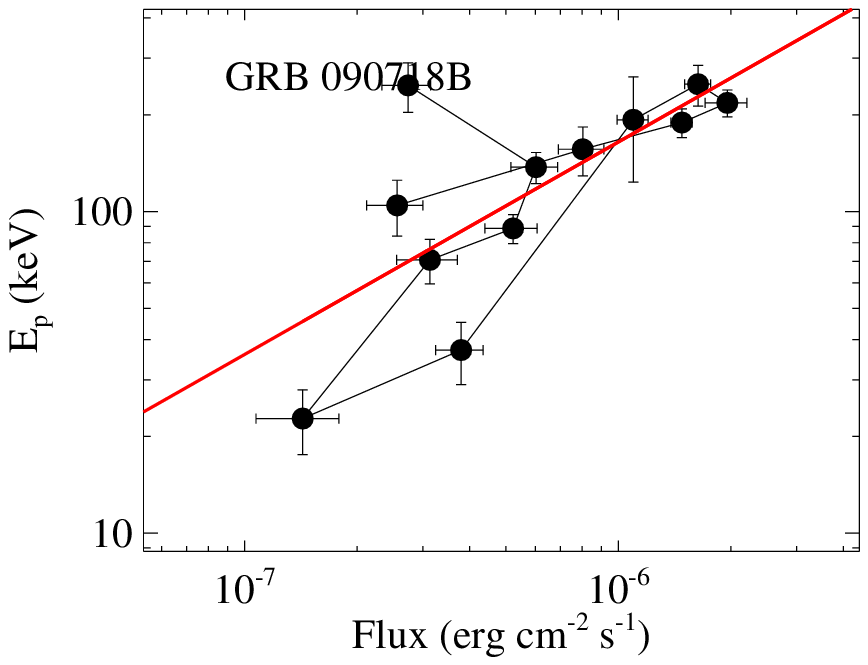}}

 \resizebox{4cm}{!}{\includegraphics{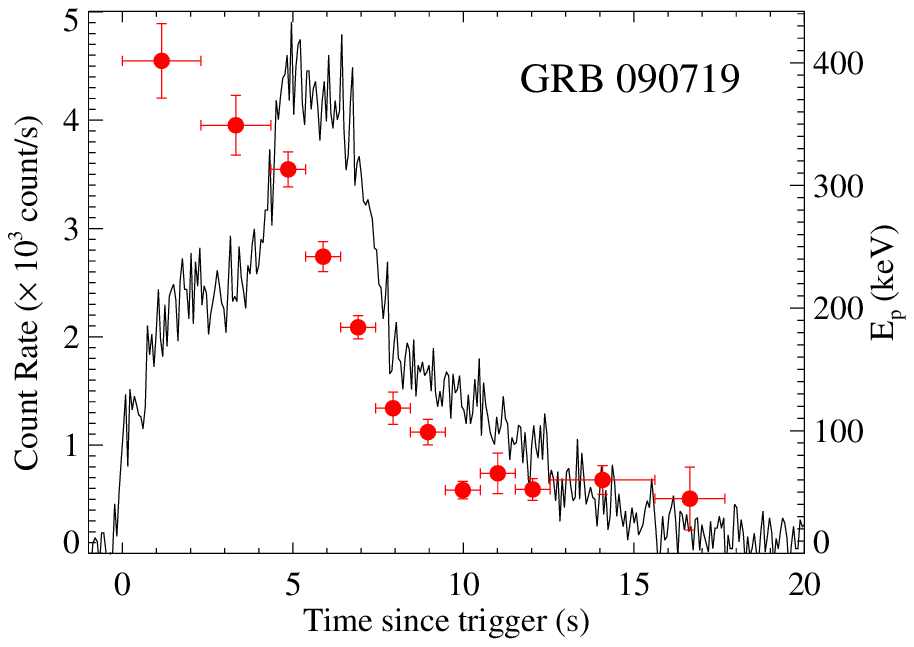}}
\resizebox{4cm}{!}{\includegraphics{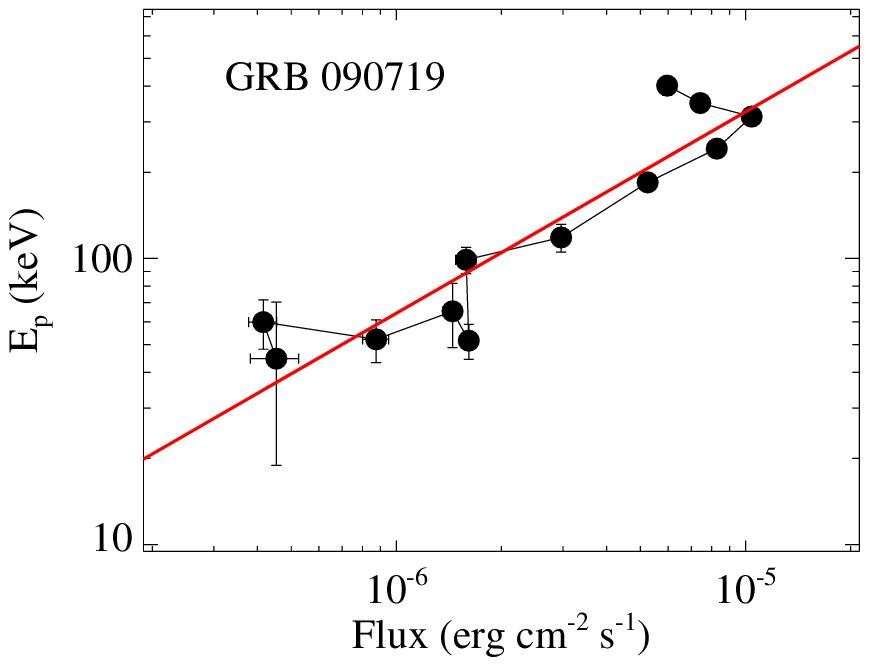}}
 \resizebox{4cm}{!}{\includegraphics{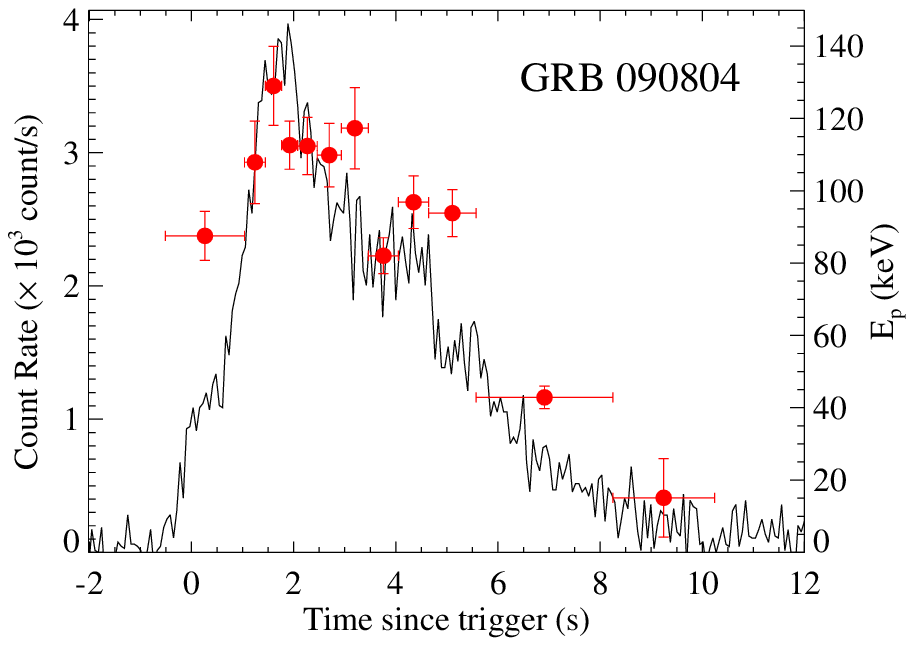}}
\resizebox{4cm}{!}{\includegraphics{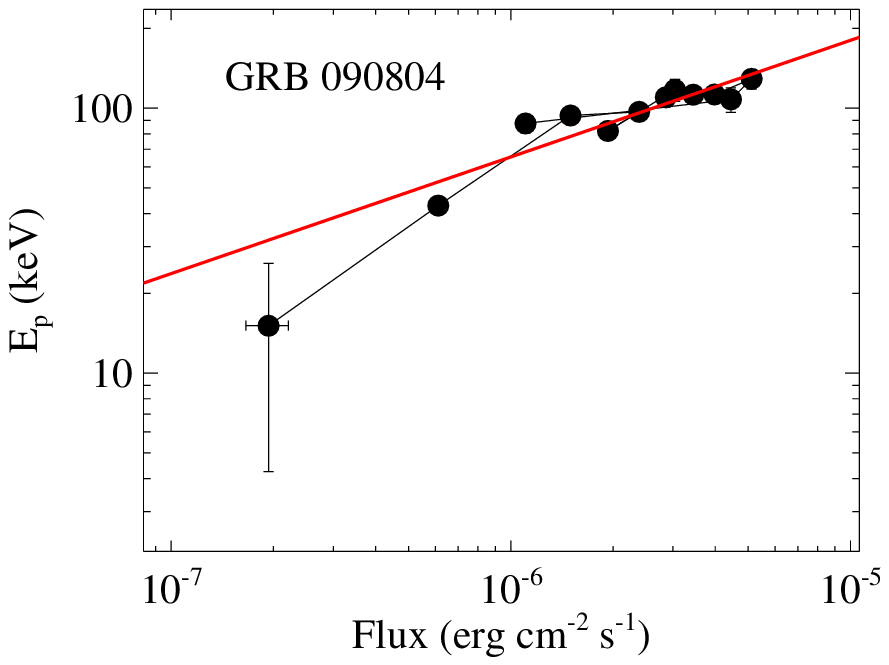}}

 \resizebox{4cm}{!}{\includegraphics{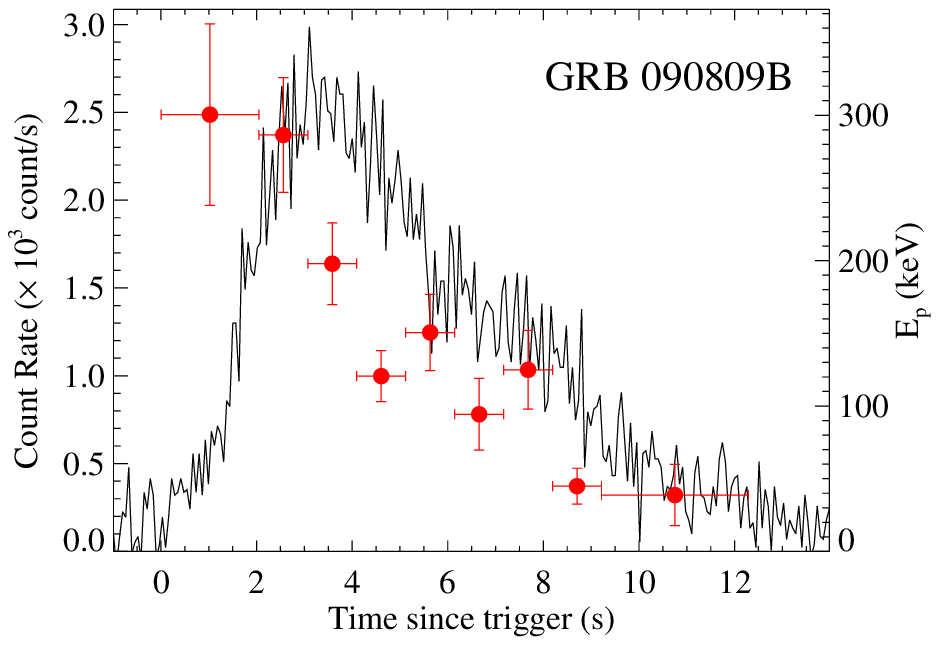}}
\resizebox{4cm}{!}{\includegraphics{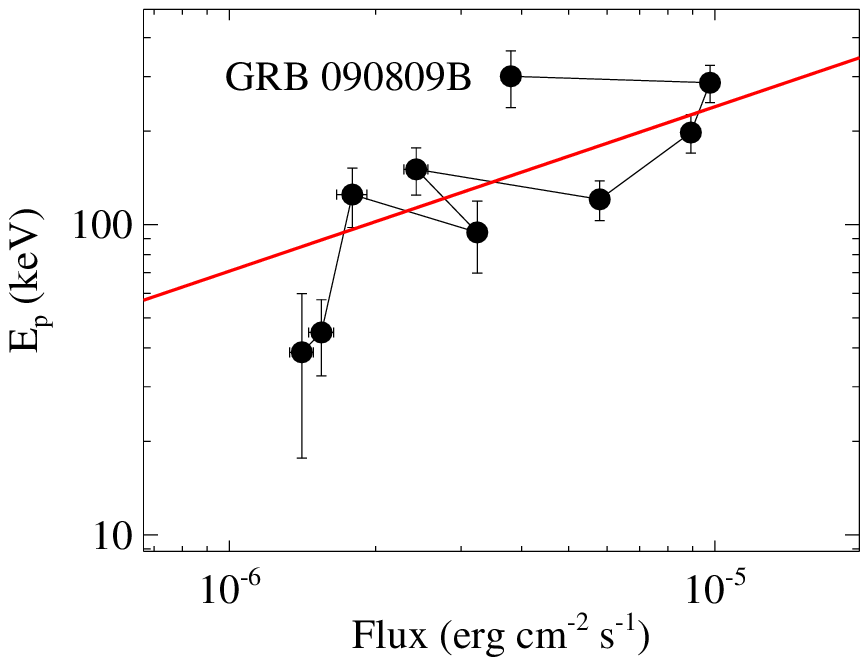}}
 \resizebox{4cm}{!}{\includegraphics{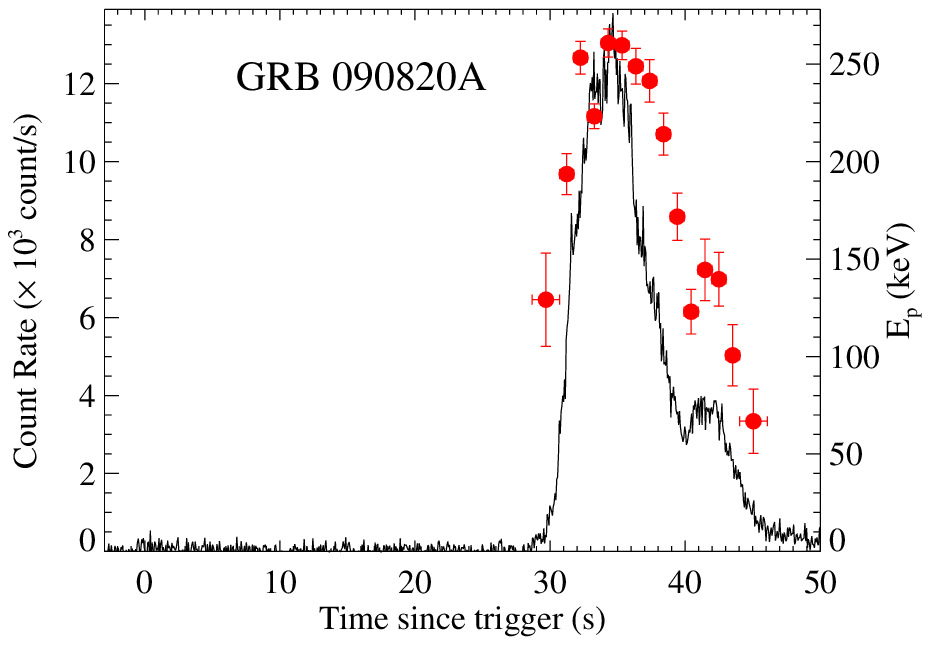}}
\resizebox{4cm}{!}{\includegraphics{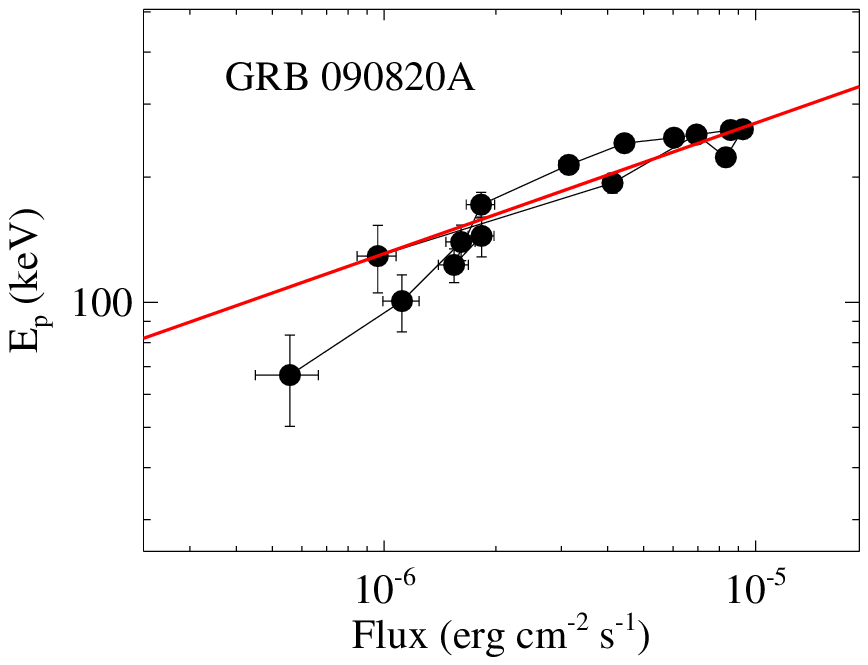}}

 \resizebox{4cm}{!}{\includegraphics{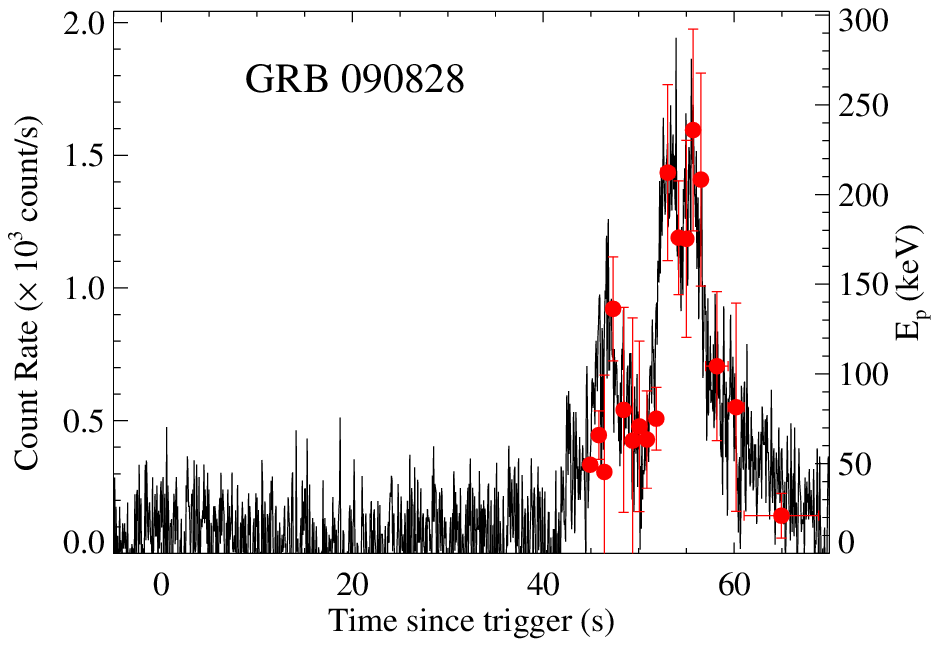}}
\resizebox{4cm}{!}{\includegraphics{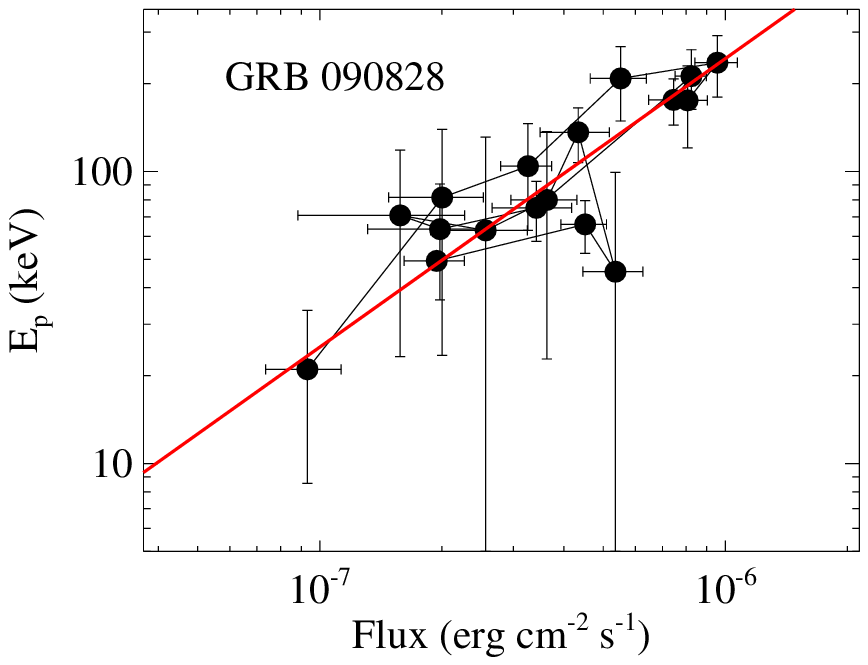}}
 \resizebox{4cm}{!}{\includegraphics{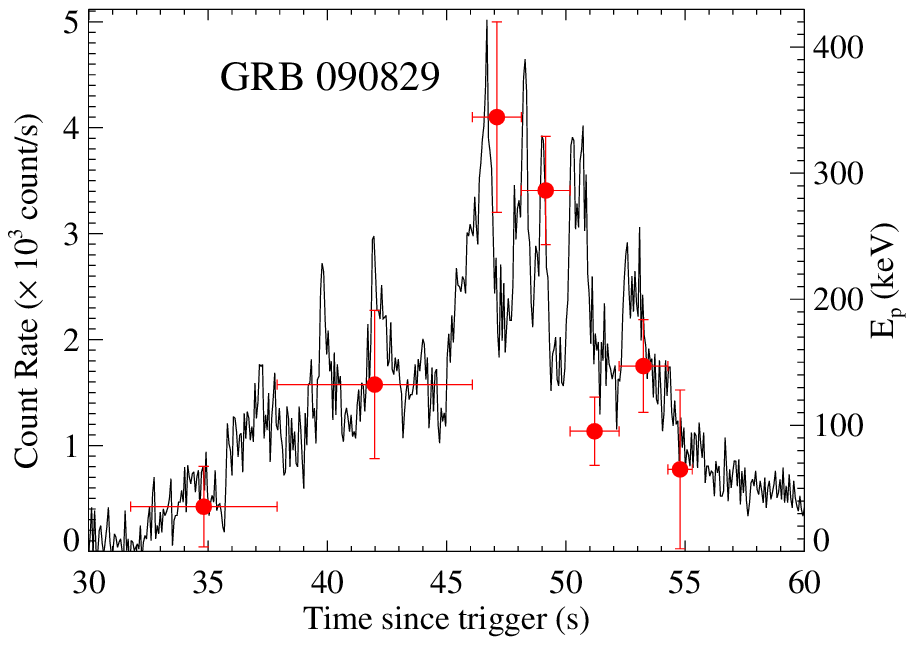}}
\resizebox{4cm}{!}{\includegraphics{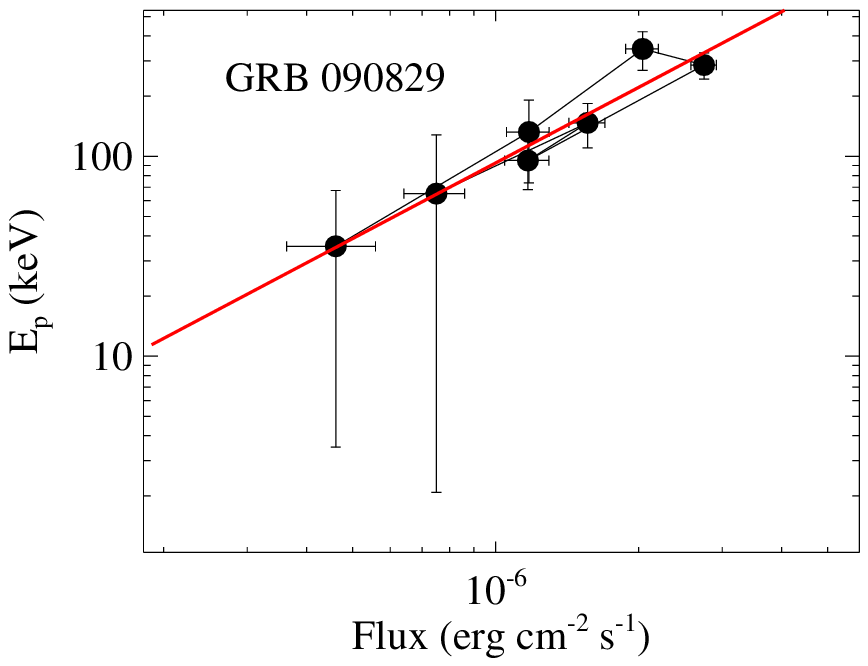}}

\resizebox{4cm}{!}{\includegraphics{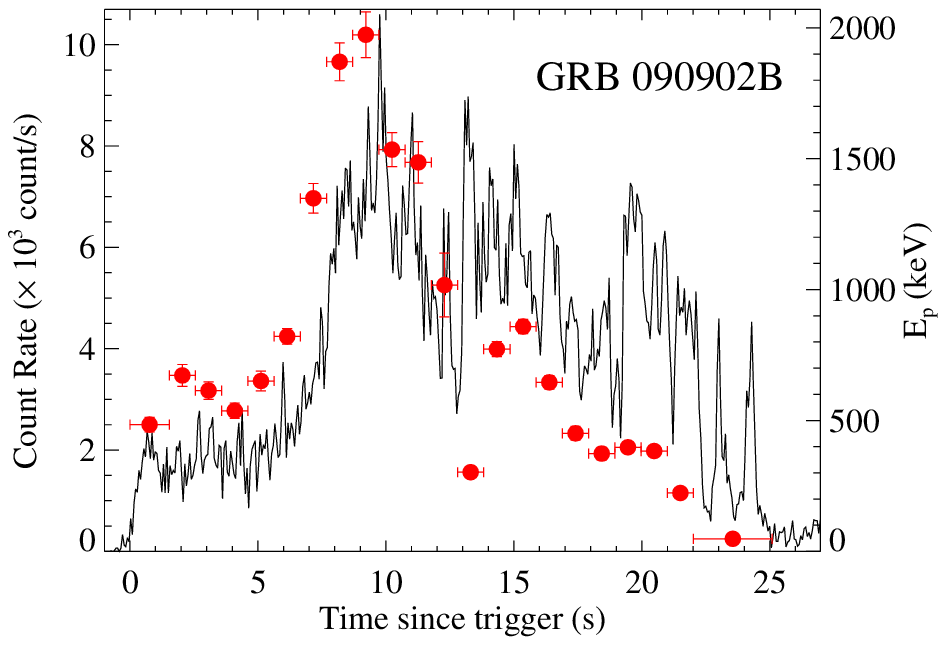}}
\resizebox{4cm}{!}{\includegraphics{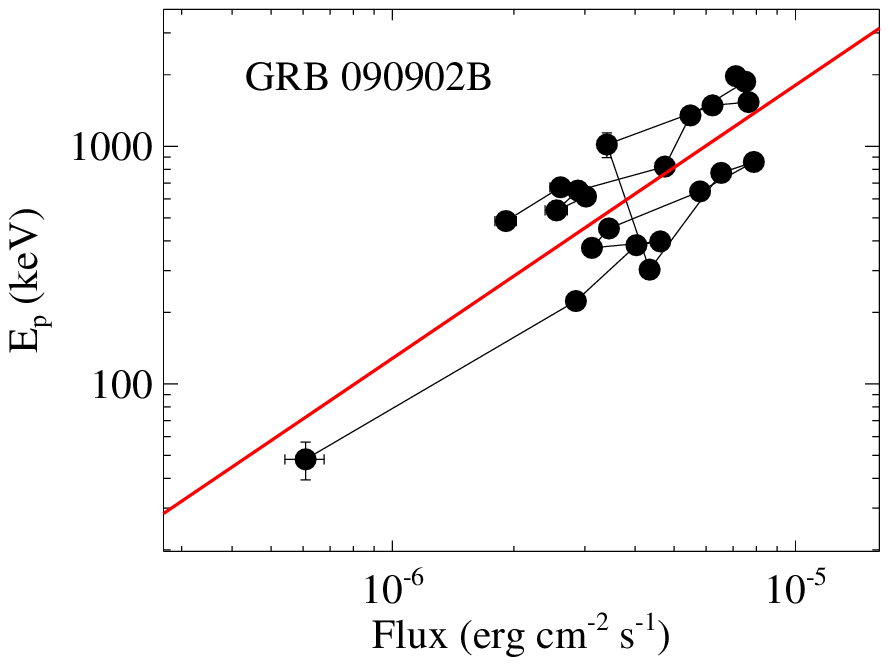}}
 \resizebox{4cm}{!}{\includegraphics{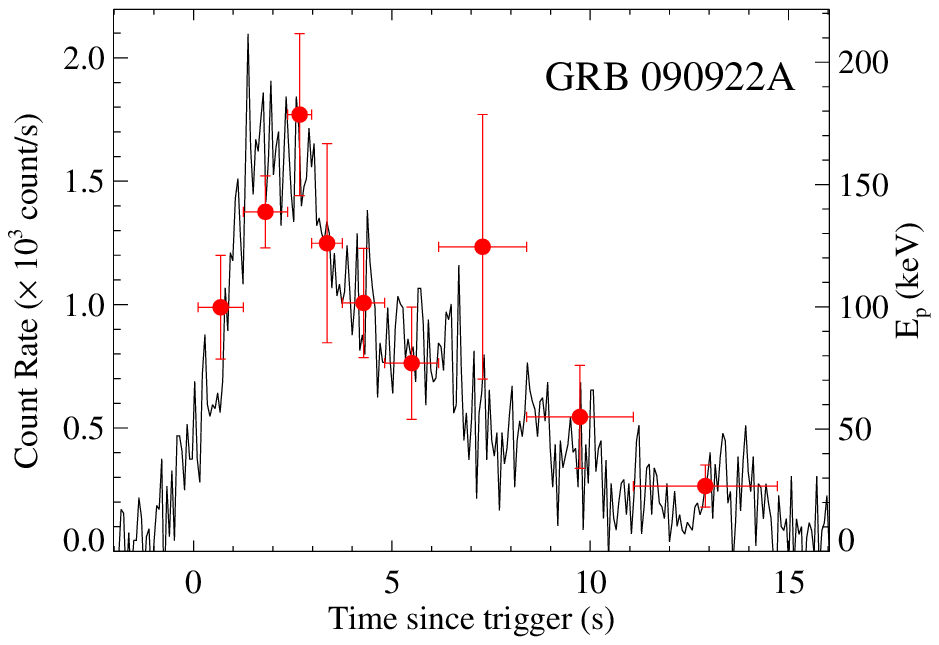}}
\resizebox{4cm}{!}{\includegraphics{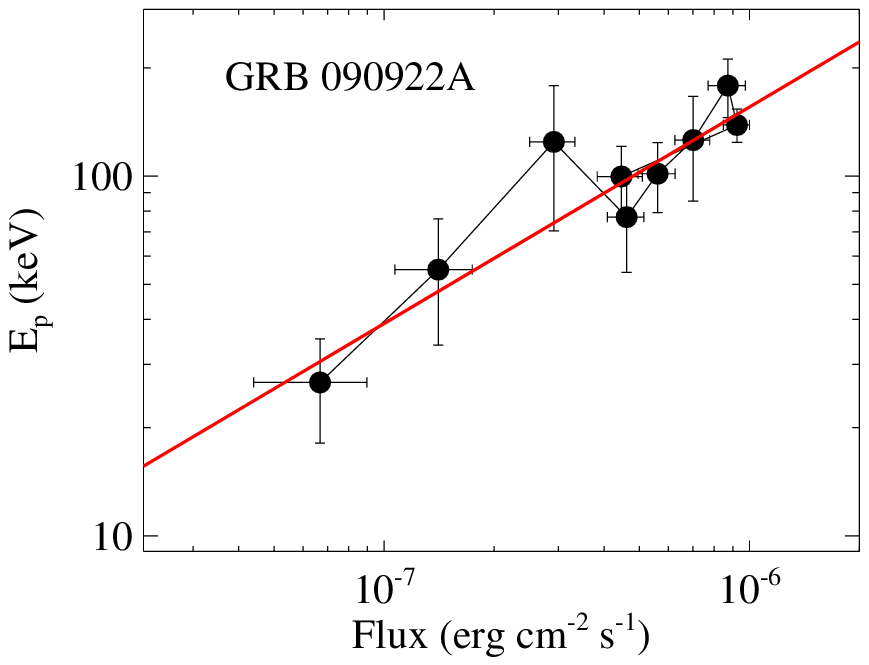}}
\end{figure*}

\addtocounter{figure}{-1}
\begin{figure*}
\caption{{\it-continued.} }
 \resizebox{4cm}{!}{\includegraphics{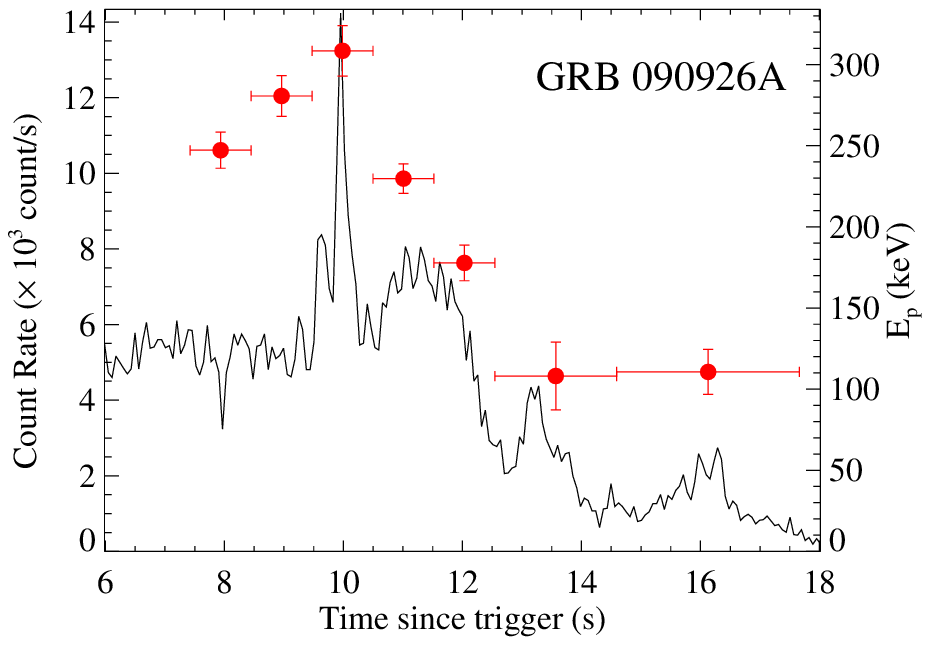}}
 \resizebox{4cm}{!}{\includegraphics{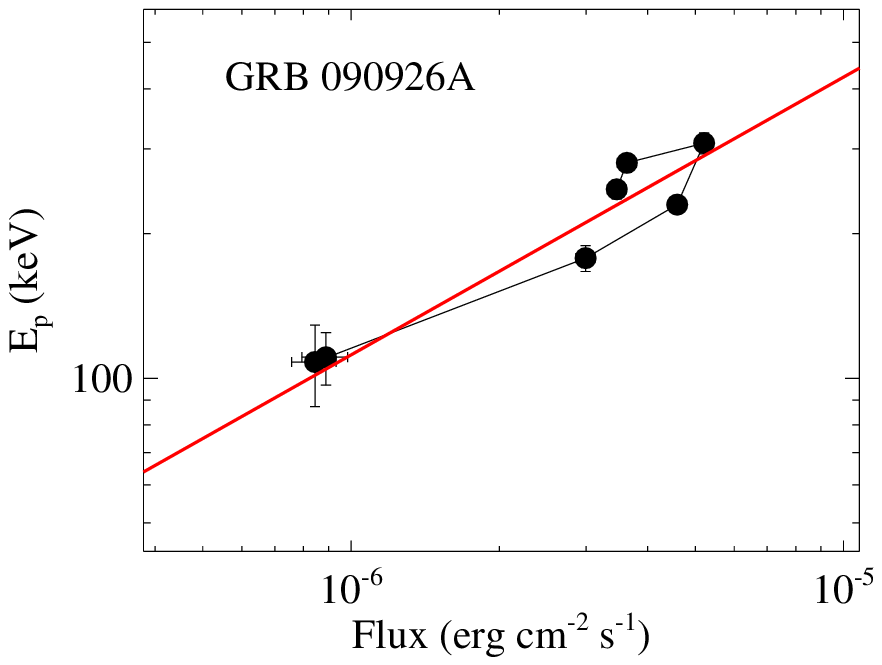}}
\resizebox{4cm}{!}{\includegraphics{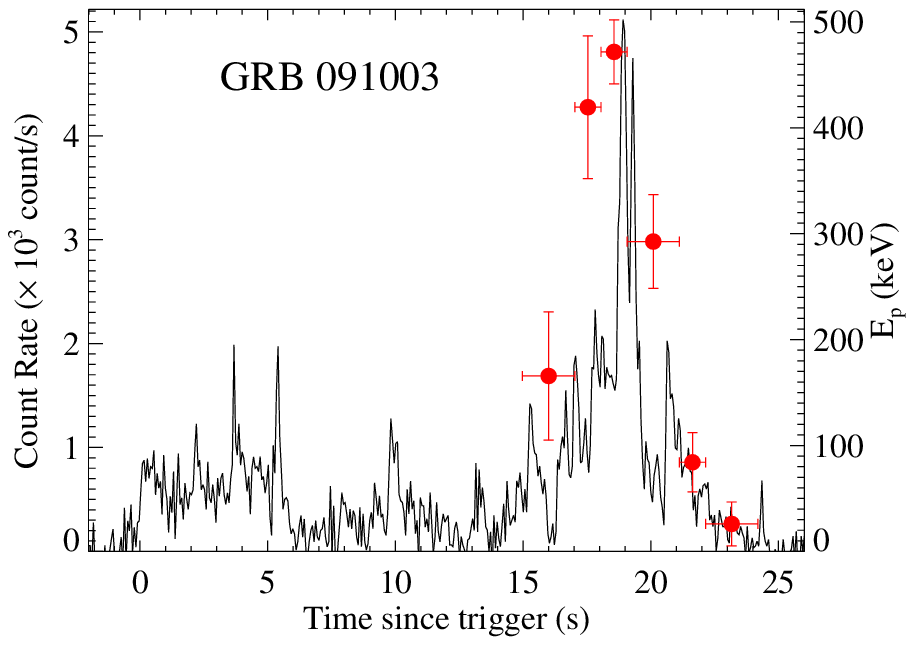}}
\resizebox{4cm}{!}{\includegraphics{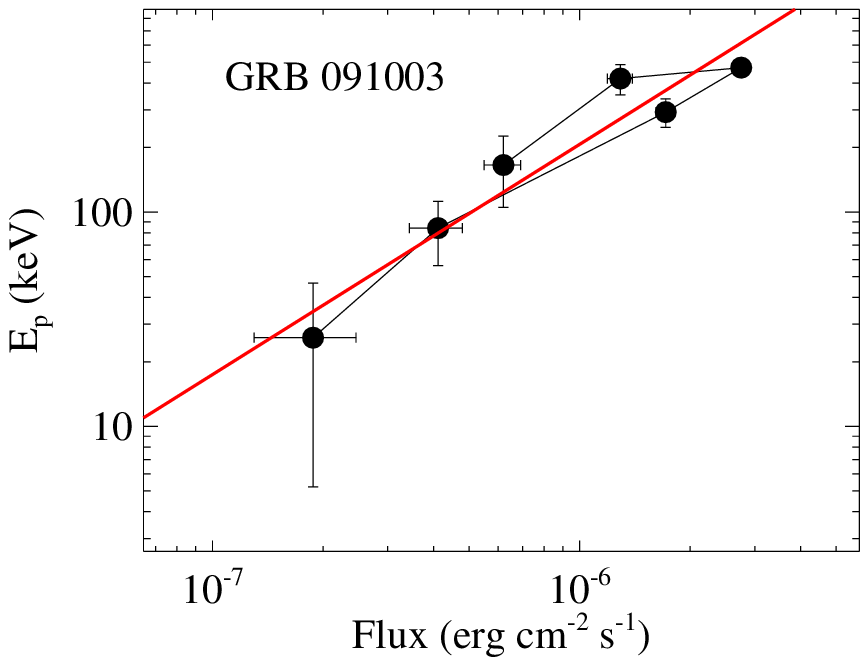}}

\resizebox{4cm}{!}{\includegraphics{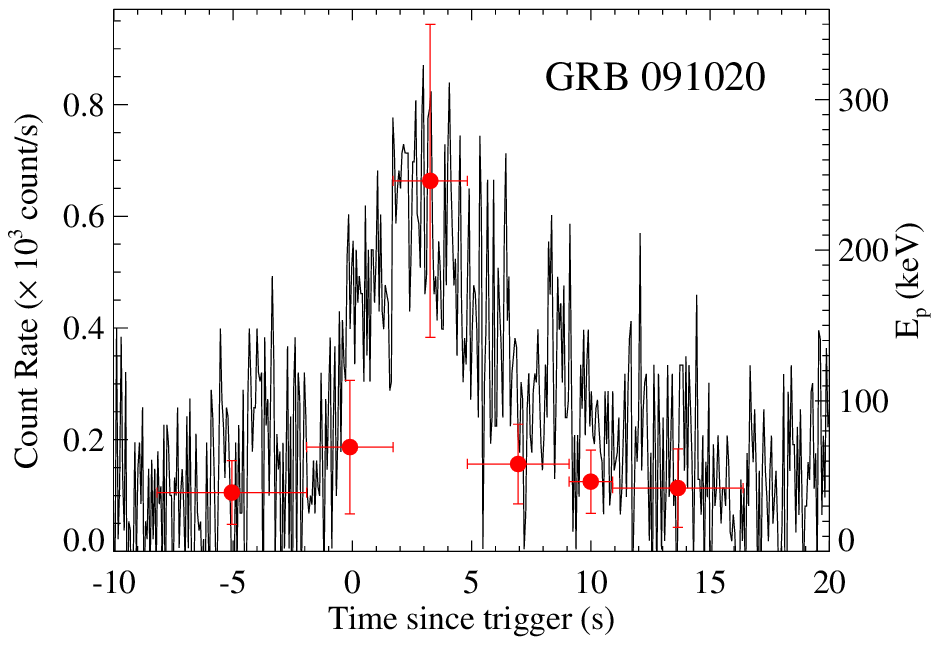}}
\resizebox{4cm}{!}{\includegraphics{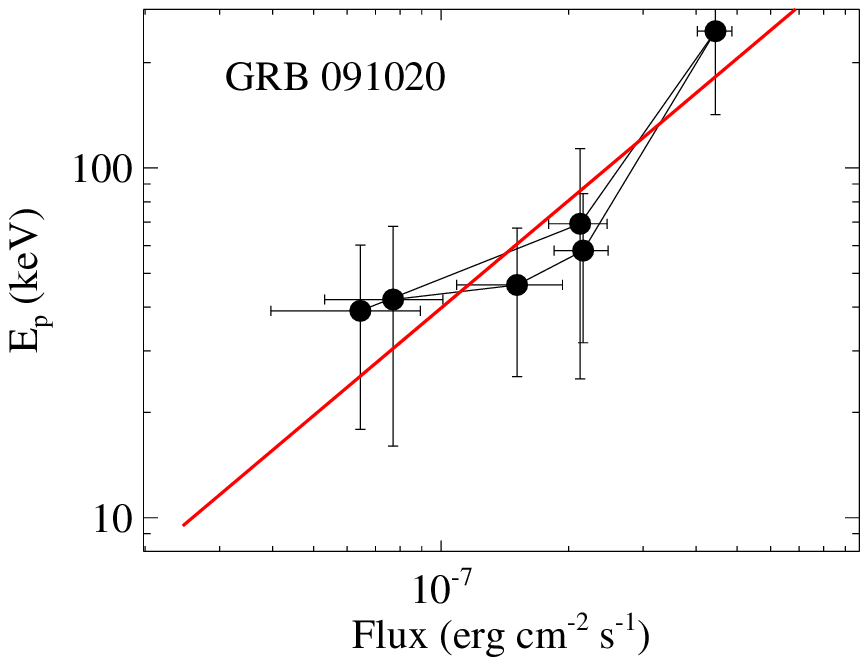}}
\resizebox{4cm}{!}{\includegraphics{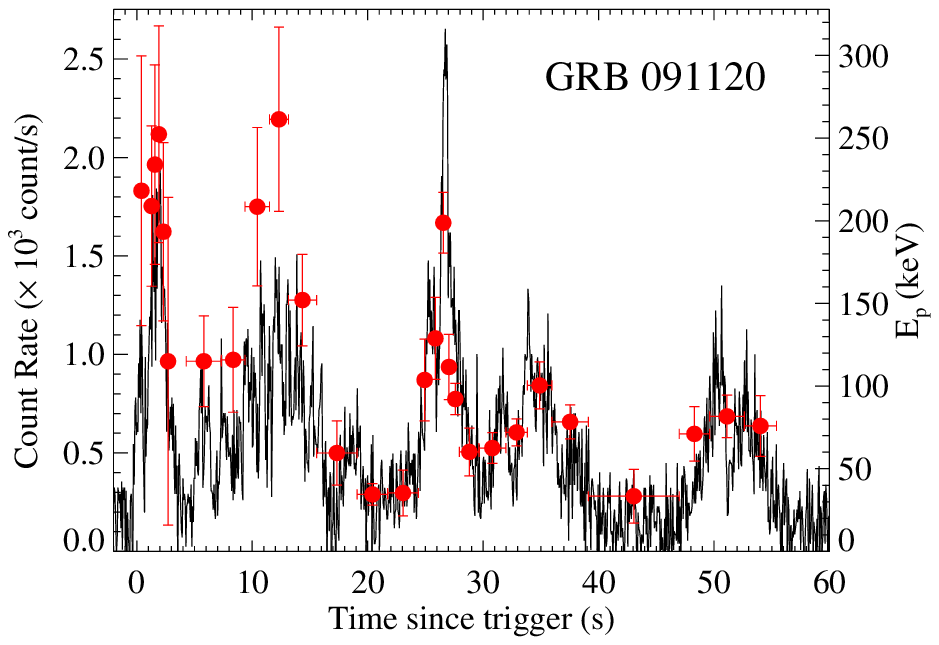}}
\resizebox{4cm}{!}{\includegraphics{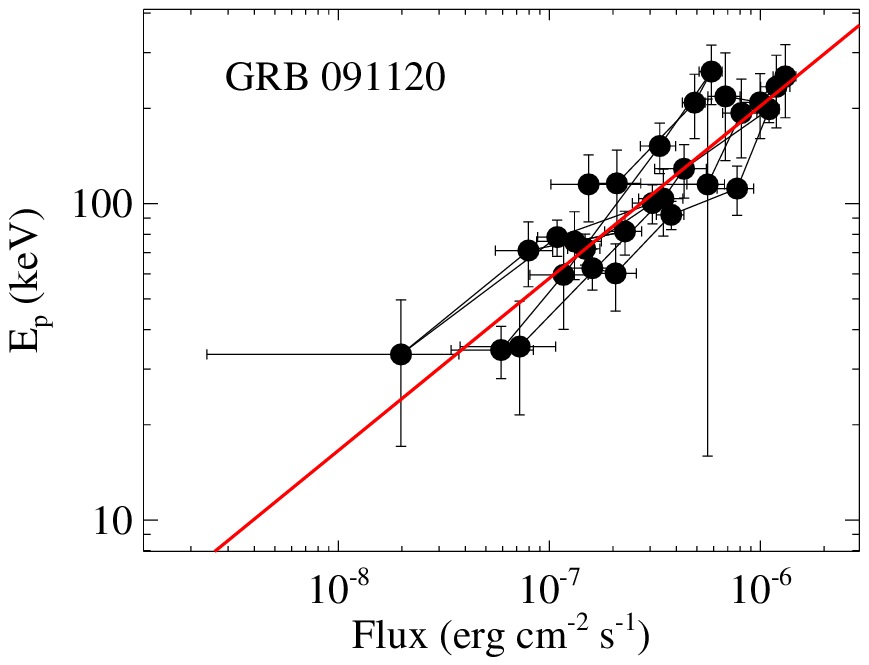}}

\resizebox{4cm}{!}{\includegraphics{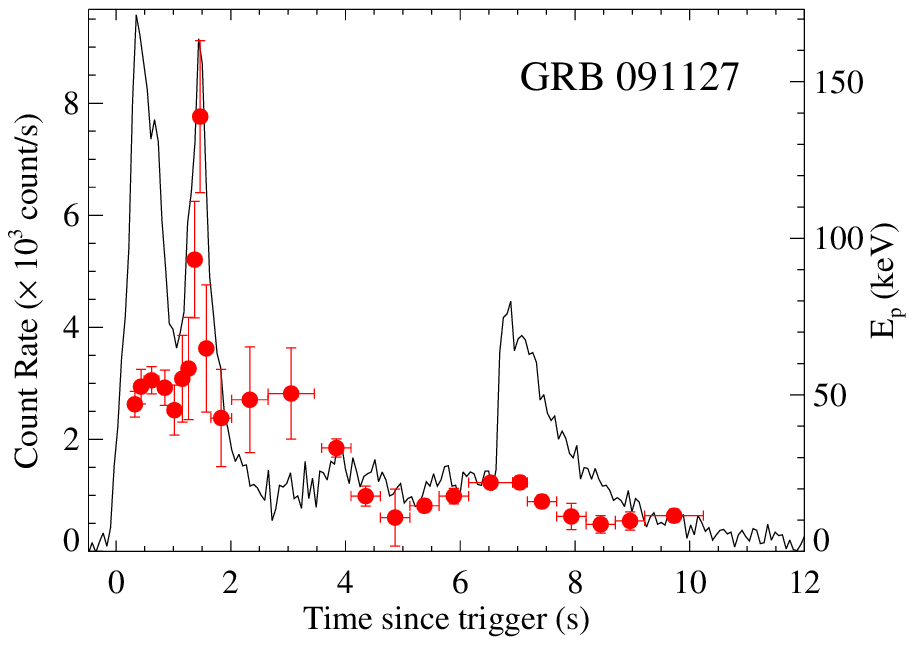}}
\resizebox{4cm}{!}{\includegraphics{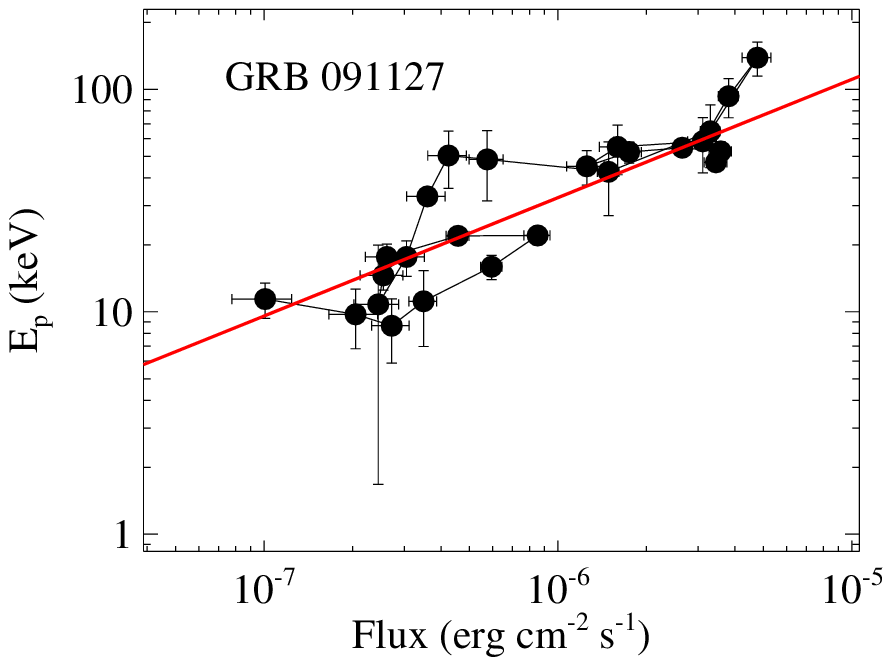}}
 \resizebox{4cm}{!}{\includegraphics{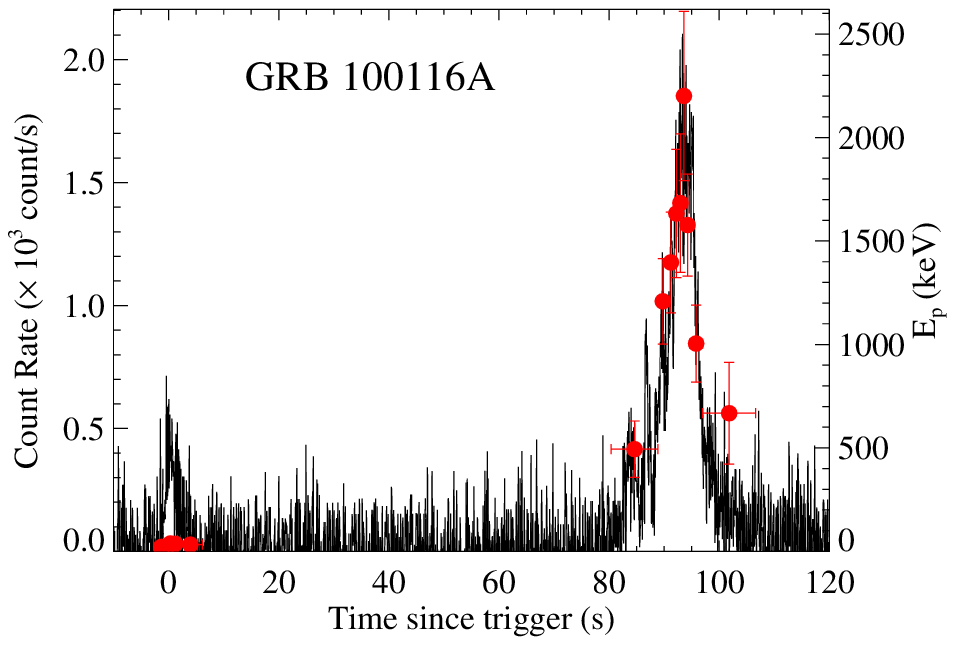}}
\resizebox{4cm}{!}{\includegraphics{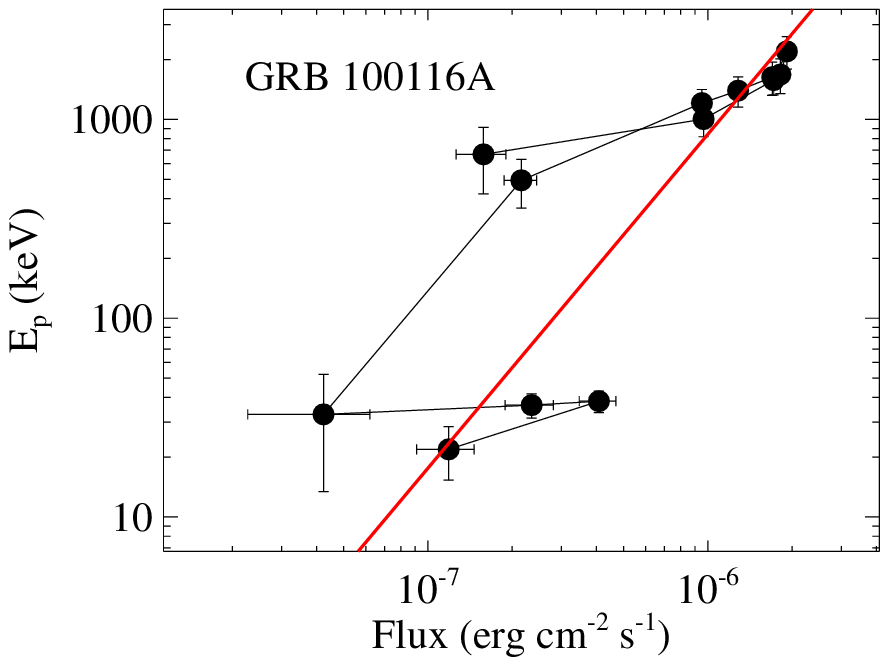}}

 \resizebox{4cm}{!}{\includegraphics{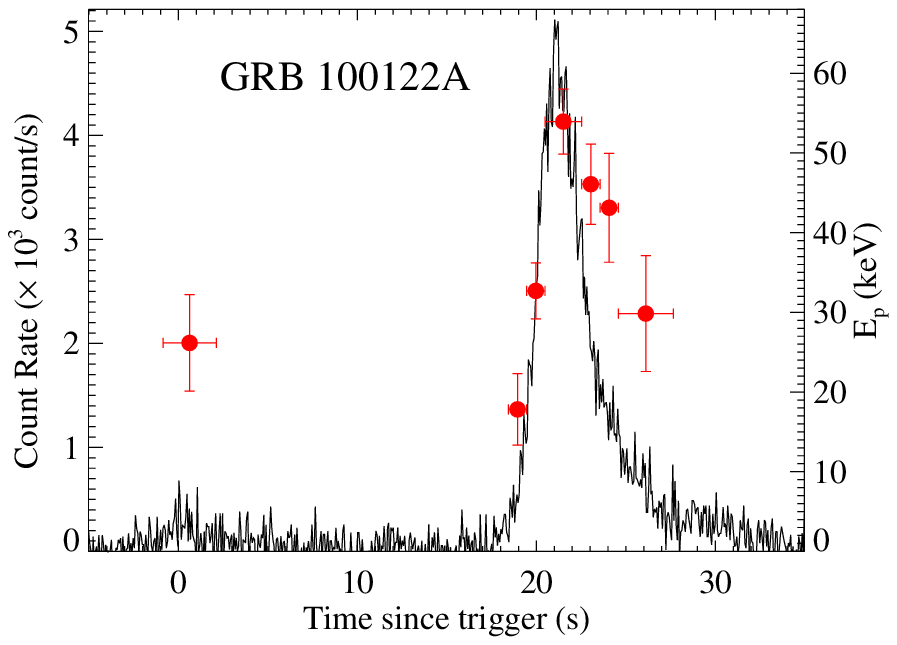}}
\resizebox{4cm}{!}{\includegraphics{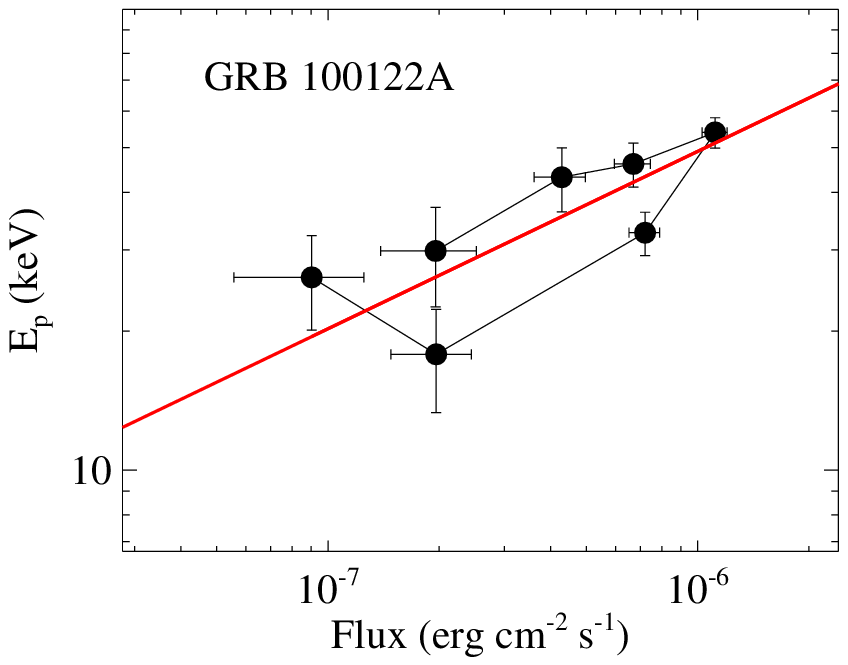}}
 \resizebox{4cm}{!}{\includegraphics{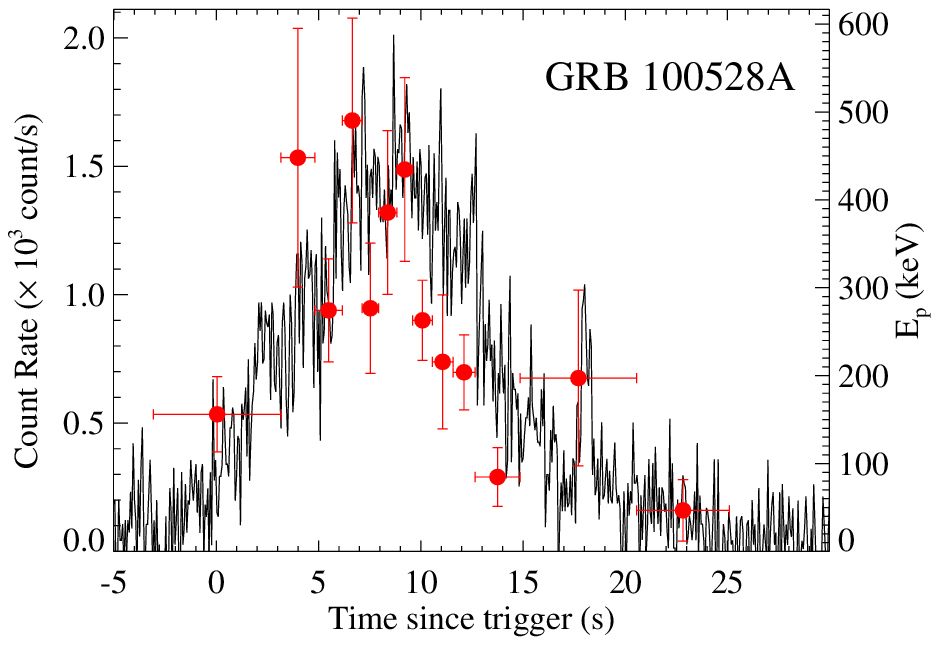}}
\resizebox{4cm}{!}{\includegraphics{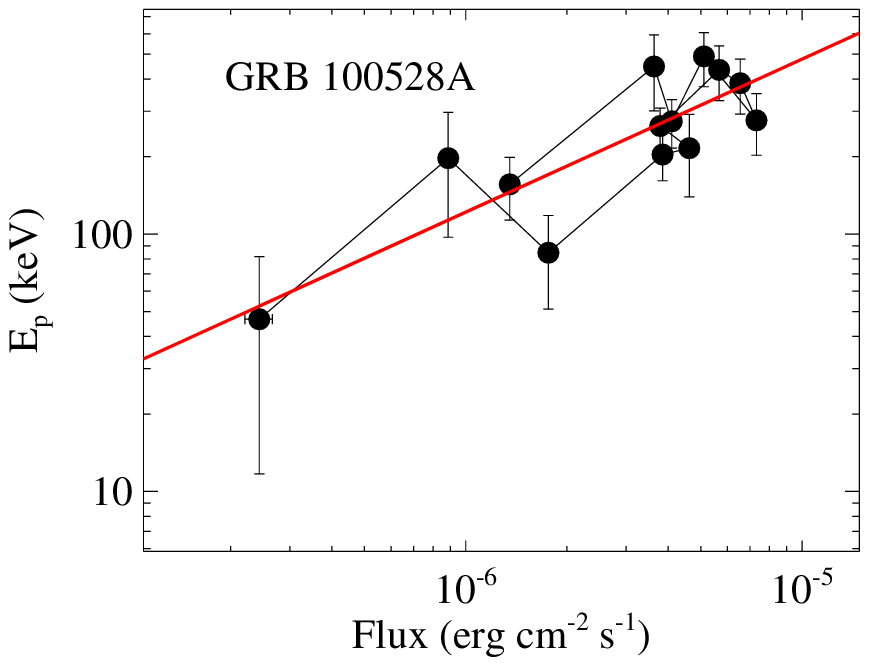}}

 \resizebox{4cm}{!}{\includegraphics{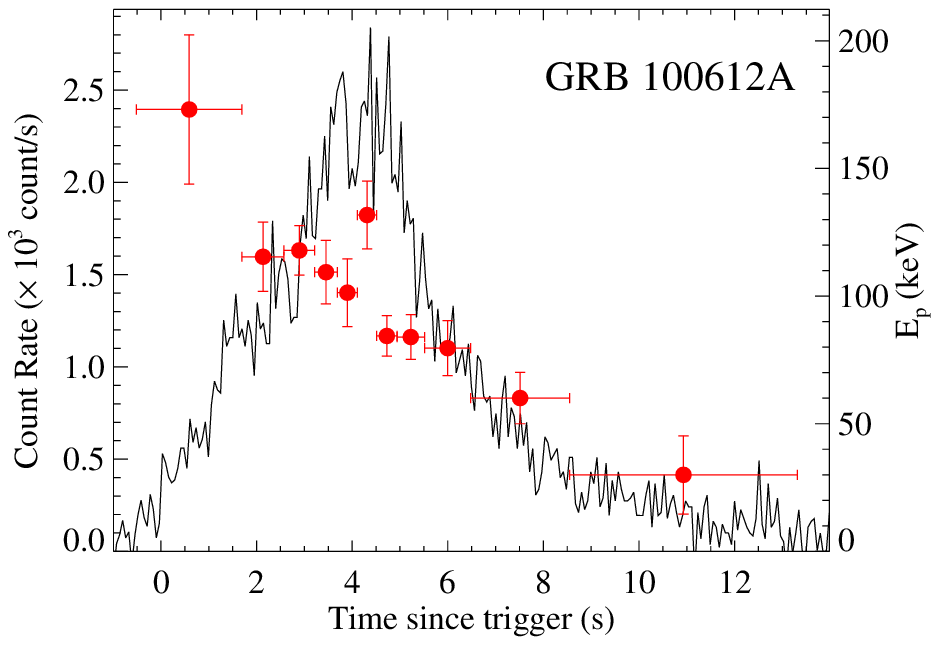}}
\resizebox{4cm}{!}{\includegraphics{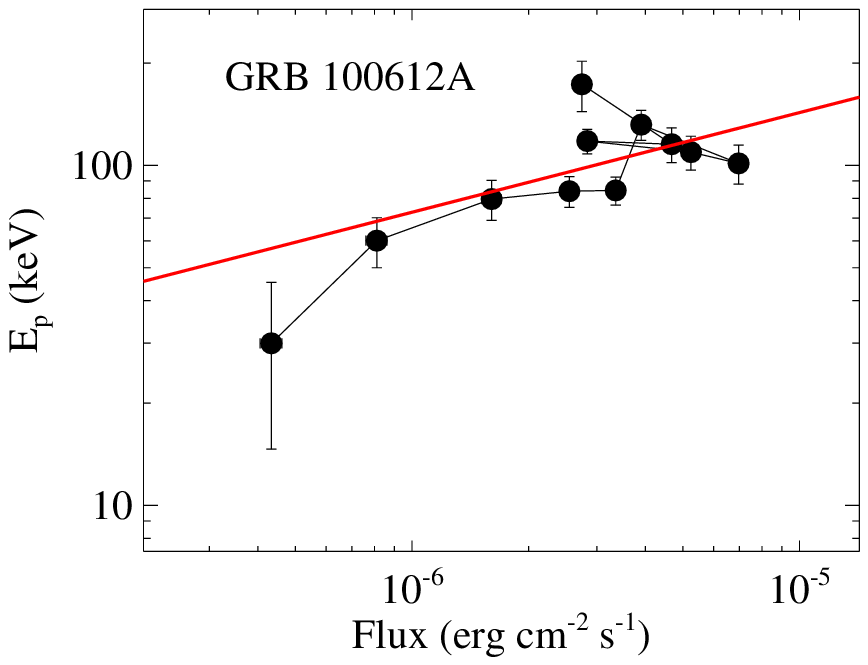}}
 \resizebox{4cm}{!}{\includegraphics{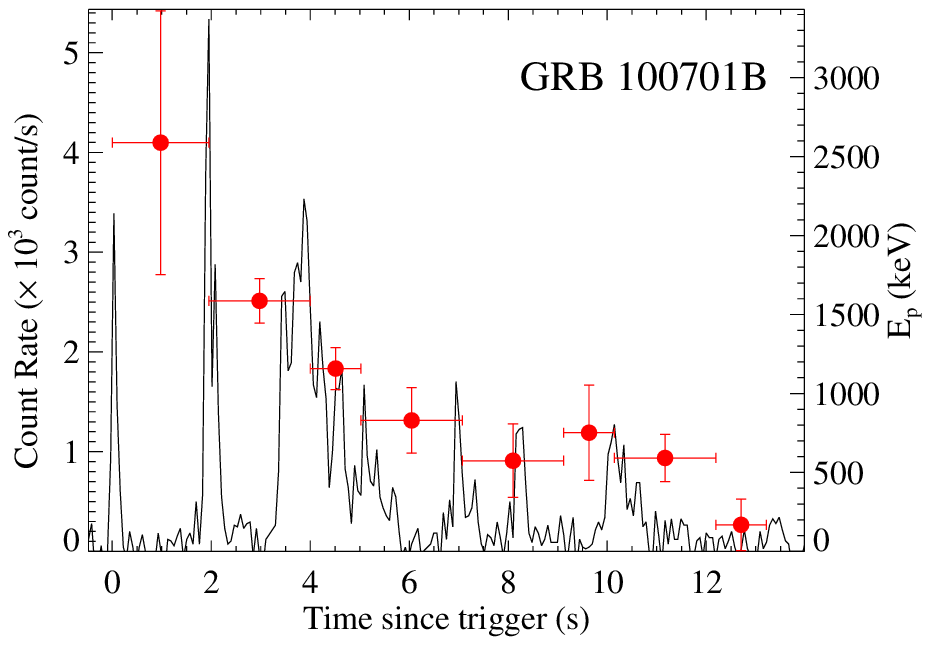}}
\resizebox{4cm}{!}{\includegraphics{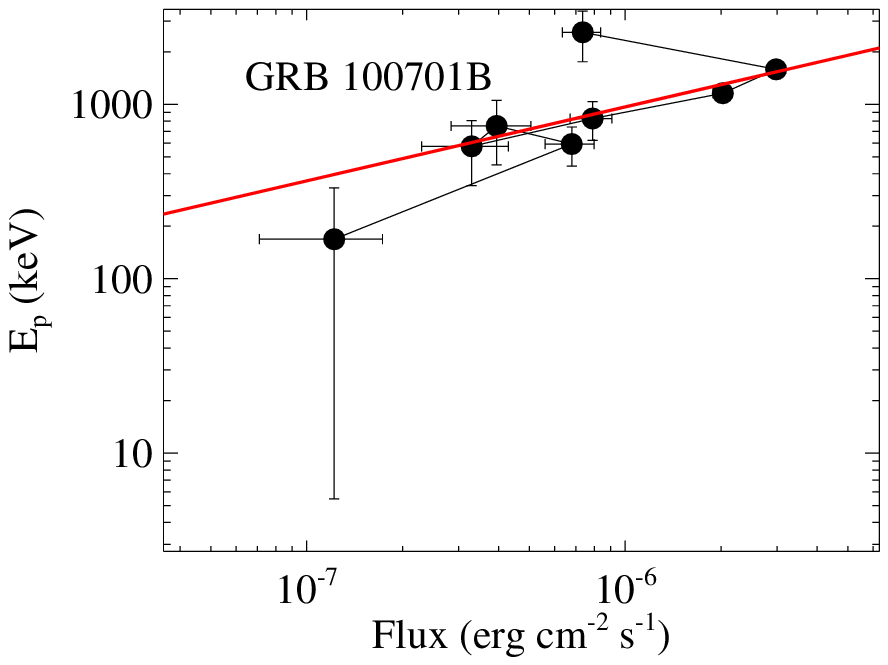}}

 \resizebox{4cm}{!}{\includegraphics{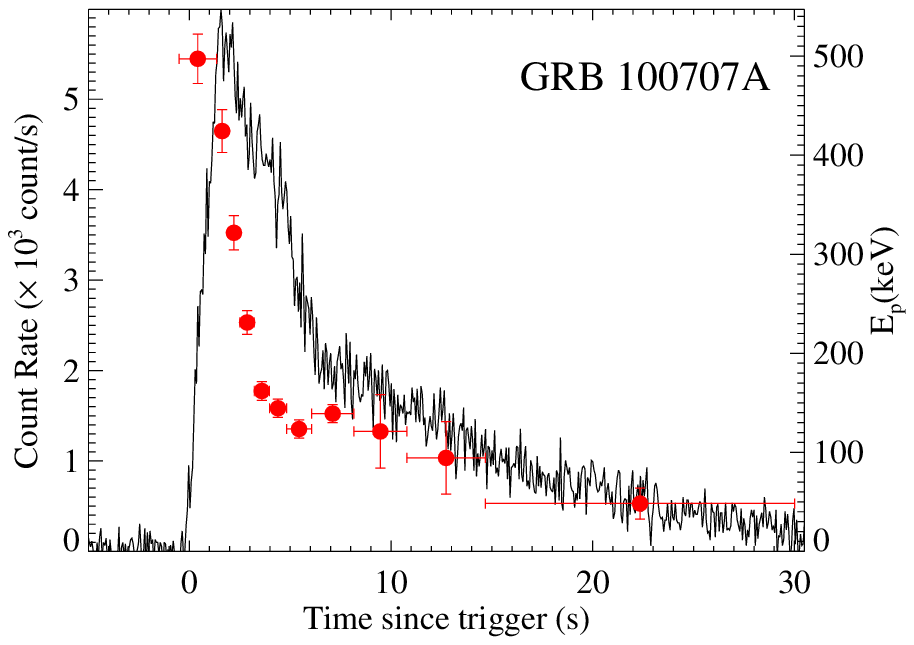}}
\resizebox{4cm}{!}{\includegraphics{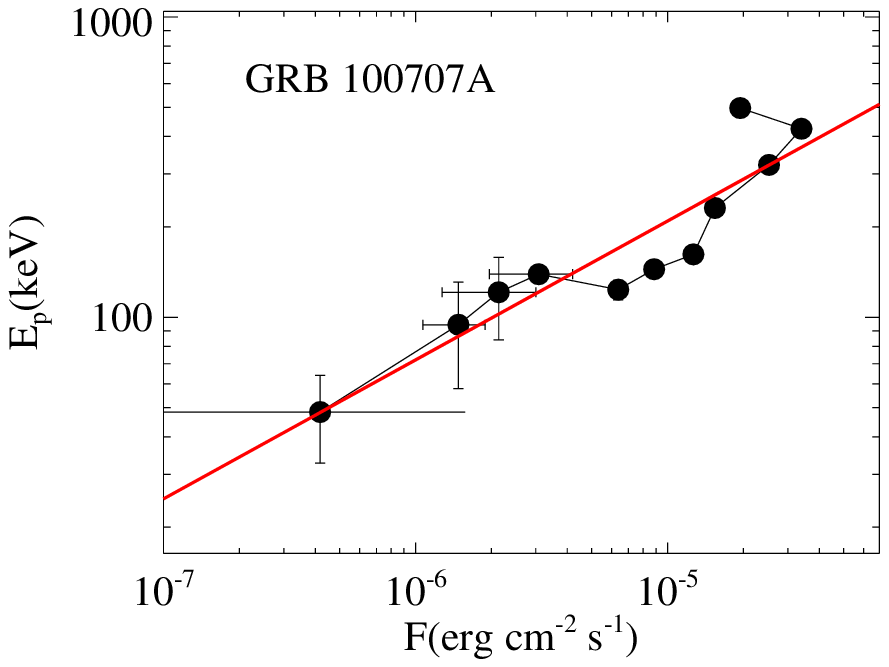}}
\resizebox{4cm}{!}{\includegraphics{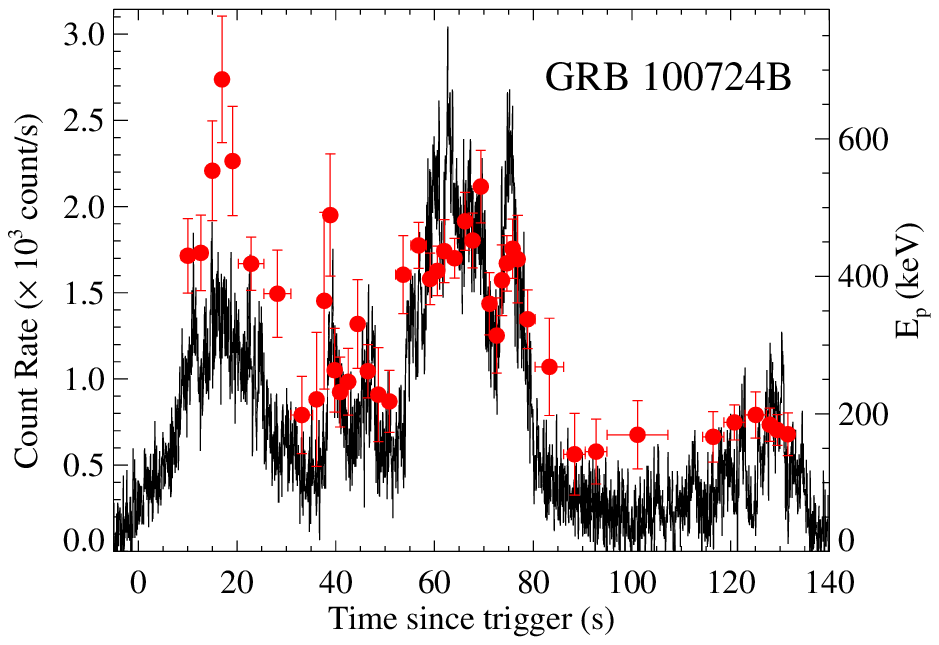}}
\resizebox{4cm}{!}{\includegraphics{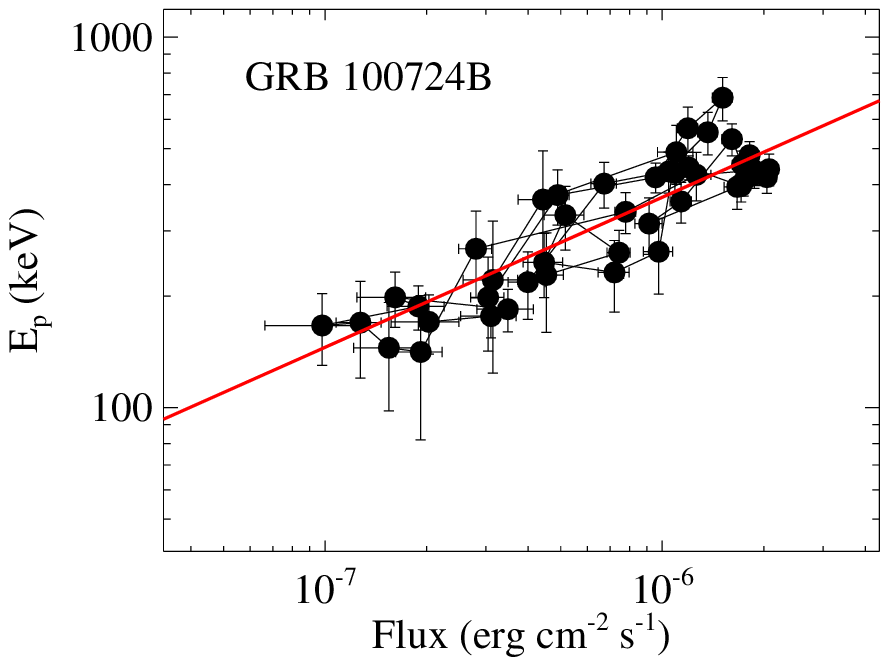}}

\resizebox{4cm}{!}{\includegraphics{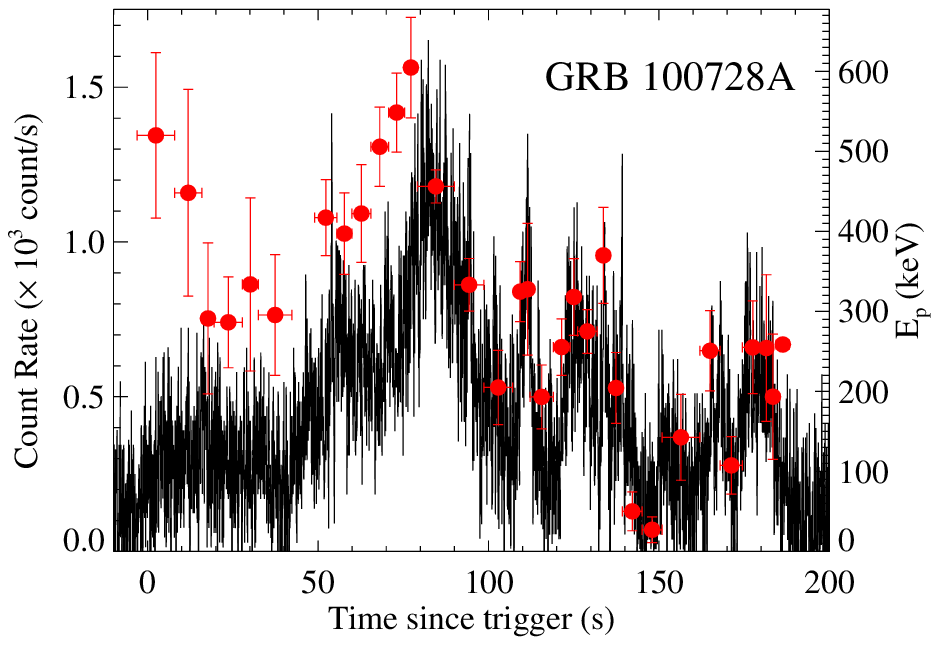}}
\resizebox{4cm}{!}{\includegraphics{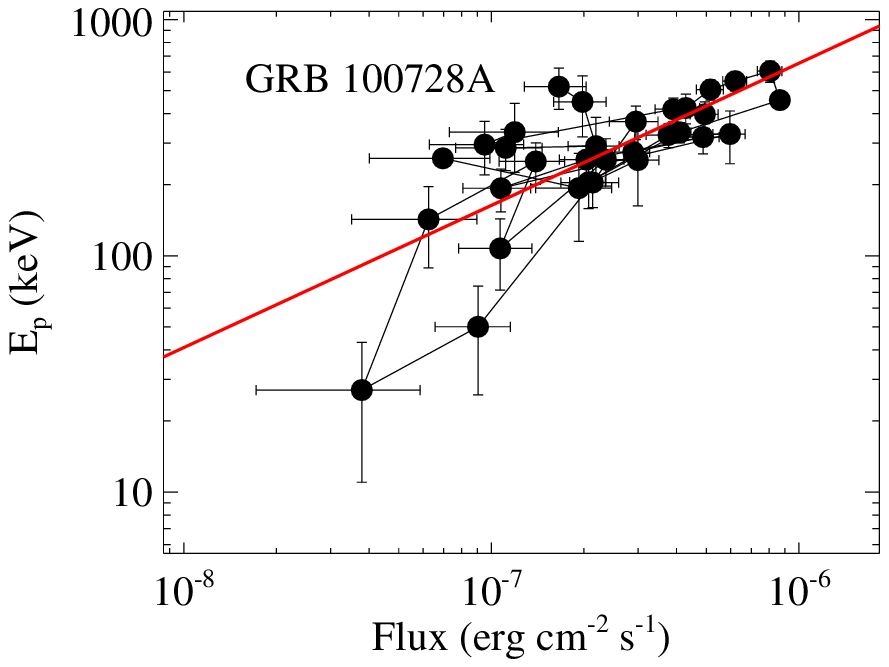}}
 \resizebox{4cm}{!}{\includegraphics{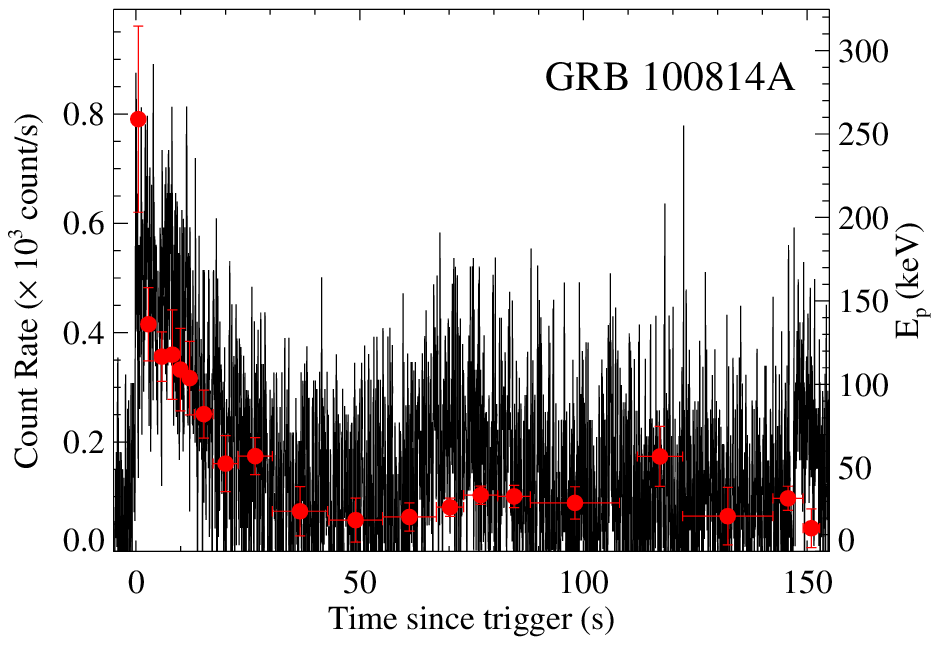}}
 \resizebox{4cm}{!}{\includegraphics{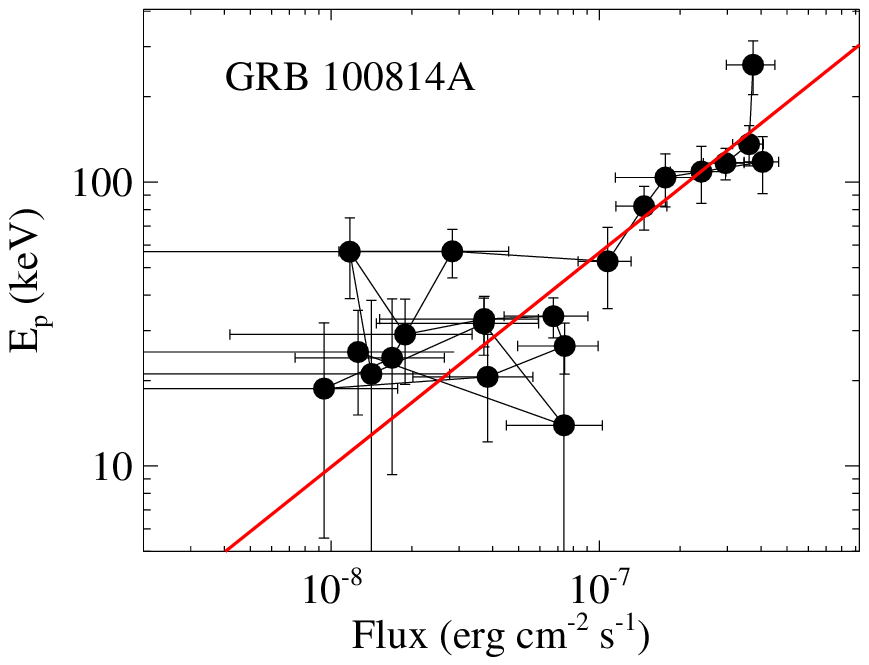}}
\end{figure*}

\addtocounter{figure}{-1}
\begin{figure*}
\caption{{\it-continued.} }
 \resizebox{4cm}{!}{\includegraphics{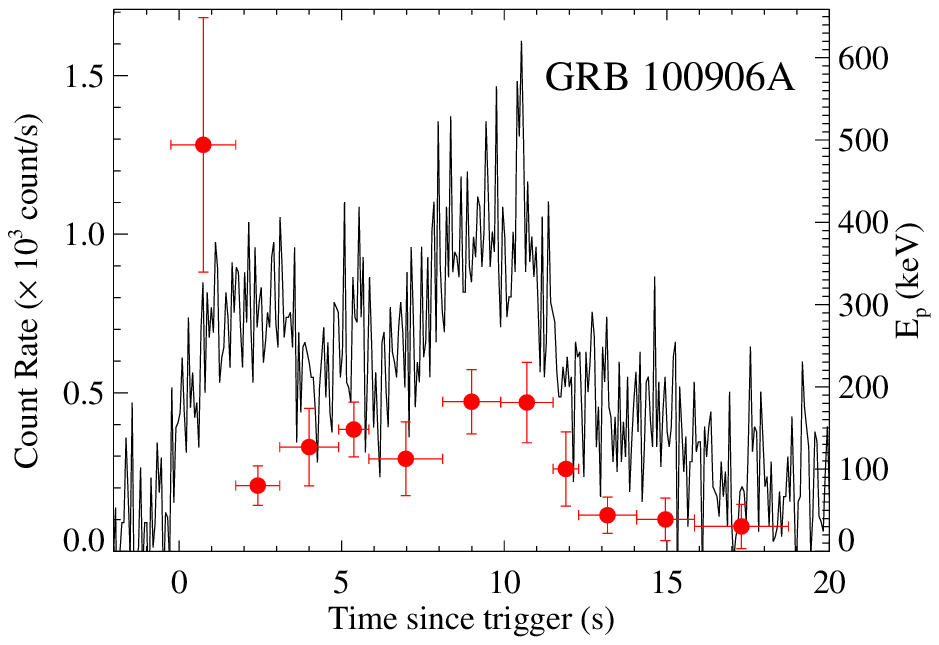}}
 \resizebox{4cm}{!}{\includegraphics{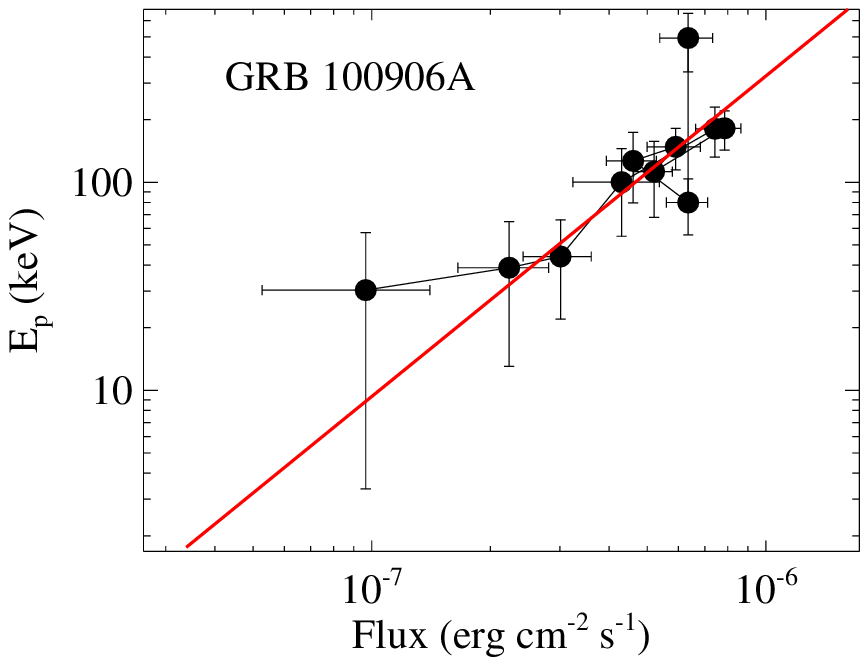}}
  \resizebox{4cm}{!}{\includegraphics{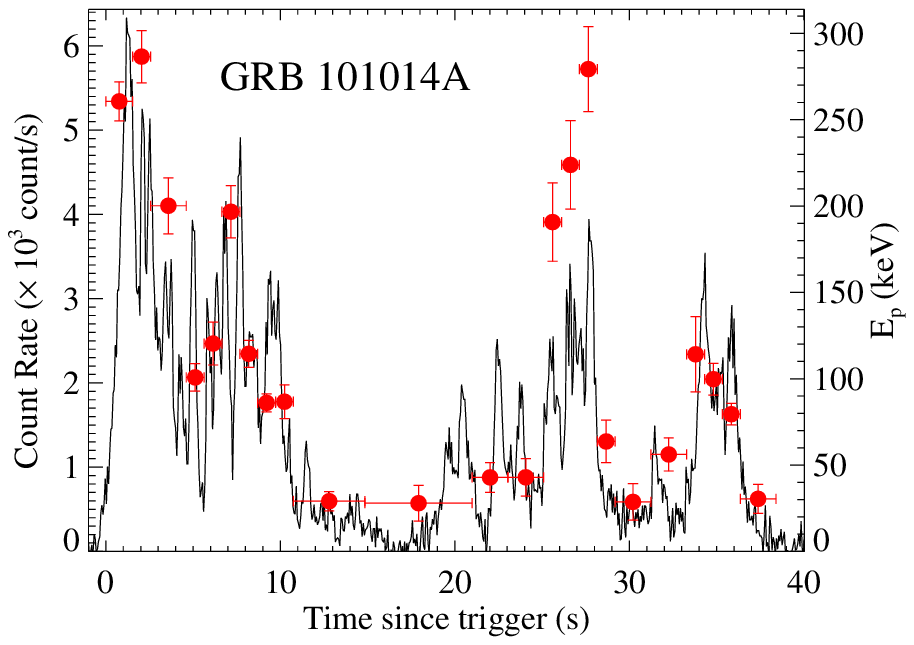}}
\resizebox{4cm}{!}{\includegraphics{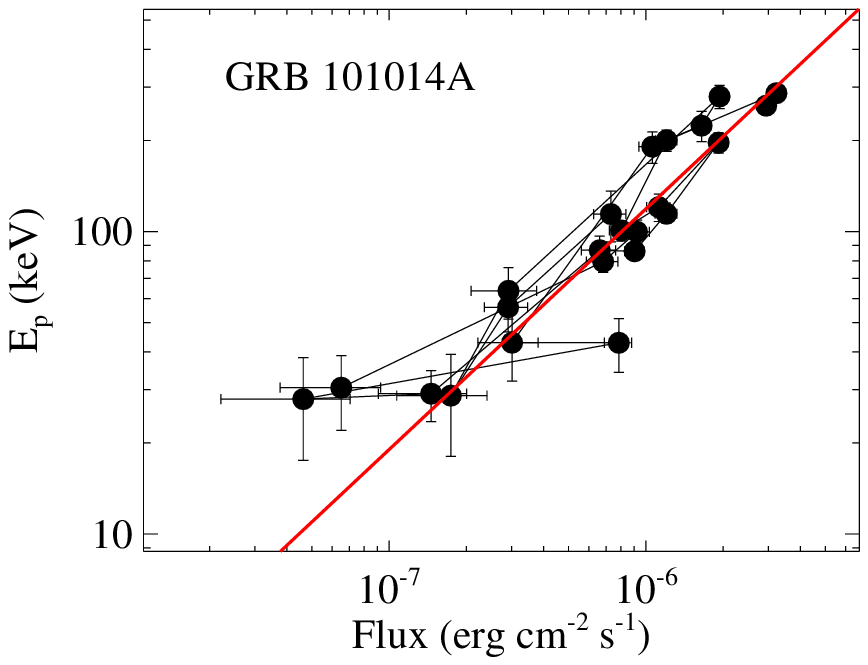}}

 \resizebox{4cm}{!}{\includegraphics{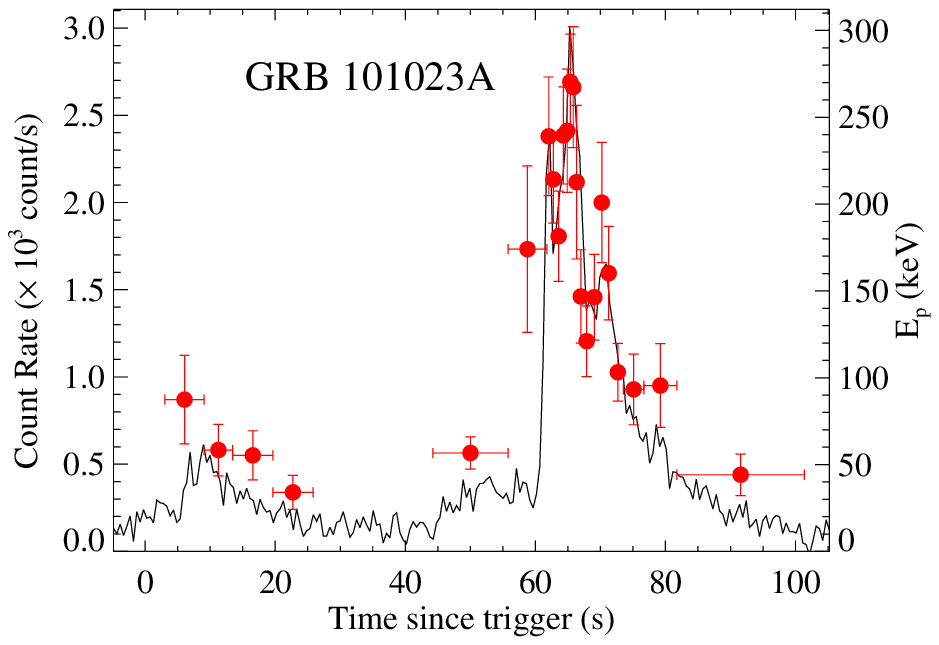}}
\resizebox{4cm}{!}{\includegraphics{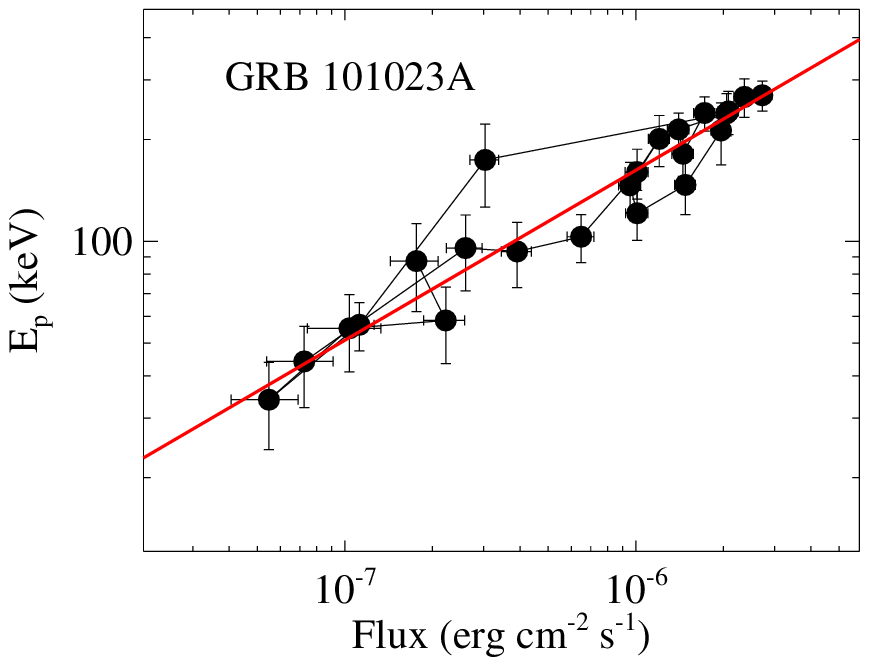}}
\resizebox{4cm}{!}{\includegraphics{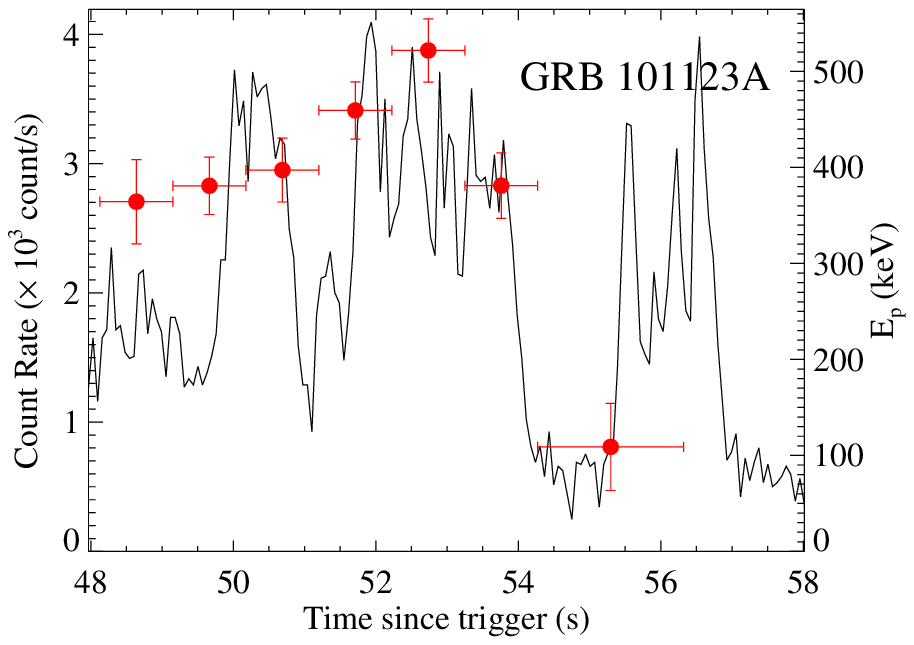}}
\resizebox{4cm}{!}{\includegraphics{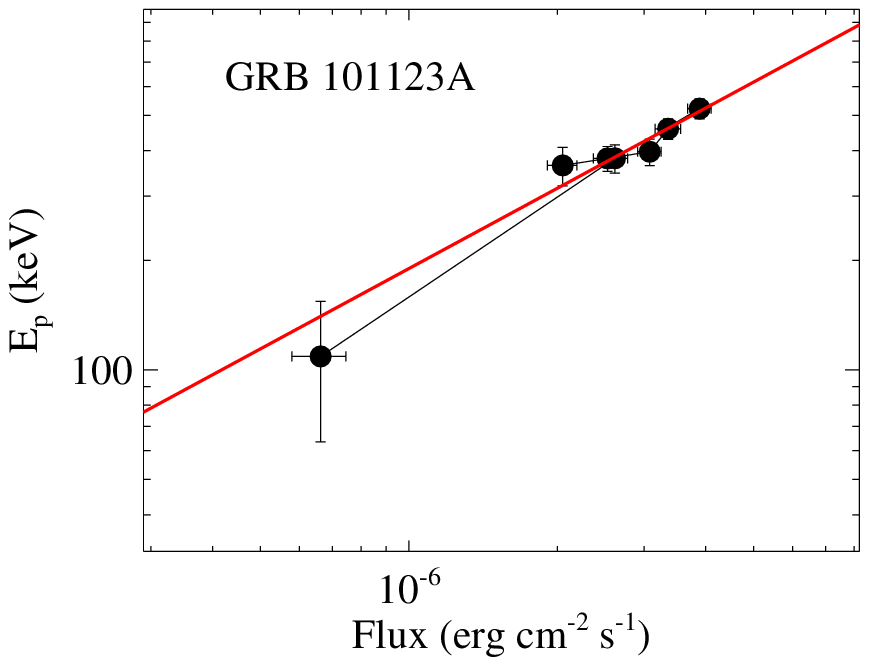}}

\resizebox{4cm}{!}{\includegraphics{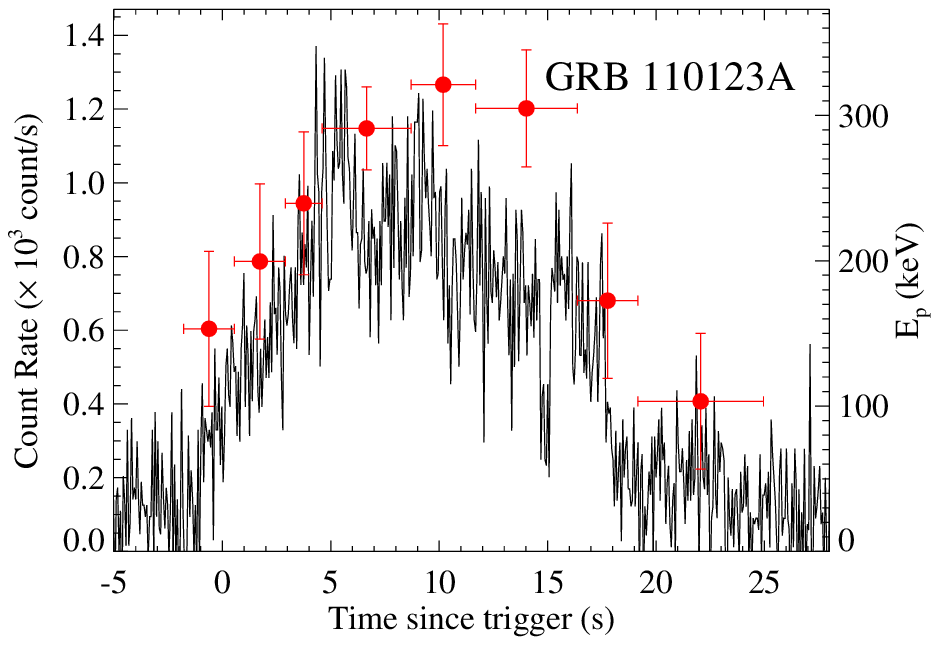}}
\resizebox{4cm}{!}{\includegraphics{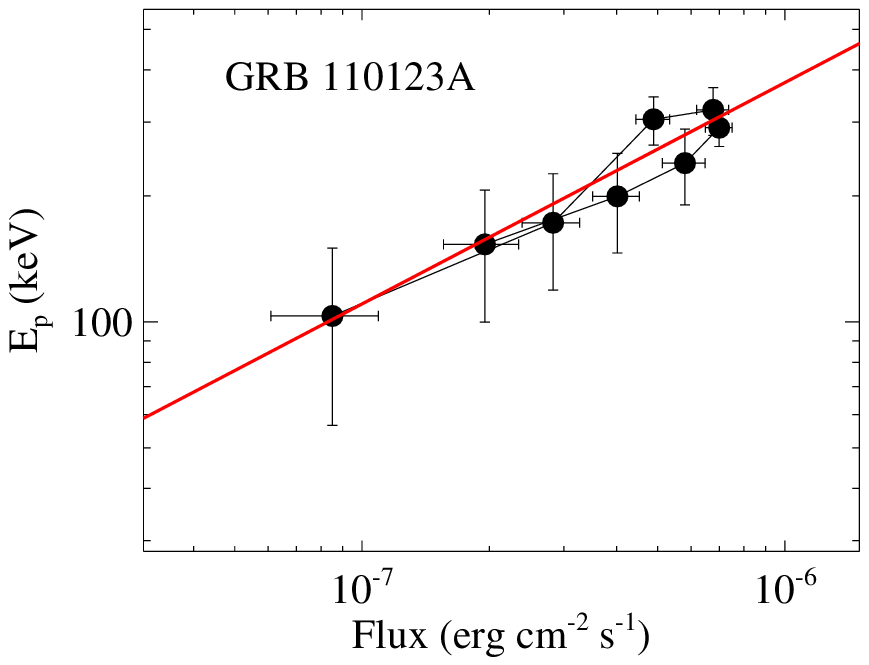}}
\resizebox{4cm}{!}{\includegraphics{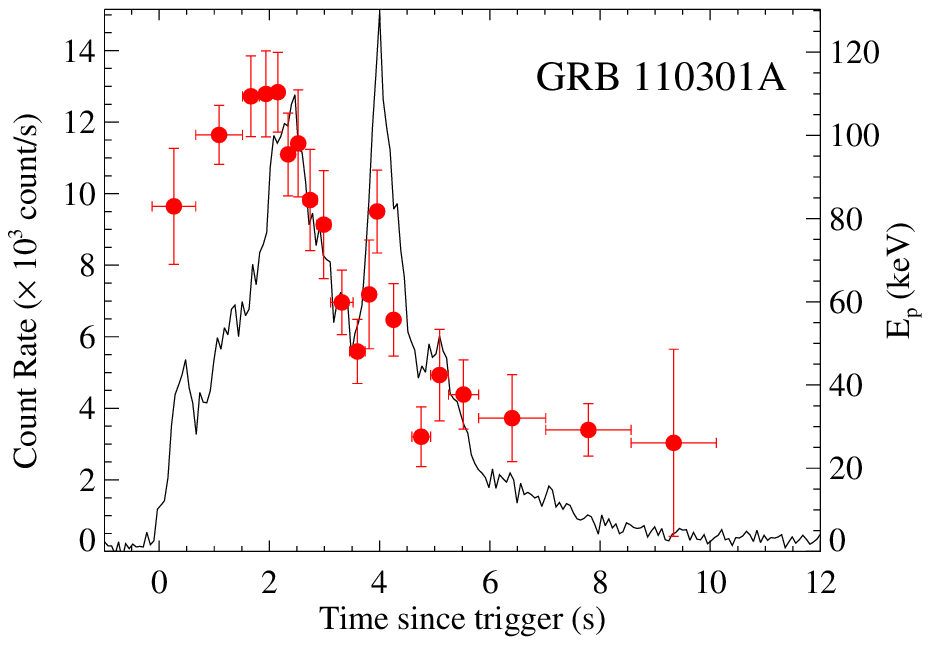}}
\resizebox{4cm}{!}{\includegraphics{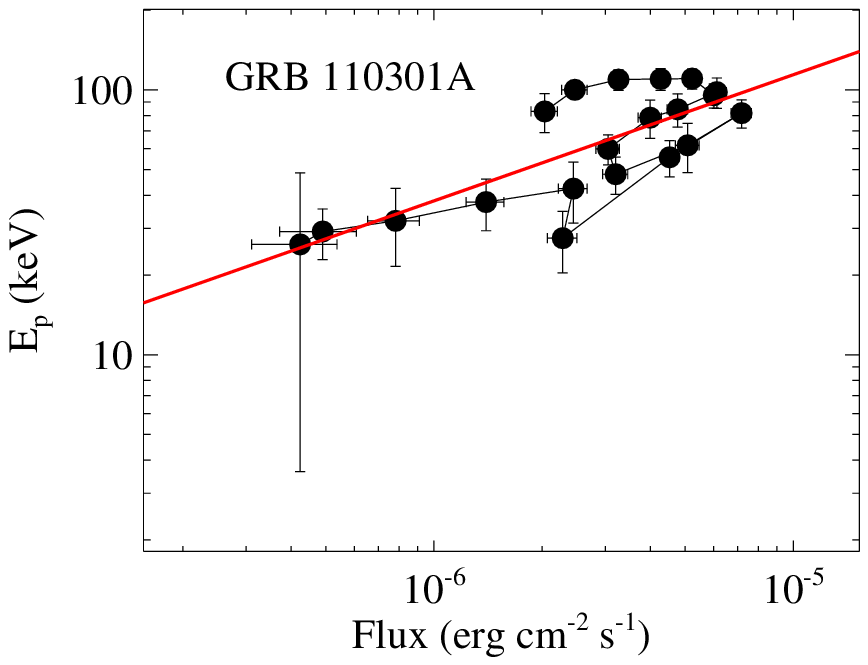}}

 \resizebox{4cm}{!}{\includegraphics{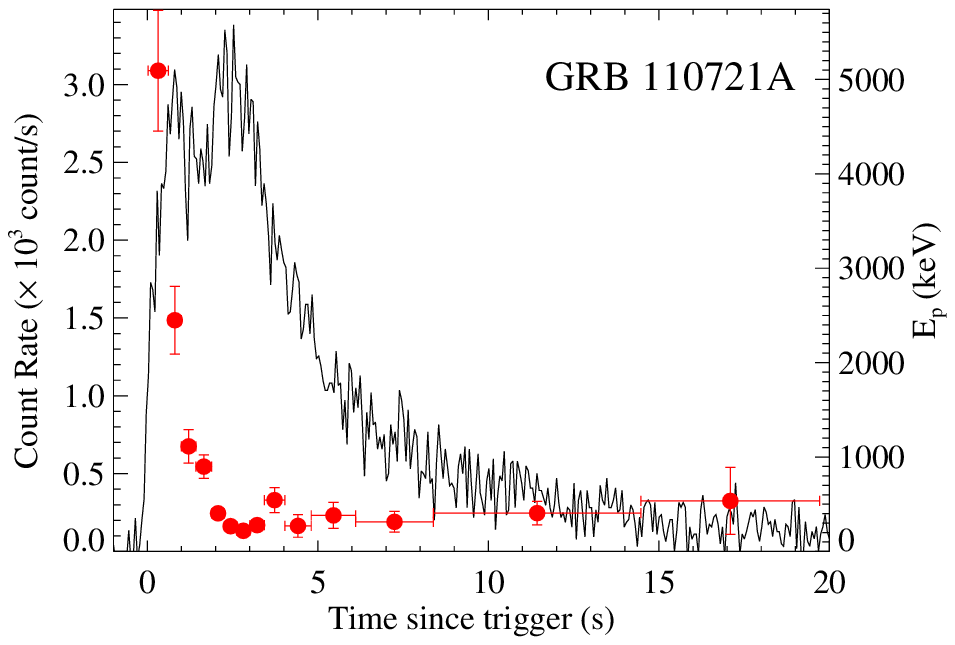}}
\resizebox{4cm}{!}{\includegraphics{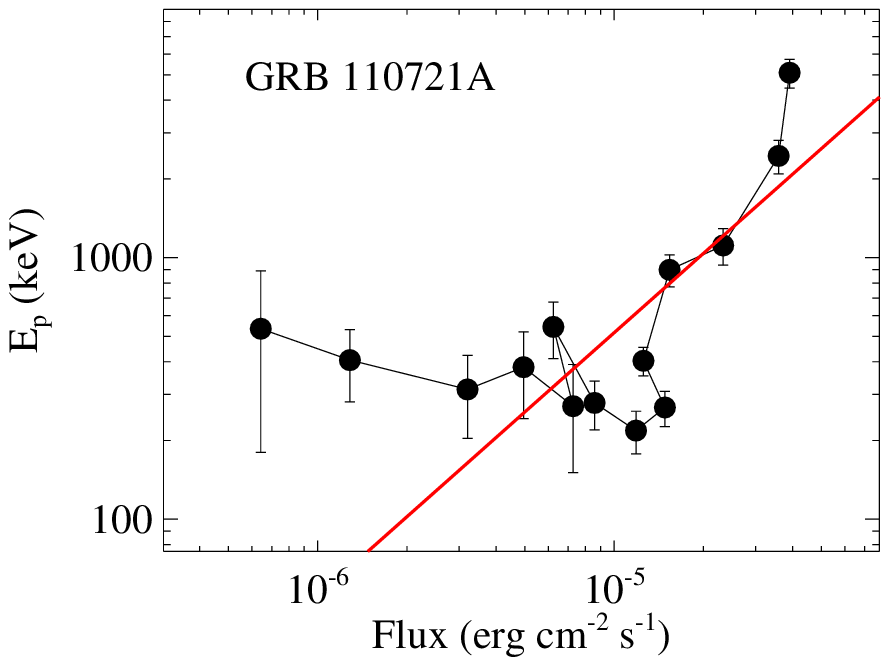}}
 \resizebox{4cm}{!}{\includegraphics{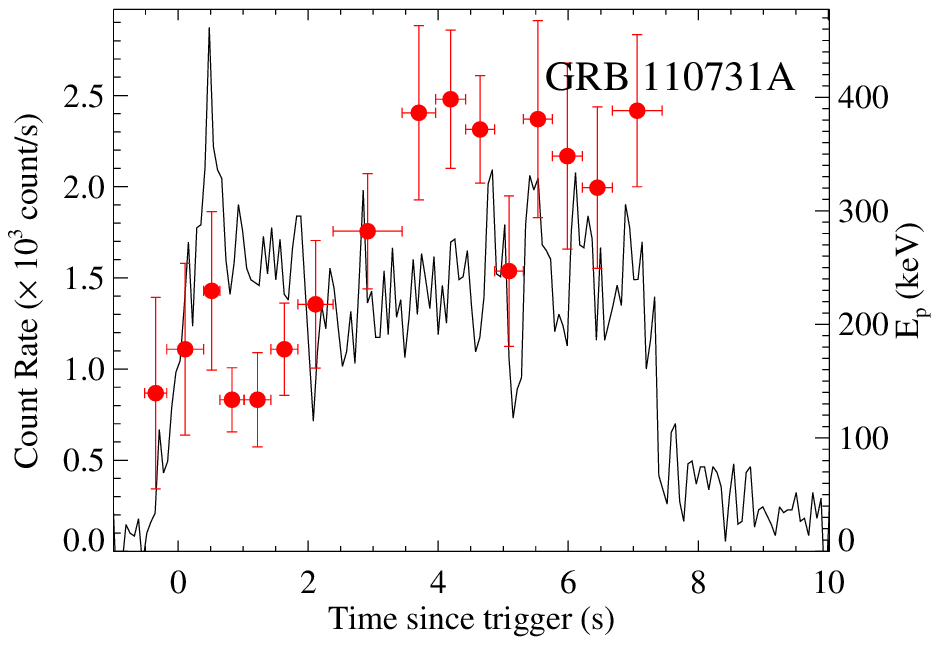}}
\resizebox{4cm}{!}{\includegraphics{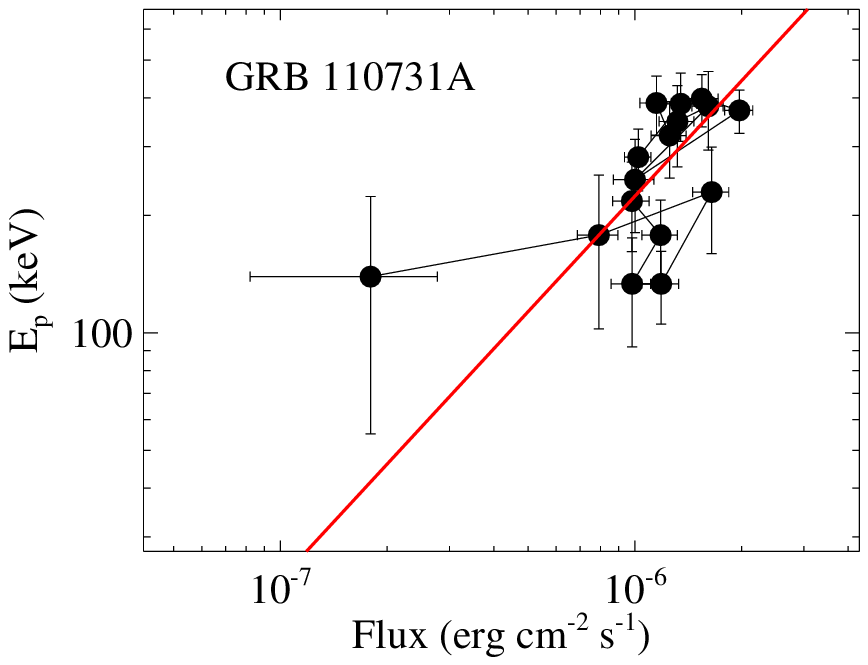}}

 \resizebox{4cm}{!}{\includegraphics{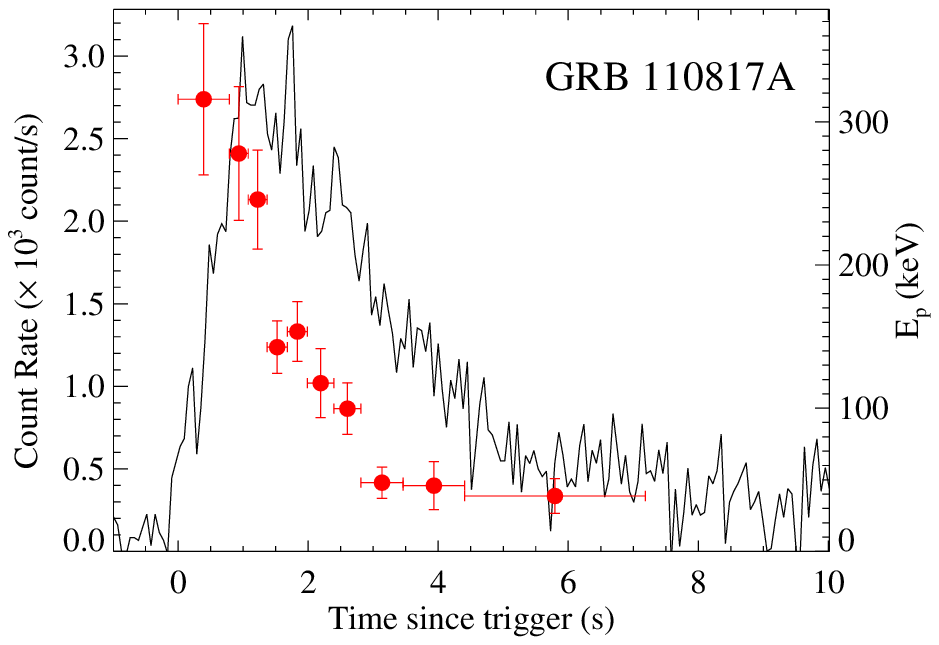}}
\resizebox{4cm}{!}{\includegraphics{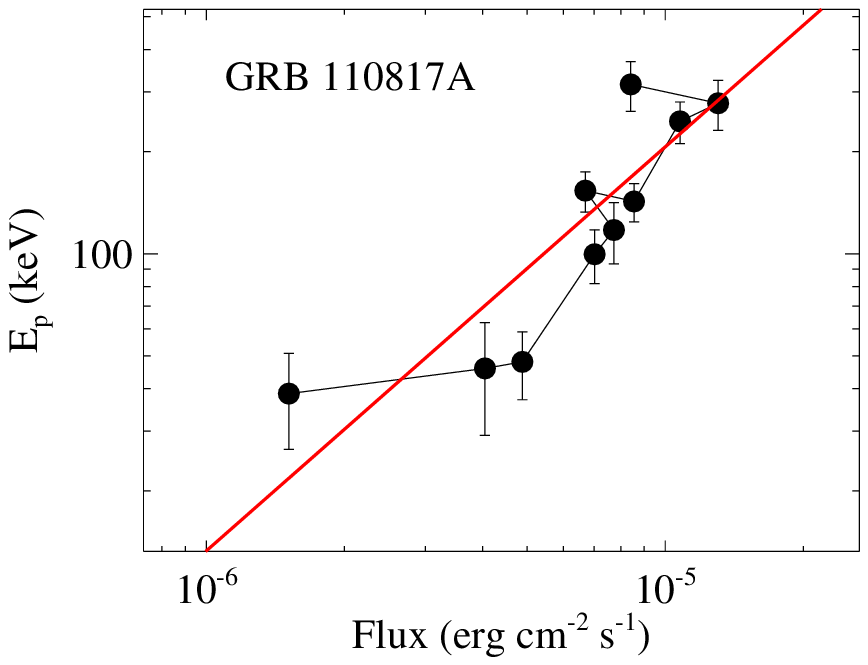}}
 \end{figure*}

\begin{figure*}
\resizebox{4cm}{!}{\includegraphics{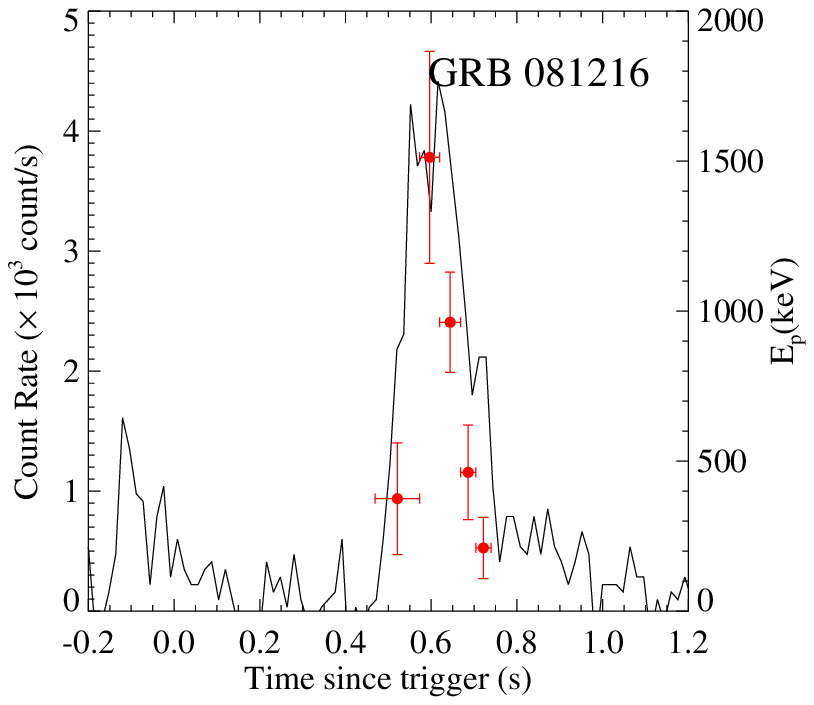}}
\resizebox{4cm}{!}{\includegraphics{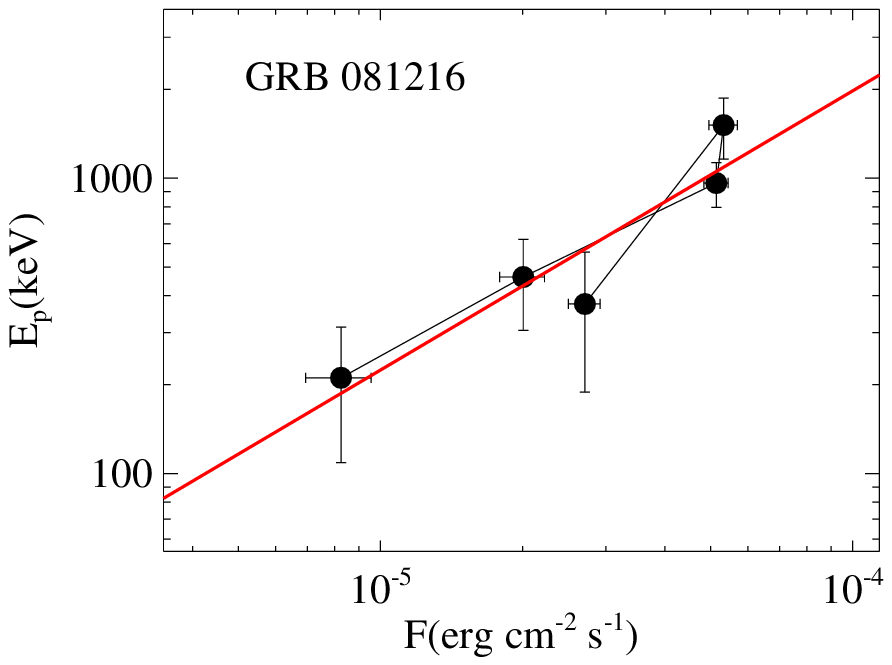}}
\resizebox{4cm}{!}{\includegraphics{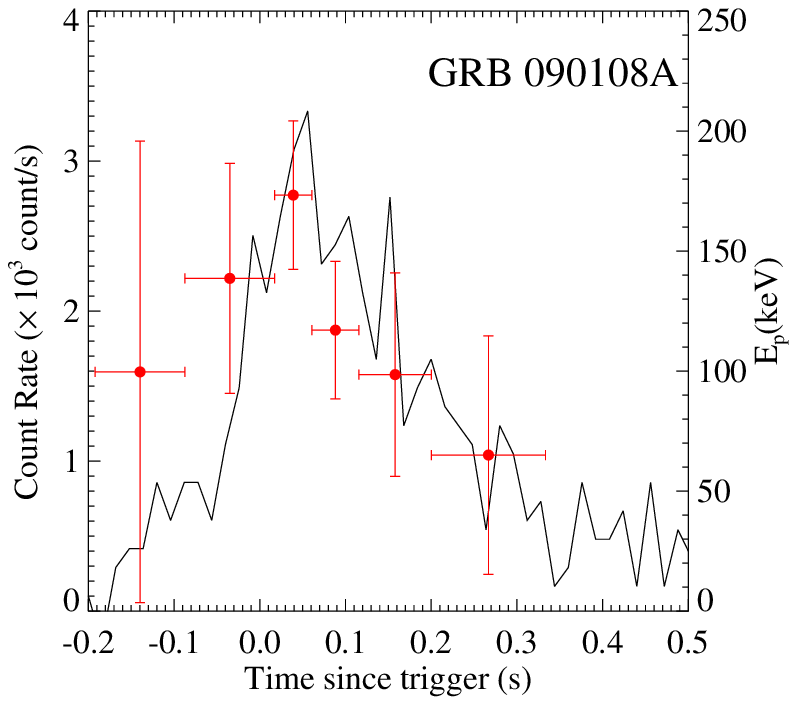}}
\resizebox{4cm}{!}{\includegraphics{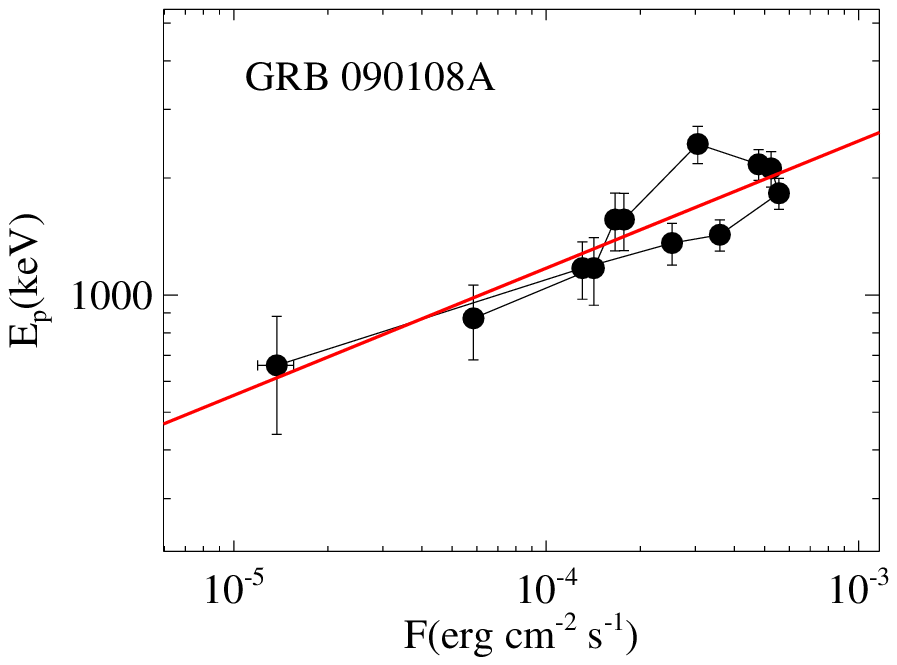}}

\resizebox{4cm}{!}{\includegraphics{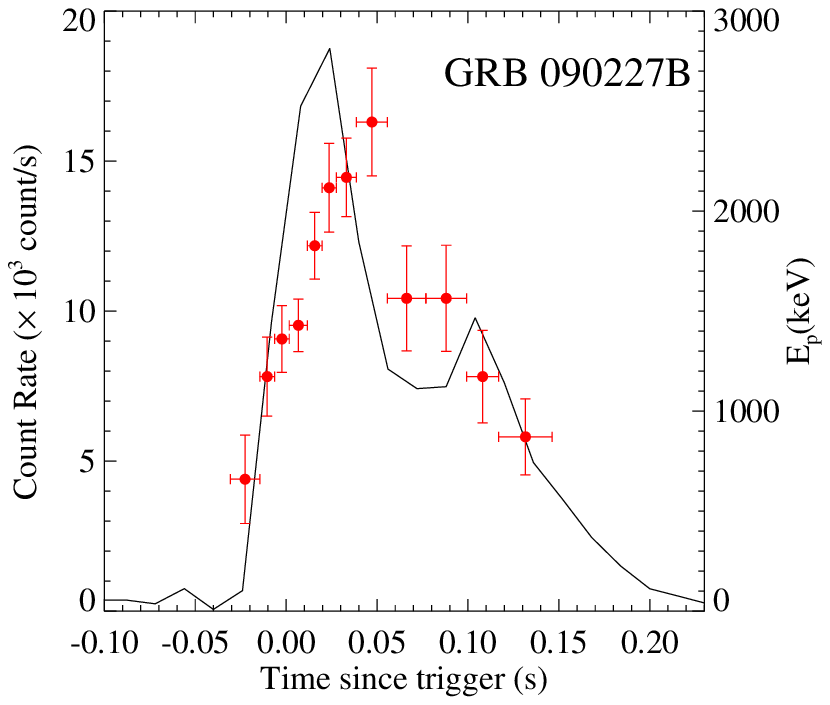}}
\resizebox{4cm}{!}{\includegraphics{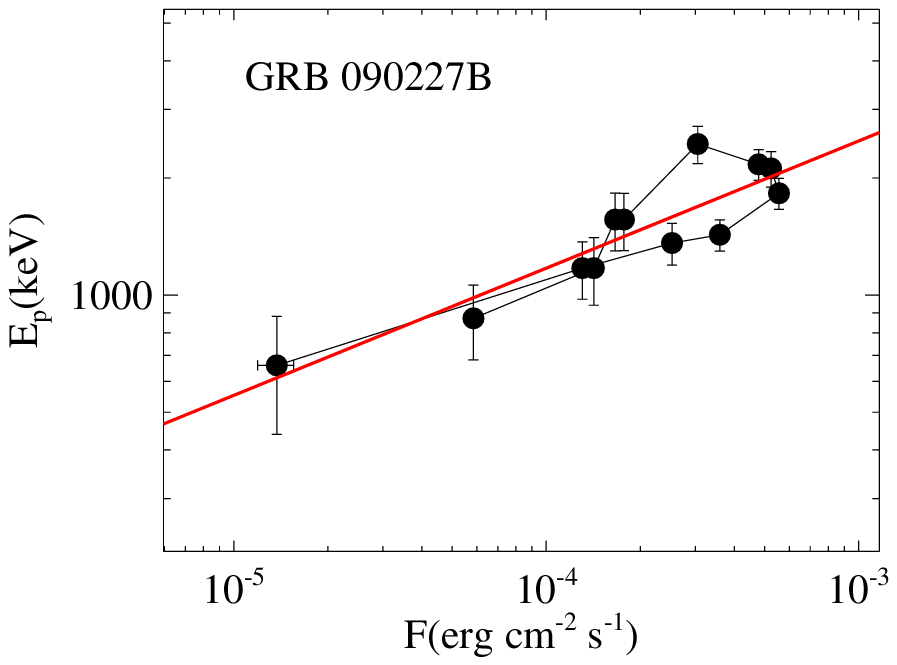}}
\resizebox{4cm}{!}{\includegraphics{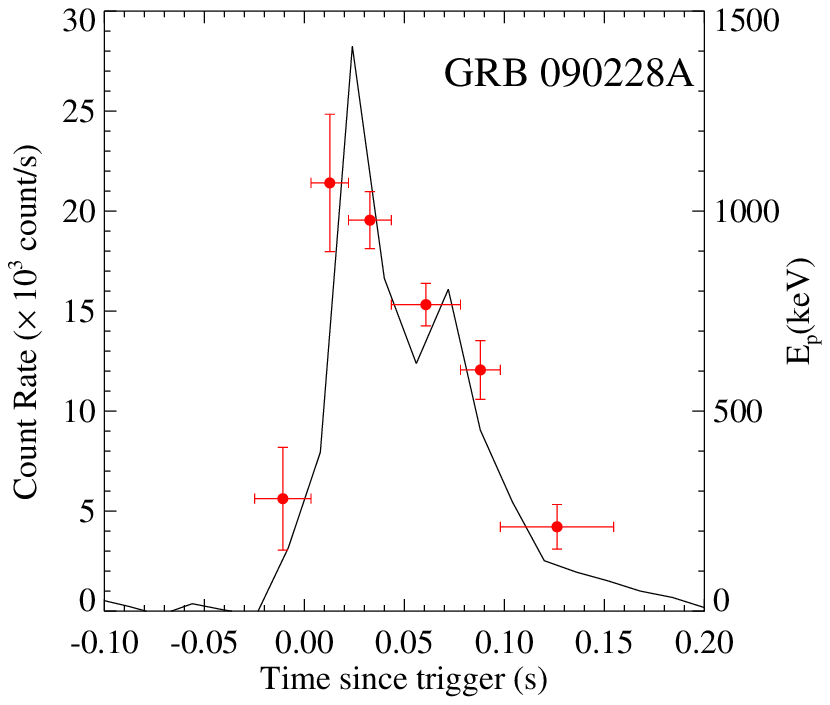}}
\resizebox{4cm}{!}{\includegraphics{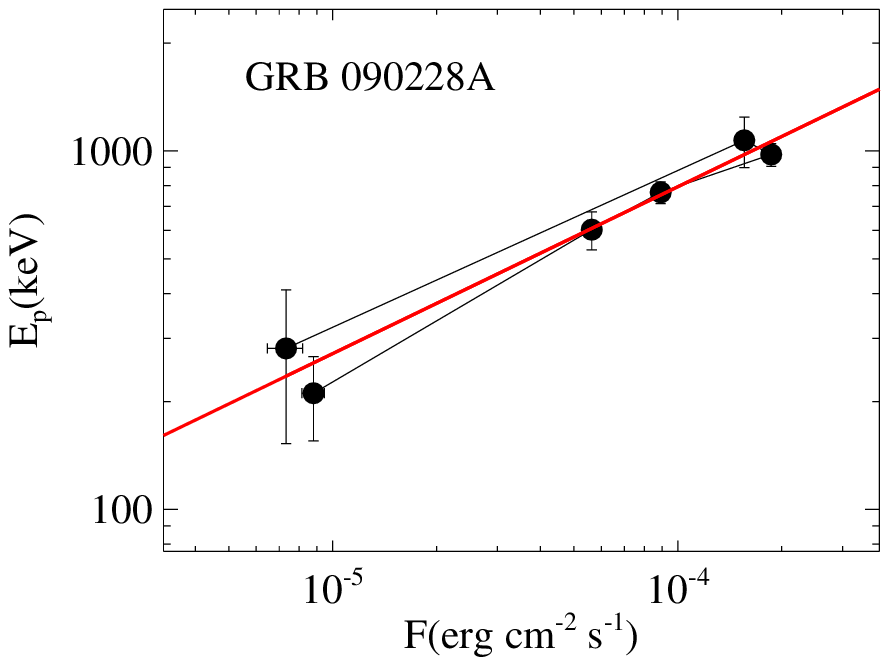}}

\resizebox{4cm}{!}{\includegraphics{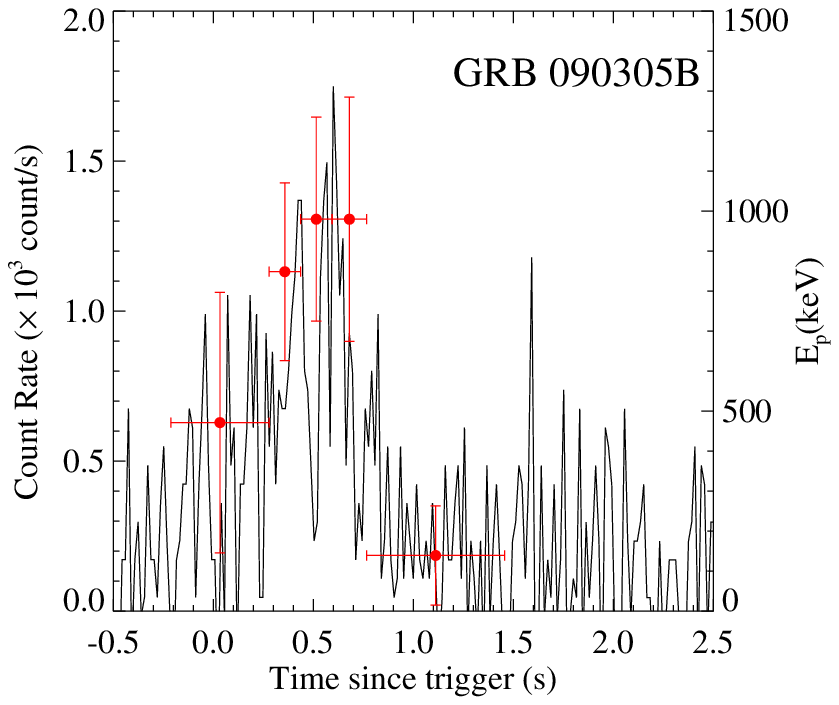}}
\resizebox{4cm}{!}{\includegraphics{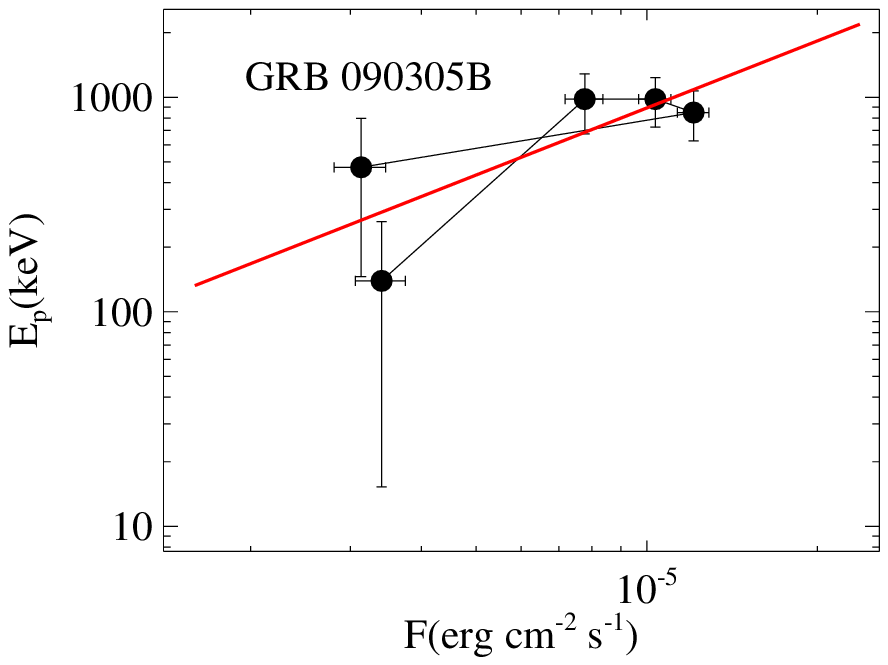}}
\resizebox{4cm}{!}{\includegraphics{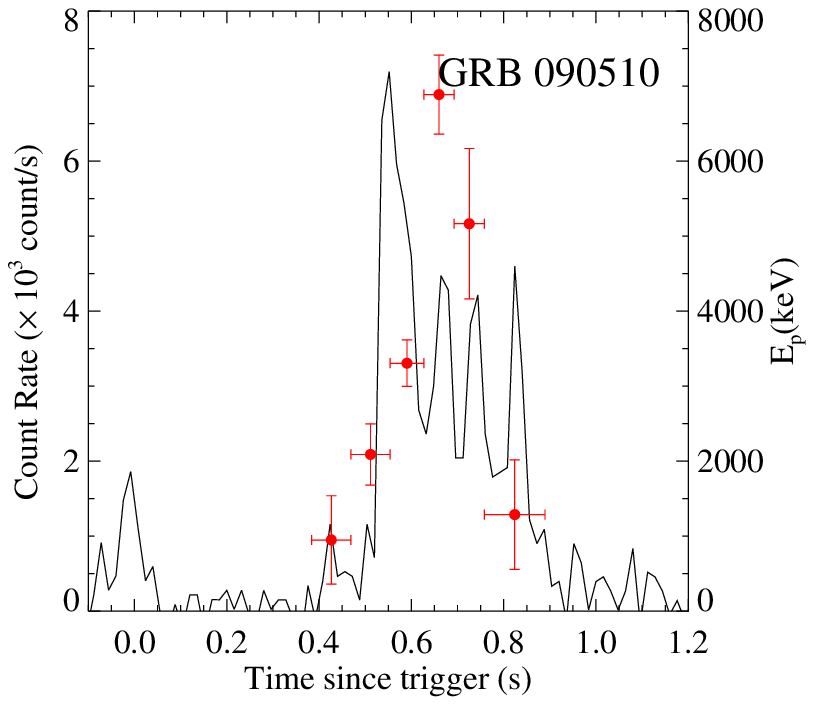}}
\resizebox{4cm}{!}{\includegraphics{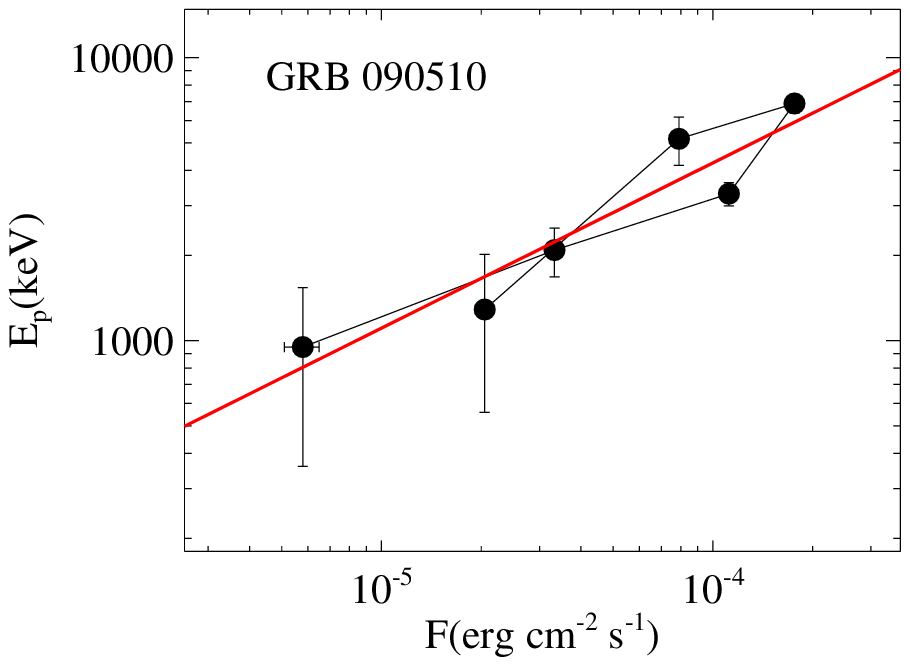}}

\resizebox{4cm}{!}{\includegraphics{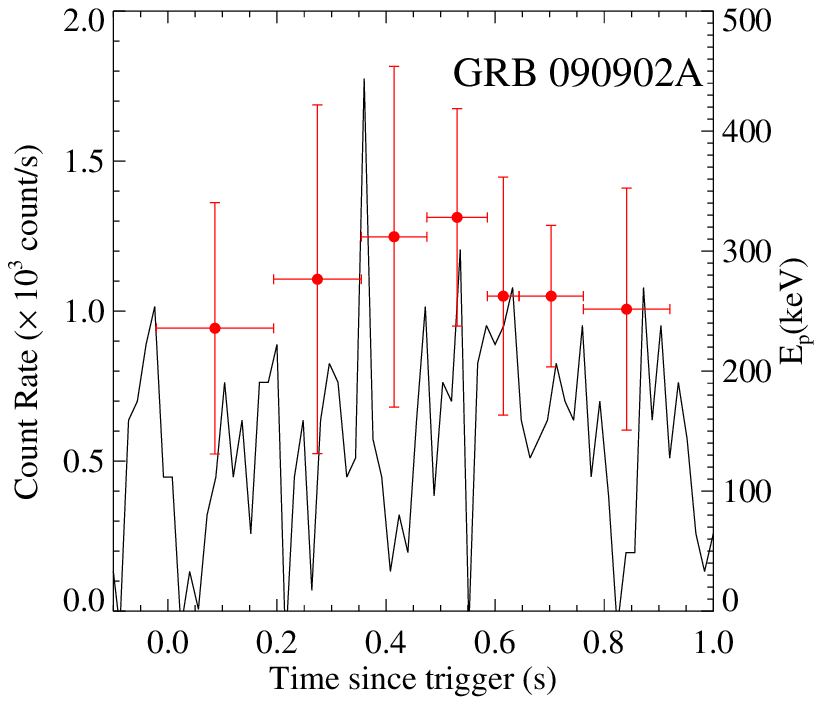}}
\resizebox{4cm}{!}{\includegraphics{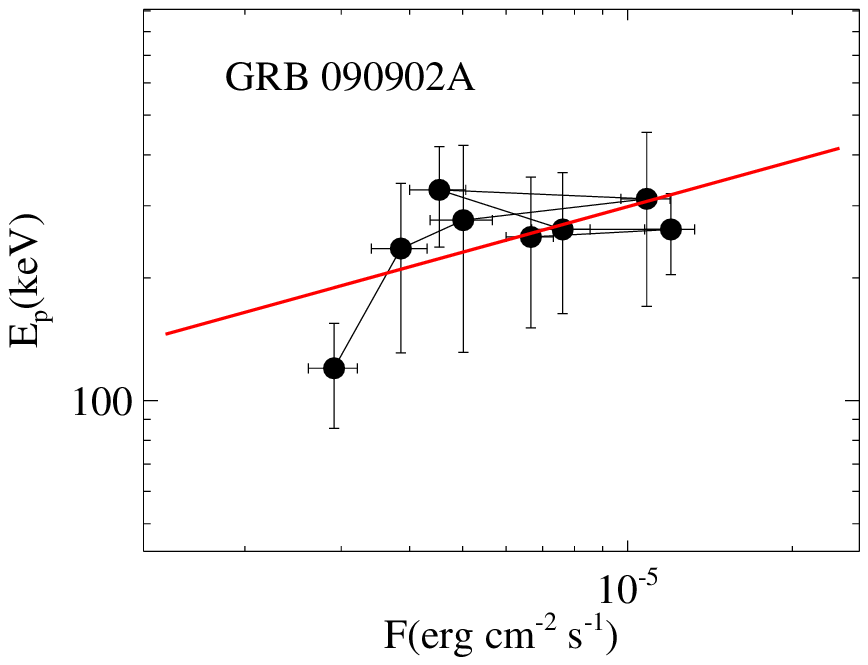}}
\resizebox{4cm}{!}{\includegraphics{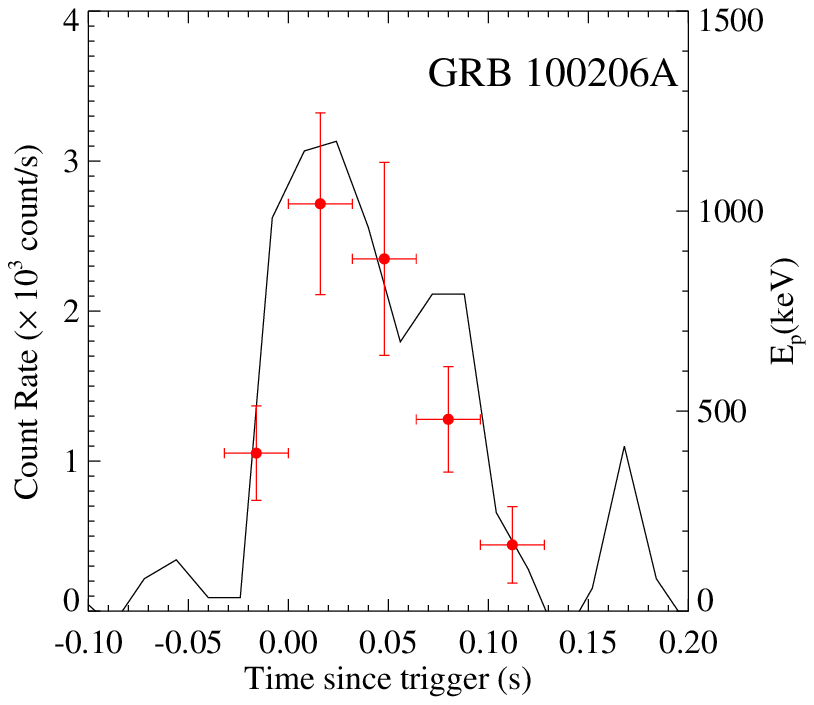}}
\resizebox{4cm}{!}{\includegraphics{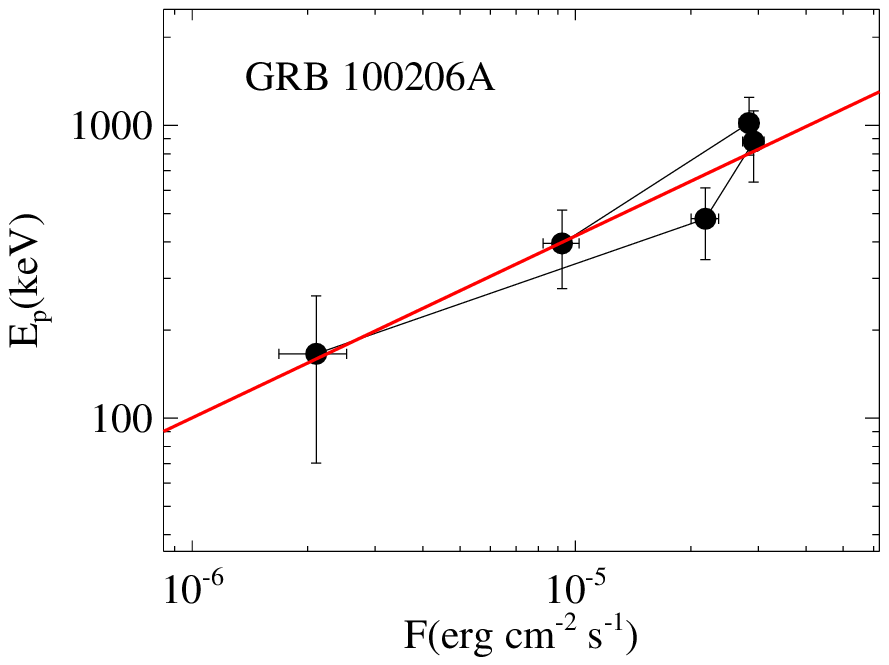}}

\resizebox{4cm}{!}{\includegraphics{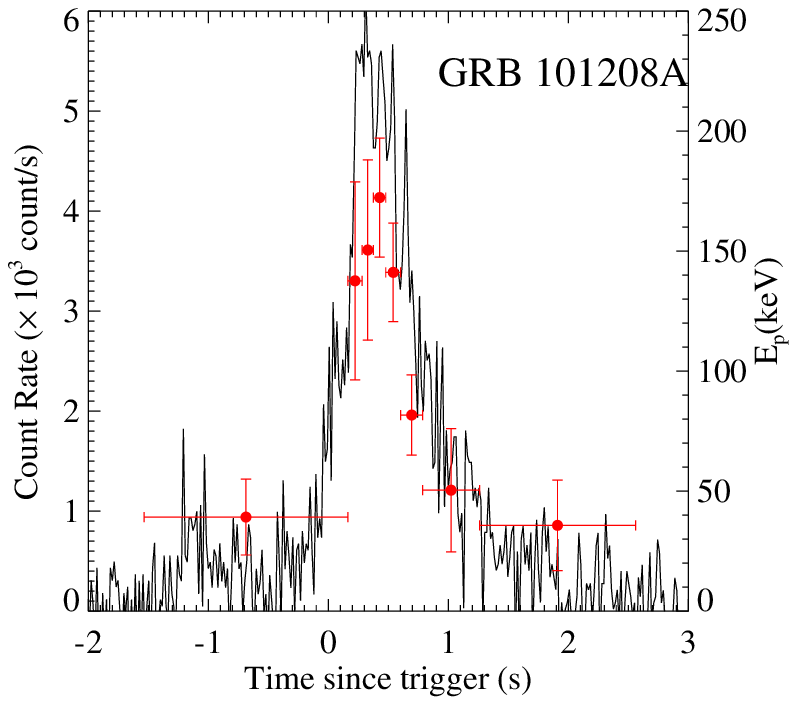}}
\resizebox{4cm}{!}{\includegraphics{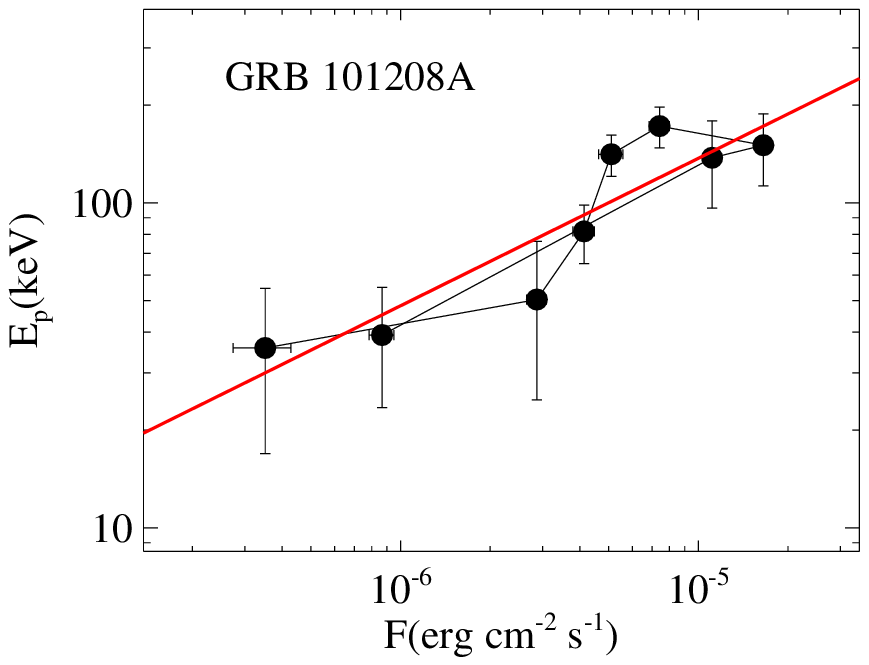}}
\resizebox{4cm}{!}{\includegraphics{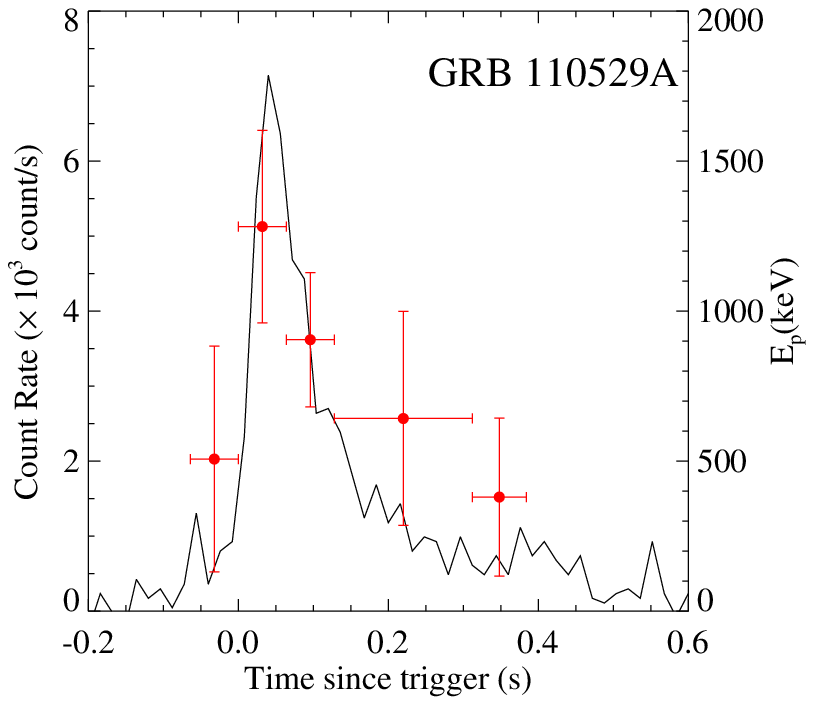}}
\resizebox{4cm}{!}{\includegraphics{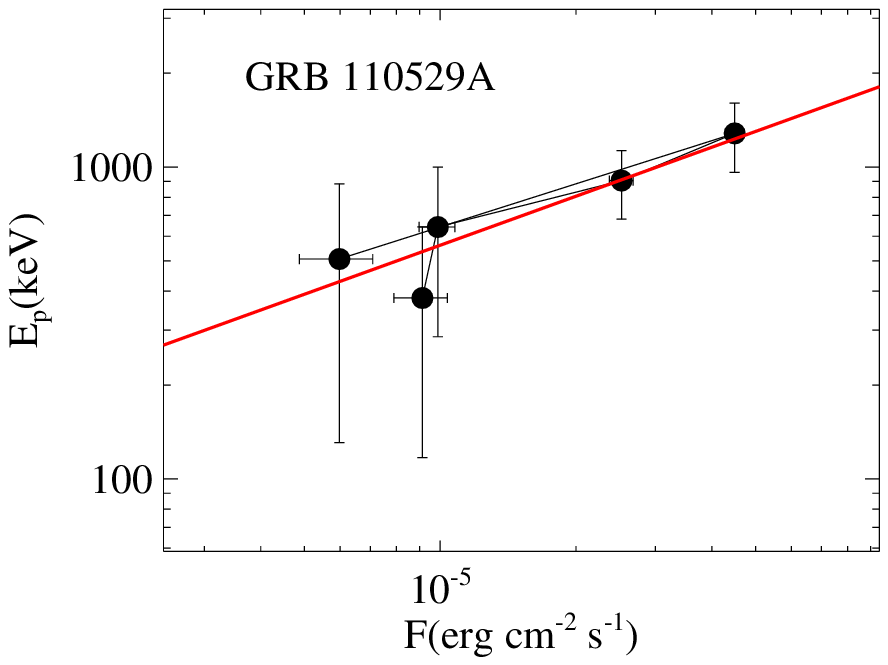}}

\resizebox{4cm}{!}{\includegraphics{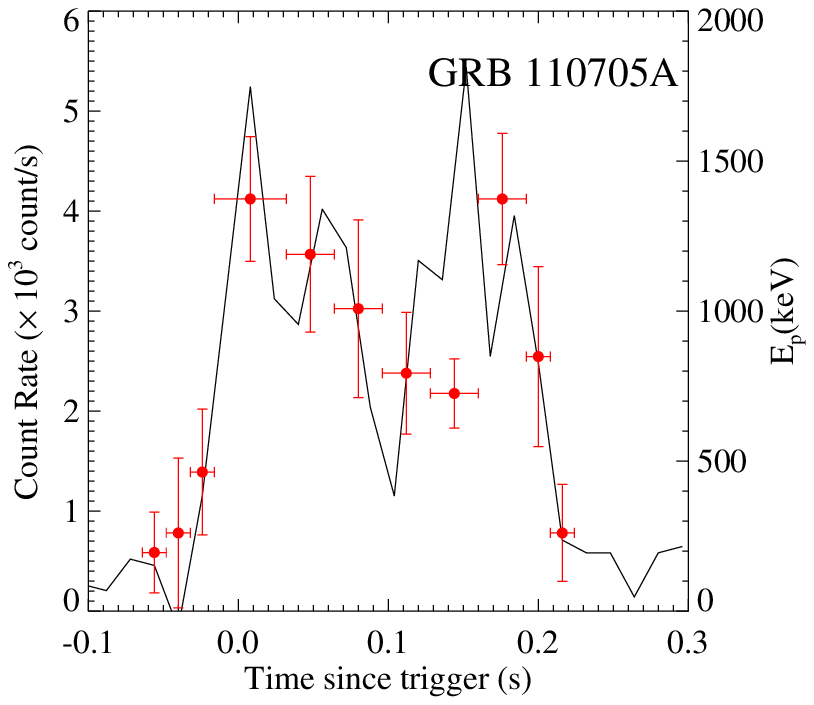}}
\resizebox{4cm}{!}{\includegraphics{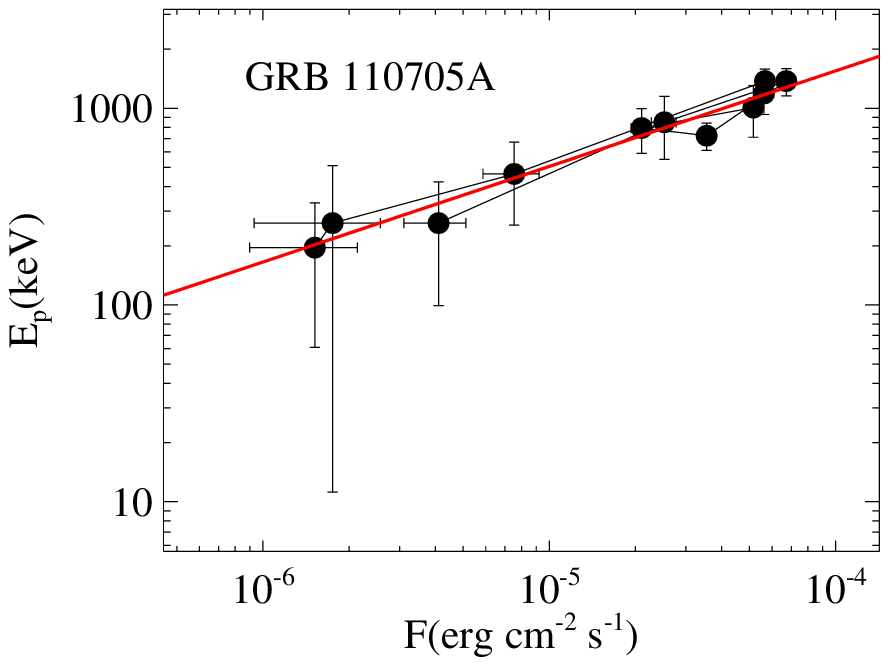}}%

\caption{The
same as Fig. \ref{long}, but for the short GRBs in our sample.}
\label{short}
\end{figure*}

\begin{figure*}
\resizebox{16cm}{!}{\includegraphics{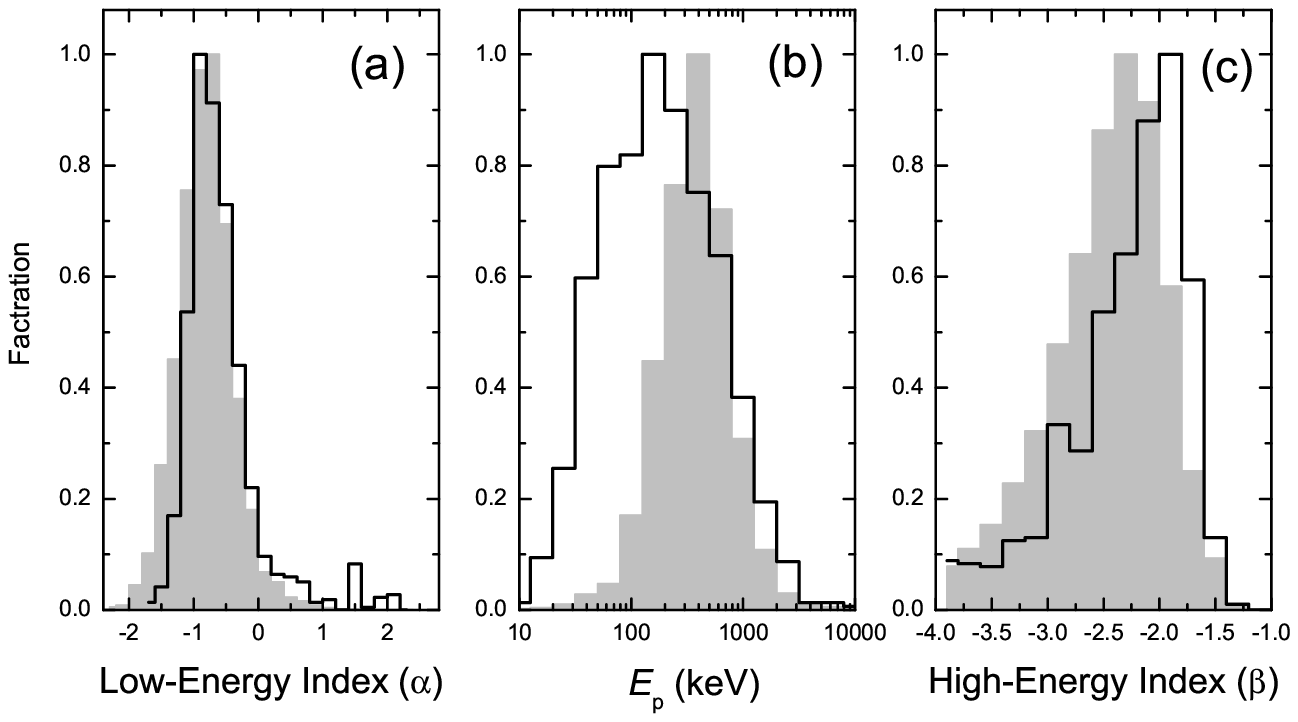}} \caption{A comparison
of time-resolved Band spectral parameters, i.e. low-energy spectral
index (a), peak energy (b), and high-energy spectral index (c),
between the 350 bright {\em CGRO}/BATSE GRBs with 8459 time-resolved
burst spectra (gray shapes, Data from Kaneko et al. 2006) and the
Fermi GRBs included in our sample (solid
lines).}\label{Distribution}
\end{figure*}

\begin{figure}
\includegraphics[angle=0,scale=1.2]{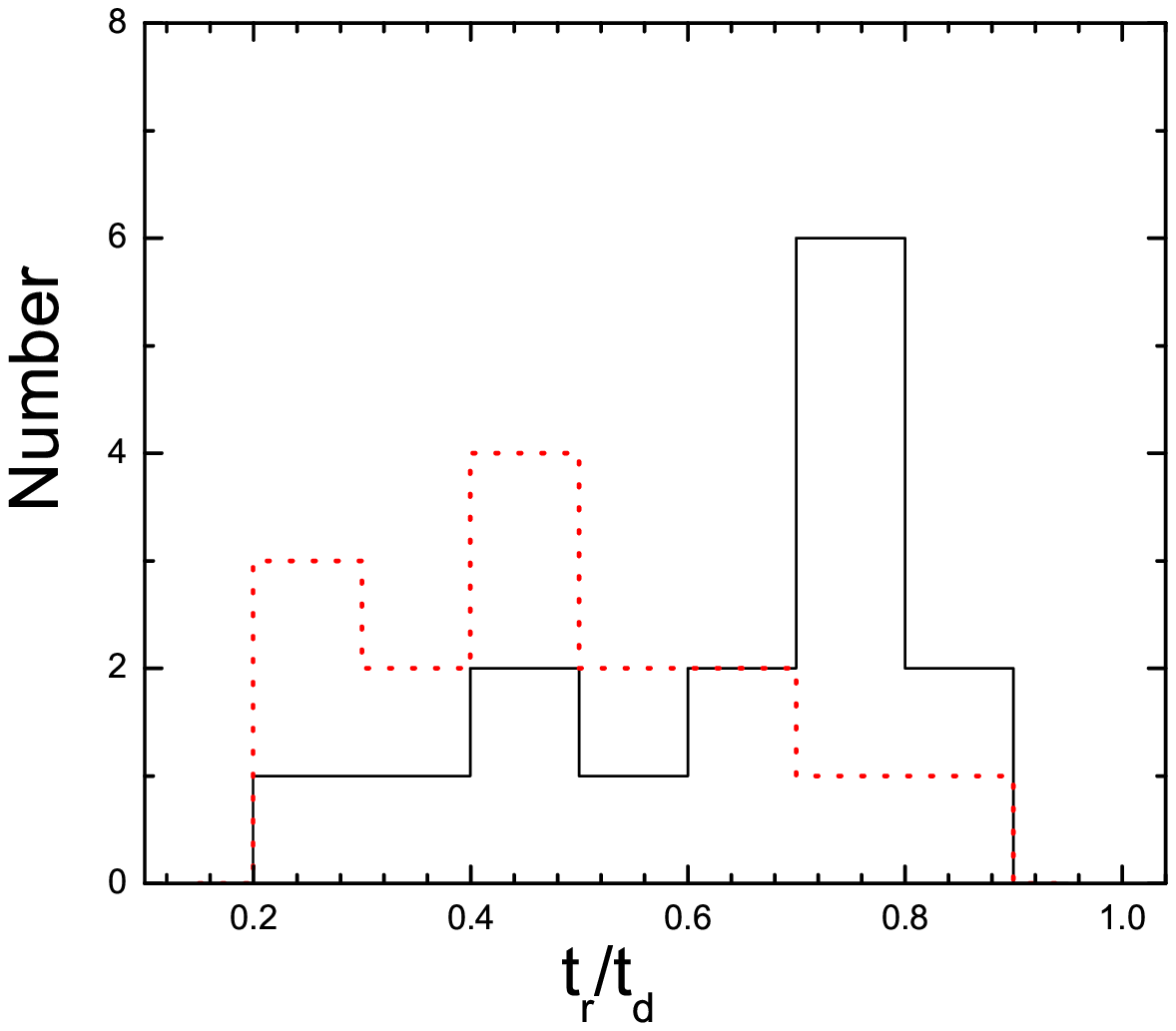}
\caption{Distributions of the ratio between the rising time scale
($t_r$) and the fall time scale ($t_d$) for the tracking pulses (the
solid line) and hard-to-soft pulses (the dotted line).}\label{trtd}
\end{figure}

\begin{figure}
\resizebox{5cm}{!}{\includegraphics{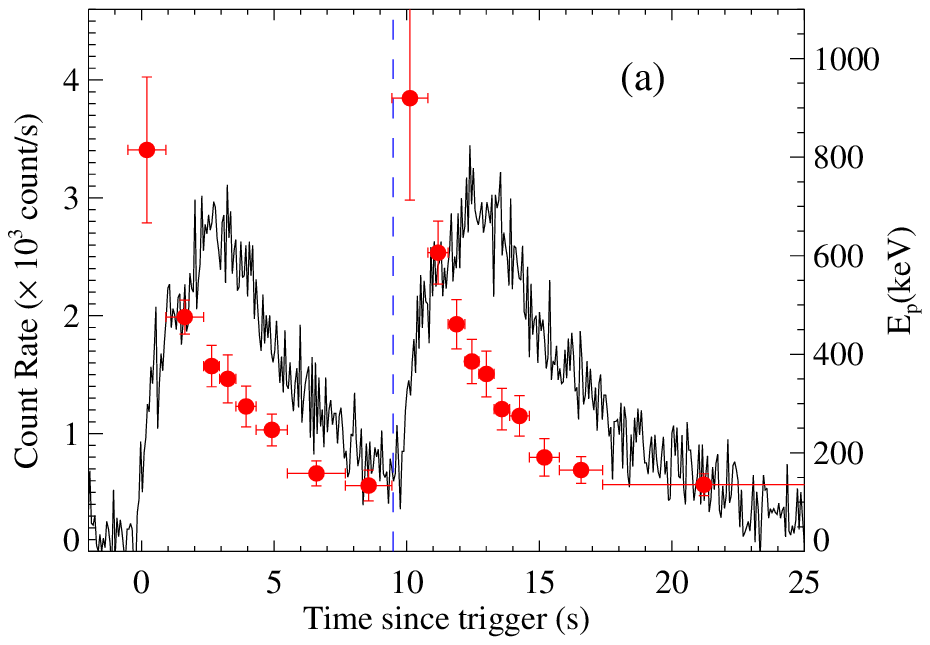}}
\resizebox{5cm}{!}{\includegraphics{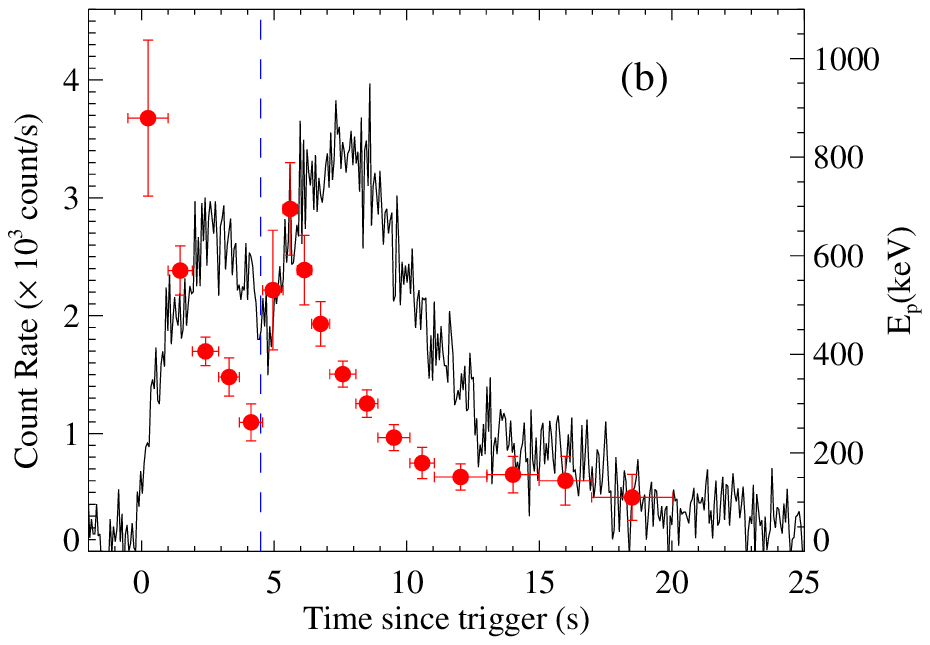}}
\resizebox{5cm}{!}{\includegraphics{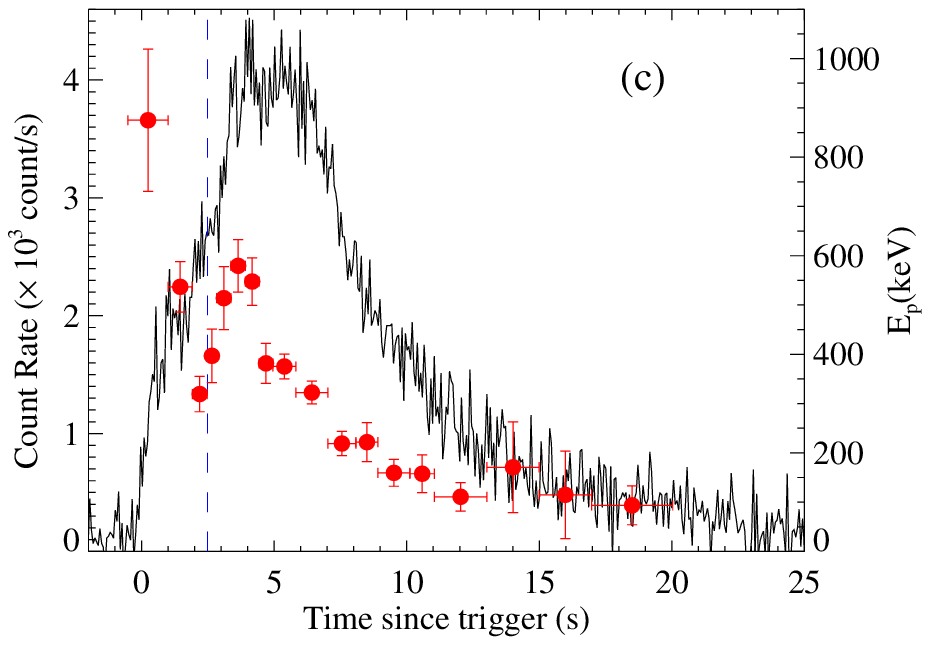}}
\resizebox{5cm}{!}{\includegraphics{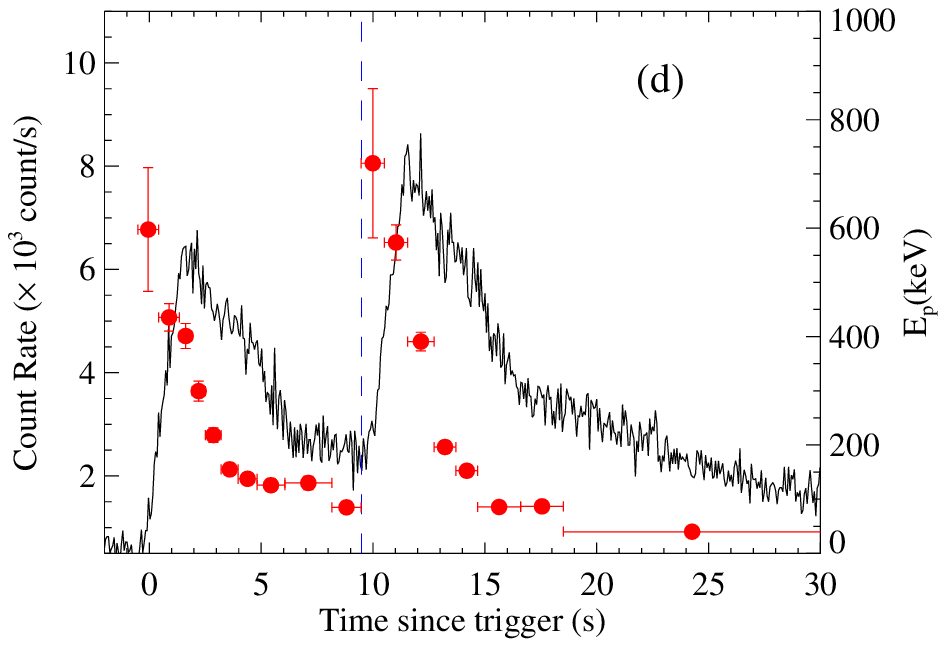}}
\resizebox{5cm}{!}{\includegraphics{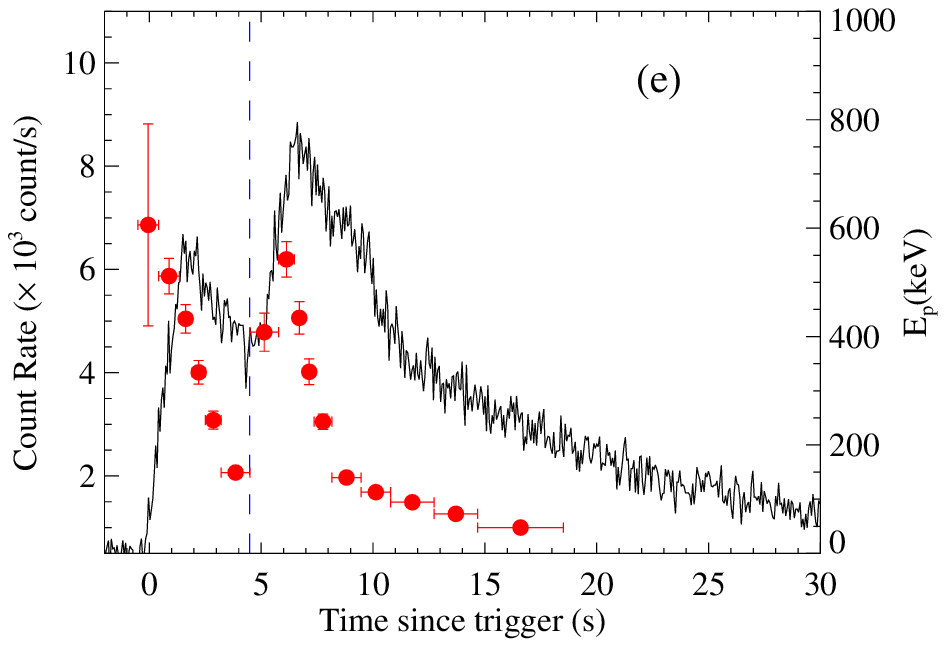}}
\resizebox{5cm}{!}{\includegraphics{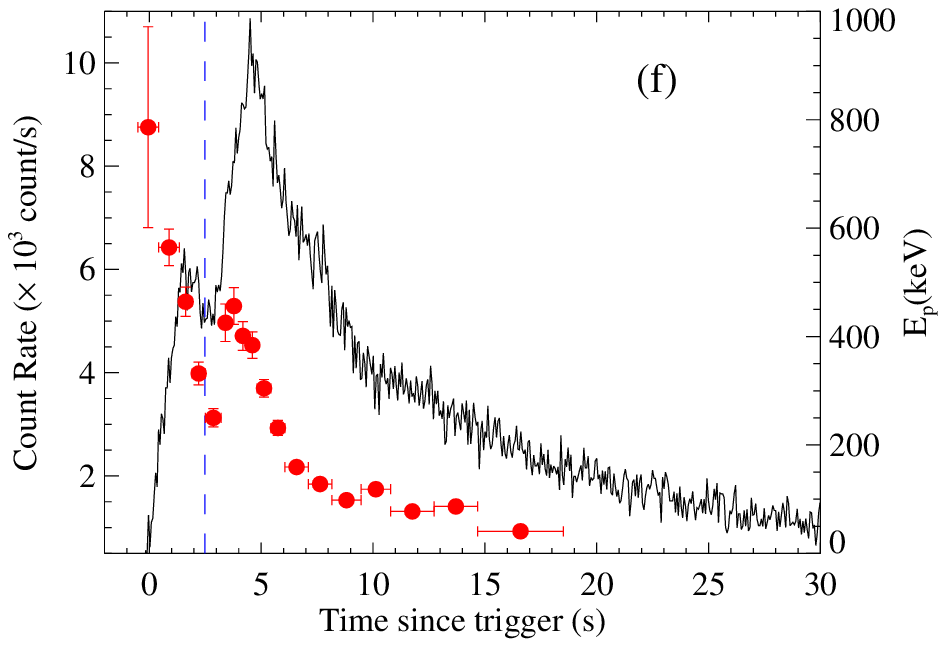}}\caption{A simulation
tests of overlapping hard-to-soft pulses for GRBs 081224 (first row)
and 100707A (second row) are taken as template. The six panels show
the $E_{\rm p}$ evolution pattern of the simulated GRB with
different pulse separations: 10 s (first column), 5 s (second
column), 3 s (third column). The vertical dash lines mark the onset
of the second pulses. The symbols are the same as Fig.
\ref{long}.}\label{mockabc}
\end{figure}

\clearpage
\begin{figure}
\centering \resizebox{10cm}{!}{\includegraphics{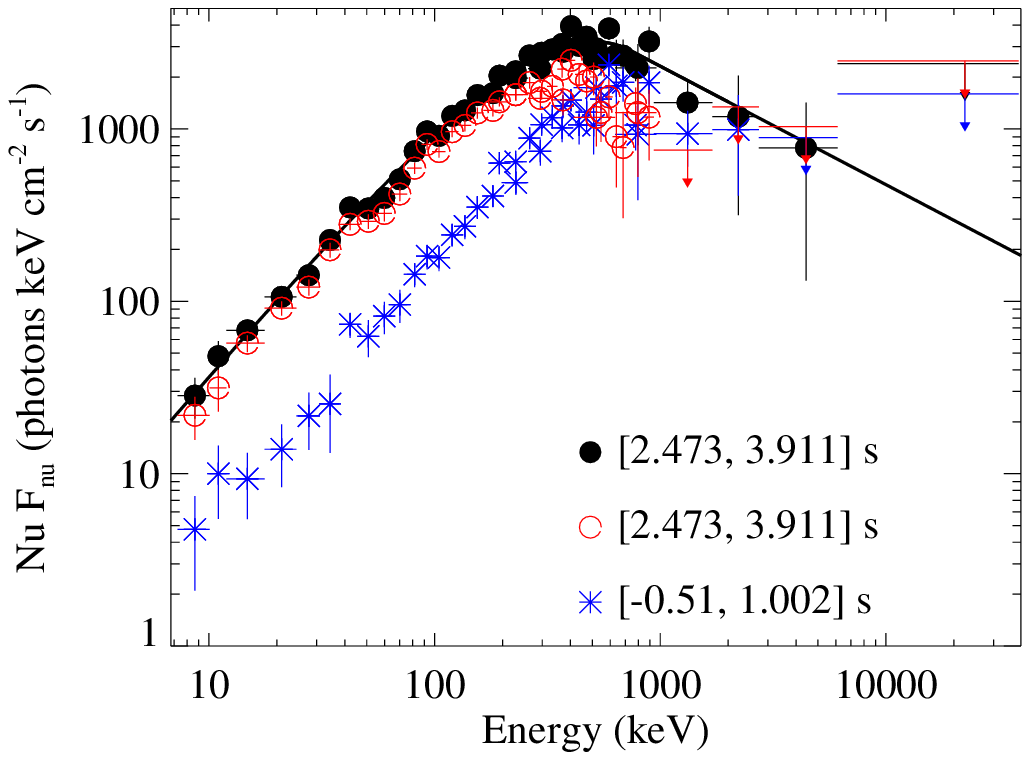}}
\caption{Illustrations of the superposition effect on the spectral
shape for the simulated GRB 081224. These three mock spectra
correspond to that of the onset time bin of the second pulse of the
mock GRB as shown in Fig. \ref{mockabc}. Every mock spectrum in the
onset time bin of the second pulse of the mock GRB ({\em solid
dots}) is roughly the superimposition of the observed GRB 081224
spectra in the two different time bins as marked in the legends.
}\label{mock_c}
\end{figure}

\begin{figure}
\centering \resizebox{14cm}{!}{\includegraphics{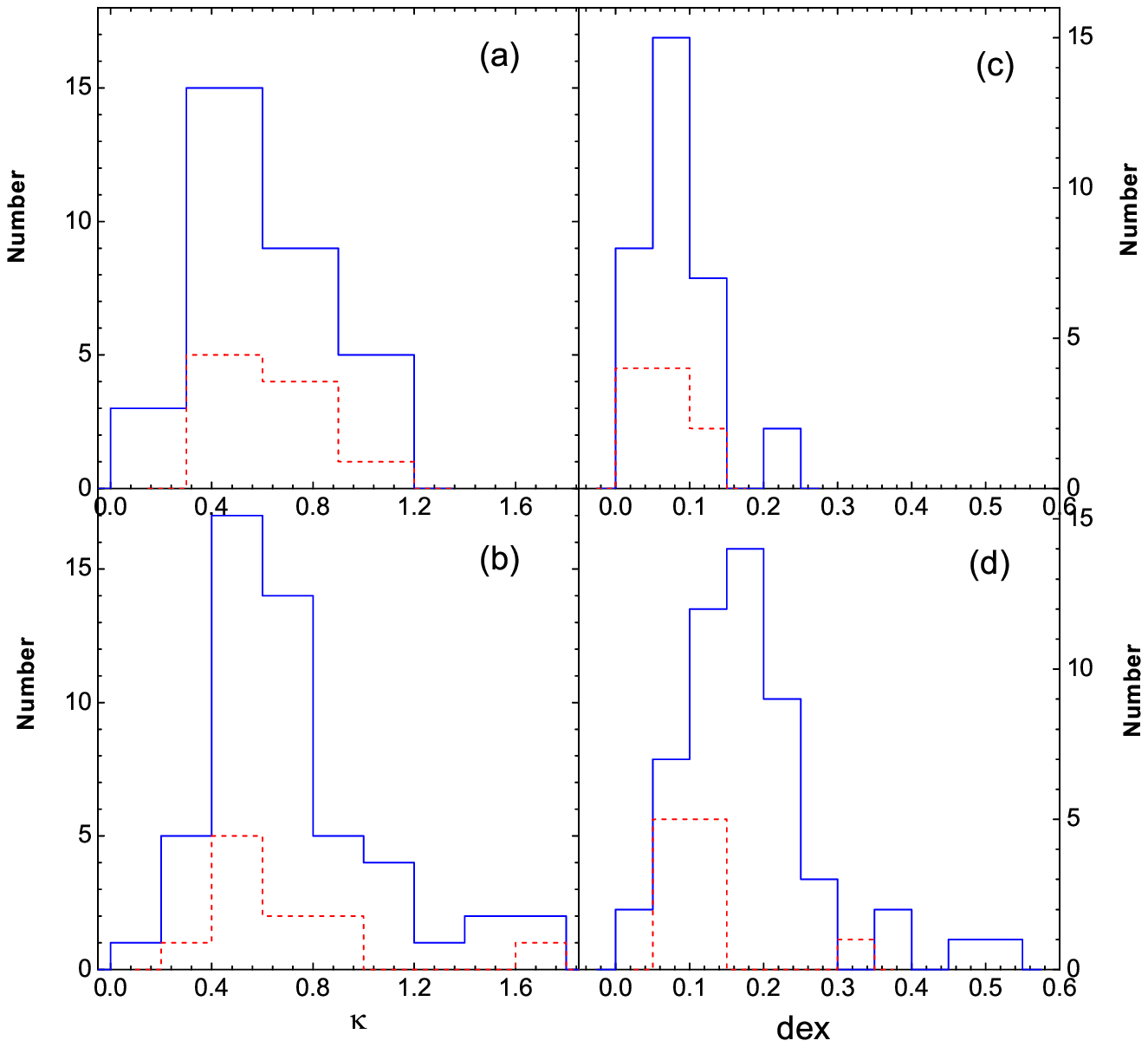}}
\caption{Distributions of the correlation power law index $\kappa$
(a, b) of the $E_{\rm p}-F$ correlation and the scatter (dex) of
data points around the best $E_{\rm p}-F$ correlation (c, d) in the
decay phase of the pulses (the first row panels) and for the entire
burst (the second row panels). The solid and dashed lines denote the
long and short GRBs in our sample, respectively.}\label{kd}
\end{figure}

\begin{figure}
\centering \resizebox{14cm}{!}{\includegraphics{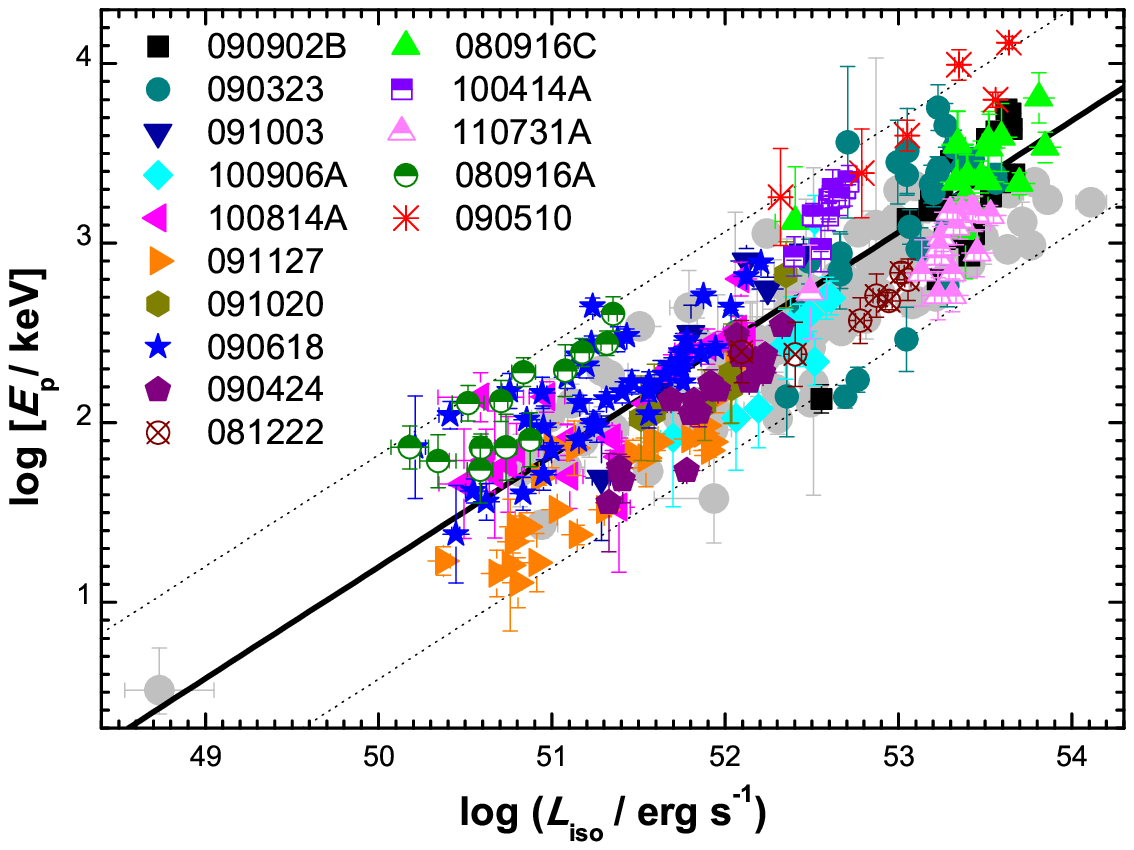}} \caption{A
comparison of the time resolved $E_{\rm p}-L_{\gamma, \rm iso}$
correlation for 15 Fermi GRBs with known redshifts in our sample
(marked in color individually) with the time integrated $E_{\rm
p}-L_{\gamma, \rm iso}$ correlation for the 101 GRBs in Yonetoku et
al. (2010) (gray filled circles). The solid line is the best fit to
the time resolved spectra, while the two dot lines represent its
$2\sigma$ dispersion around the best fit.}\label{LEp}
\end{figure}

\include{table1}

\end{document}

%% file: table1.tex
\clearpage
\begin{deluxetable}{lllllllll}
%\begin{longtable}{lllllllll}
\tablewidth{400pt}
\tabletypesize{\tiny} \tablecaption{Time-resolved Spectral fits with the Band
function for the GRBs in our sample. The fluence is in the 10-1000 keV band.} \tablenum{1}
\tablehead{\colhead{GRB}&\colhead{Slice\#}&\colhead{$t_{\rm
start}$(s)}&\colhead{$t_{\rm end}$(s)}&\colhead{$E_{\rm
p}$(keV)}&\colhead{$\alpha$}&\colhead{$\beta$}&\colhead{fluence\tablenotemark{a}}&\colhead{
$\chi_r^{2}$}
} \startdata
080825C &   1   &   -3.456  &   0.004   &   104.0   $\pm$   64.4    &   -0.71   $\pm$   1.12    &   -4.41   $\pm$   0.00    &   0.29    $\pm$   0.11    &   0.94     \\
080825C &   2   &   0.004   &   1.664   &   231.4   $\pm$   44.9    &   -0.62   $\pm$   0.15    &   -2.08   $\pm$   0.17    &   4.18    $\pm$   0.17    &   1.12     \\
080825C &   3   &   1.664   &   3.712   &   372.0   $\pm$   55.5    &   -0.83   $\pm$   0.08    &   -2.26   $\pm$   0.21    &   14.47   $\pm$   0.37    &   1.10     \\
080825C &   4   &   3.712   &   4.736   &   191.8   $\pm$   33.2    &   -0.67   $\pm$   0.15    &   -2.43   $\pm$   0.37    &   17.15   $\pm$   0.15    &   0.87     \\
080825C &   5   &   4.736   &   5.760   &   130.3   $\pm$   25.6    &   -0.34   $\pm$   0.26    &   -1.94   $\pm$   0.09    &   20.09   $\pm$   0.13    &   1.32     \\
080825C &   6   &   5.760   &   6.784   &   68.9    $\pm$   20.6    &   0.47    $\pm$   0.94    &   -1.84   $\pm$   0.09    &   21.66   $\pm$   0.11    &   0.92     \\
080825C &   7   &   6.784   &   9.856   &   152.3   $\pm$   58.4    &   -0.87   $\pm$   0.43    &   -4.41   $\pm$   0.00    &   22.97   $\pm$   0.29    &   0.93     \\
080825C &   8   &   9.856   &   12.928  &   107.5   $\pm$   31.6    &   0.02    $\pm$   0.54    &   -1.76   $\pm$   0.07    &   29.15   $\pm$   0.33    &   0.87     \\
\enddata
\tablenotetext{a}{In units of $10^{-6}$ erg cm$^{-2}$}
\end{deluxetable}